\documentclass[11pt]{article}

\usepackage{amsfonts}
\usepackage{bm}
\usepackage{amsmath}
\usepackage{epsfig}
\usepackage{amsmath,amsfonts,amssymb}
\usepackage[small,bf,hang]{caption}

\usepackage[usenames,dvipsnames]{color}
\usepackage{amssymb}
\usepackage[parfill]{parskip}
\usepackage{breqn}

\renewcommand{\bar}[1]{\overline{#1}}

\newcommand{\bmat}{\left(\begin{array}}
\newcommand{\emat}{\end{array}\right)}

\def\gtrsim{\mathrel{\raise.3ex\hbox{$>$\kern-.75em\lower1ex\hbox{$\sim$}}}}

\def\vec#1{\textbf{\emph{#1}}}

\def\-{\hphantom{-}}

\def\s2{\frac{1}{\sqrt2}}

\def\dim{{\rm dim \,}}

\def\mg{m_{3/2}}
\def\mg2{m^2_{3/2}}

\def\Dsl{\,\raise.15ex\hbox{/}\mkern-13.5mu D} 

\def\be{\begin{equation}}
\def\ee{\end{equation}}
\def\bea{\begin{eqnarray}}
\def\eea{\end{eqnarray}}

\newcommand{\nn}{\nonumber}

\begin{document}

\pagestyle{plain}

\makeatletter
\@addtoreset{equation}{section}
\makeatother
\renewcommand{\theequation}{\thesection.\arabic{equation}}
\pagestyle{empty}
\begin{center}
\ \

\vskip 2cm

\LARGE{\LARGE\bf Second order higher-derivative corrections in\\ Double Field Theory \\[10mm]}
\vskip 0.5cm
\large{Eric Lescano and Diego Marqu\'es
 \\[6mm]}
{\small\it Instituto de Astronom\'ia y F\'isica del Espacio \\ IAFE-CONICET-UBA \\
Buenos Aires, Argentina. \\ [.3 cm]}

{\small \verb"{elescano, diegomarques}@iafe.uba.ar"}\\[2cm]

\small{\bf Abstract} \\[0.5cm]\end{center}

{\small HSZ Double Field Theory is a higher-derivative theory of gravity with exact and manifest T-duality symmetry. The first order corrections in the massless sector were shown to be governed solely by Chern-Simons deformations of the three-form field strength. We compute the full action with up to six derivatives ${\cal O} (\alpha'{}^2)$ for the universal sector containing the metric, two-form and dilaton fields. The Green-Schwarz transformation of the two-form field remains uncorrected to second order. In addition to the expected Chern-Simons-squared and Riemann-cubed terms the theory contains a cubic Gauss-Bonnet interaction, plus other six-derivative unambiguous terms involving the three-form field strength whose presence indicates that the theory must contain further higher-derivative corrections.
}

\newpage
\setcounter{page}{1}
\pagestyle{plain}
\renewcommand{\thefootnote}{\arabic{footnote}}
\setcounter{footnote}{0}

\tableofcontents

\section{Introduction} \label{SEC:Intro}

In recent years there has been renewed interest in understanding how dualities constrain higher-derivative interactions in string effective actions. The most convenient frameworks to address this challenge are those in which dualities appear as manifest symmetries. Double Field Theory (DFT) \cite{Siegel:1993xq}-\cite{Hull:2009mi} formulates the two-derivative universal massless sector of string theory in a way that T-duality symmetry can be anticipated before dimensional reduction (for reviews see \cite{reviews}). In this context, the question of how T-duality constrains higher-derivative terms simply translates into the question of what the most general consistent higher-derivative deformations of DFT are. Recent works discussing the duality structure of the first order corrections to the heterotic string are \cite{Bedoya:2014pma}. In \cite{Marques:2015vua}, a two parameter family of first order consistent deformations of DFT was introduced, containing in a unified framework the corrections to bosonic \cite{Metsaev:1987zx} and heterotic strings \cite{Metsaev:1987zx,Bergshoeff:1988nn}.

Higher-derivative theories of gravity are always defined up to field redefinitions. There are two different set of fields that are important when dealing with duality invariant theories. On the one hand, there are gauge and diffeomorphism covariant fields, whose duality transformations receive higher-derivative corrections. On the other hand, there are duality covariant fields, whose T-duality transformations are exactly those given by the Buscher rules or more generally the action of $O(D,D,\mathbb{R})$, without higher-derivative corrections. The latter are non-covariant with respect to gauge and diffeomorphism transformations: to lowest order in a derivative expansion they transform tensorially, but the gauge transformations are deformed by higher-derivative corrections. Both set of fields turn out to be connected through higher-derivative field redefinitions that are neither gauge nor duality covariant.

The standard way of writing a higher-derivative theory of gravity is in terms of gauge covariant fields, such that gauge and diffeomorphism invariance is manifest. If the action turns out to be duality invariant, then the duality symmetries are hidden in this formulation. On the contrary, the fields that are preferred by duality invariant frameworks like DFT are the duality covariant fields, so the actions obtained from these constructions are manifestly duality invariant but the gauge symmetries are hidden.  When the action is written in terms of gauge covariant fields, duality transformations mix terms with different orders in derivatives. Instead, when the action is written in terms of duality covariant fields, duality invariance holds order by order, but the different orders are mixed by gauge transformations. Presumably, a duality and gauge invariant higher-derivative theory of gravity will require an infinite number of interactions with increasing number of derivatives. If such an expansion were truncated to a given order in derivatives, the action would fail to be either gauge or duality invariant, depending on which set of fields is chosen to express it.

When dealing with duality invariant frameworks such as DFT, the preferred set of fields are the duality covariant ones. If the theory were truncated to a given order in derivatives, the resulting action would be exactly duality invariant, but the gauge transformations would only be approximate symmetries to that order. If instead one first rewrites the action in terms of the gauge covariant fields, and then truncates to a given order in derivatives, the truncated action would be exactly gauge invariant, but the duality transformations would only be an approximate symmetry to that order.

Some years ago, O. Hohm, W. Siegel and B. Zwiebach introduced a gauge invariant higher-derivative theory of gravity that is both exactly and manifestly T-duality invariant \cite{Hohm:2013jaa}, henceforth called HSZ theory. It was built from a duality symmetric chiral CFT, and does not correspond to a standard string theory. The massless spectrum includes a metric, an antisymmetric tensor and a dilaton, and the full theory contains in addition two massive spin-two ghosts and massive scalars \cite{Hohm:2016lim}. The spectrum is reminiscent but qualitatively different to that in the chiral string theory investigated in \cite{Huang:2016bdd}. The truncated two-derivative theory exactly matches standard DFT \cite{Hohm:2013jaa}, so the theory corresponds to a higher derivative deformation of DFT. The deformations occur at different levels. While DFT depends on a generalized metric ${\cal H}_{M N}$ constrained to be an element of  the duality group, the fundamental field in HSZ theory is an unconstrained double metric ${\cal M}_{M N}$. To lowest order both metrics coincide, but the double metric contains extra degrees of freedom associated to massive excitations \cite{Hohm:2016lim}. The relation between the two metrics was discussed in \cite{Hohm:2015mka}. To lowest order the gauge transformations are given by the standard DFT generalized Lie derivative, but in HSZ theory they receive higher-derivative deformations that in turn induce higher-derivative terms in the action. Remarkably, the latter happens to be cubic in the double metric, and due to the strong constraint (to be defined latter) it contains at most six derivatives.

The massive modes in the full theory can be integrated out, ending with a low-energy effective theory for the generalized metric ${\cal H}_{MN}$ and generalized dilaton $d$ that contains an infinite series of higher-derivative terms \cite{Hohm:2016lim}-\cite{Hohm:2015mka}. It is then of interest to elucidate what these corrections look like in terms of the standard gauge covariant massless fields, namely the metric $g_{\mu \nu}$, two-form $B_{\mu \nu}$ and dilaton $\phi$, in a manifestly diffeomorphism and gauge invariant form expanded in powers of $\alpha'$ \footnote{Note that since this is not a standard string theory, $\alpha'$ is in principle not related to the tension of a string, but simply a dimensional parameter used to balance units. Throughout the paper we will set for convenience $\alpha' = 1$, but still refer to the different orders in the action as containing different powers of $\alpha'$. As is standard, terms with $2(i+1)$ derivatives in the action are of order ${\cal O}(\alpha'{}^{i})$. }. To first order in $\alpha'$, the deformed gauge transformations were shown to reproduce the Green-Schwarz \cite{Green:1984sg} transformation of the two-form
\be
\delta B_{\mu \nu} = \partial_{[\mu|} \partial_\rho \xi^\sigma \Gamma_{|\nu]\sigma}^\rho \ , \label{GS}
\ee
so the first order corrections to the action are governed by Chern-Simons terms $\Omega_{\mu \nu \rho}$ that correct the three-form field strength of the Kalb-Ramond field \cite{Hohm:2014eba}
 \be
\widehat H_{\mu \nu \rho} = 3\, \partial_{[\mu} B_{\nu \rho]} + 3 \Omega_{\mu \nu \rho} \ , \ \ \ \ \Omega_{\mu \nu \rho} = \Gamma_{[\mu| \sigma}^\delta \partial_{|\nu} \Gamma_{\rho] \delta}^\sigma + \frac 2 3
\Gamma_{[\mu | \sigma}^\delta \Gamma_{|\nu| \lambda}^\sigma \Gamma_{|\rho] \delta}^\lambda \ . \label{Hhat}
\ee
On shell three-point functions with vanishing dilaton were computed to all orders in $\alpha'$ in \cite{Naseer:2016izx}, leading to the conclusion that to second order the theory contains a Riemann-cubed interaction. The state of the art is that to second order, the gauge invariant action that reproduces the on-shell cubic amplitudes of HSZ theory is the following
\be
S = \int d^Dx \sqrt{-g} e^{-2 \phi}\left(R + 4 (\nabla \phi)^2  - \frac 1 {12} \widehat H^2 - \frac 1 {48}\, R_{\mu \nu}{}^{\alpha \beta} R_{\alpha \beta}{}^{\rho \sigma} R_{\rho \sigma}{}^{\mu \nu}\right) \ . \label{UsmanBartonAction}
\ee
The coefficient of the Riemann-cubed term is minus the coefficient of the same term in the bosonic string effective action. There could be extra second order terms in this action that do not contribute to the three-point functions, such as a cubic Gauss-Bonnet term that is also present in the bosonic string \cite{Metsaev:1986yb}.

In this work we compute the full second order action in terms of the standard gauge covariant fields in a form that is manifestly invariant under diffeomorphisms and the gauge transformations of the antisymmetric tensor. The procedure is to first find the functional dependence of the double metric in terms of the generalized metric to second order. Then, to find the dependence of the components of the generalized metric on (derivatives of) the standard fields. This is, to find the field redefinitions that relate the duality covariant with the gauge covariant fields. Finally, one plugs this information in the HSZ action and rewrites the result in a form that is manifestly covariant under gauge symmetries. This is our main result, and we display the final form of the full second order action in Subsection \ref{SUBSEC::FullAction}. Let us summarize some of our main findings:
\begin{itemize}
\item The Green-Schwarz transformation of the two-form (\ref{GS}) remains uncorrected to second order in $\alpha'$.
\item The first order corrections to the action are exactly those present in (\ref{UsmanBartonAction}). These were anticipated in \cite{Hohm:2014eba} and we find that there are no extra terms such as total derivatives nor interactions that could have been allowed by gauge symmetries and required by duality. This is also consistent with the results in \cite{Marques:2015vua}.
\item Up to covariant field redefinitions and boundary terms, ${\cal O}(\alpha'{}^i)$ corrections with $i$ even/odd contain even/odd powers of the two-form field. This was known to be true to first order, and explains why Riemann-squared interactions are forbidden. It also implies that the second order corrections can only contain even powers of the two-form.
\item The terms in the full second order action that contribute to on shell three-point functions with vanishing dilaton are in precise agreement with the findings in \cite{Naseer:2016izx}, and exactly coincide with (\ref{UsmanBartonAction}).
\item When the dilaton and three-form field strength are taken to vanish, the second order Lagrangian (up to field redefinitions and boundary terms) is given by
\be
L^{(2)} =  - \frac{3}{4}\, {\Omega}^{\mu \nu \rho} {\Omega}_{\mu \nu \rho} - \frac 1 {48}\, R_{\mu \nu}{}^{\alpha \beta} R_{\alpha \beta}{}^{\rho \sigma} R_{\rho \sigma}{}^{\mu \nu} - \frac 1 {24} G_3  \ ,
\ee
where $\Omega_{\mu \nu \rho}$ is the Chern-Simons three-form defined in (\ref{Hhat}), and $G_3$ is the cubic Gauss-Bonnet interaction
\be
G_3 = R_{\mu \nu}{}^{\alpha \beta} R_{\alpha \beta}{}^{\rho \sigma} R_{\rho \sigma}{}^{\mu \nu} - 2\, R^{\mu \nu \alpha \beta} R_{\nu \lambda \beta \gamma} R^\lambda{}_\mu{}^\gamma{}_\alpha \ .
\ee
Remarkably, as with the Riemann-cubed term, the coefficient of this term is also minus the coefficient of the same term in the bosonic string.
\item The more general case with vanishing dilaton $\phi = 0$ but non-vanishing metric and two-form is (again, up to field redefinition and integration by parts) given by
\bea
L^{(2)} \!\!\!&=& \!\!\!  - \frac{3}{4}\, {\Omega}^{\mu \nu \rho} {\Omega}_{\mu \nu \rho} - \frac 1 {48}\, R_{\mu \nu}{}^{\alpha \beta} R_{\alpha \beta}{}^{\rho \sigma} R_{\rho \sigma}{}^{\mu \nu} - \frac 1 {24} G_3
+ \frac{1}{32} {\widehat{H}}^2_{\rho \sigma} R^{\rho \alpha \beta \gamma} R^{\sigma}{}_{\alpha \beta \gamma} \nn \\
&& - \frac{1}{64} R_{\alpha \beta \mu \rho} R_{\nu}{}^{\rho} {\widehat{H}}_{\gamma}{}^{\alpha \beta} {\widehat{H}}^{\gamma \mu \nu}
+ \frac{1}{128} R_{\alpha \beta \sigma \mu} R_{\gamma \nu} {\widehat{H}}^{\alpha \beta \gamma} {\widehat{H}}^{\sigma \mu \nu} - \frac{1}{128} R_{\alpha \beta \gamma \sigma} {\widehat{H}}^{2 \alpha \gamma} {\widehat{H}}^{2 \beta \sigma}  \\ && - \frac{1}{768} R_{\alpha \beta} {\widehat{H}}^{2 \alpha}{}_{\mu}  {\widehat{H}}^{2 \beta \mu} - \frac{1}{256} R_{\alpha \beta} {\widehat{H}}^{\alpha}{}_{\rho \gamma} {\widehat{H}}^{\beta \rho}{}_{\nu} {\widehat{H}}^{2 \gamma \nu}
- \frac{1}{128} \nabla_{\sigma} {\widehat{H}}_{\alpha \mu \nu} \nabla^{\sigma} {\widehat{H}}_{\beta}{}^{\mu \nu} {\widehat{H}}^{2 \alpha \beta}\ , \nn
\eea
where we have defined $\widehat H^2_{\mu \nu} = \widehat H_{\mu \rho \sigma} \widehat H_\nu{}^{\rho \sigma}$. Although further field redefinitions can be used to eliminate $\widehat H_{\mu \nu}^2$ from the action, the dependence on $\widehat H_{\mu \nu \rho}$ can not be totaly removed. This implies that the full action must at least contain up to ${\cal O}(\alpha'{}^{4})$ terms,  confirming the expectation that the second order action cannot by itself be exactly duality invariant.
\end{itemize}

The paper is organized as follows. We begin with a review of HSZ theory in Section \ref{SEC:HSZ}. In Subsection \ref{SUBSEC::DoubleMetric} we introduce the gauge transformations and action for the double metric. In Subsection \ref{SUBSEC::GeneralizedMetric} we review how to relate the double metric with the generalized metric. In Subsection \ref{SUBSEC:Parameterizations} we show the general procedure that has to be followed in order to relate the duality covariant components of the generalized metric with the gauge covariant fields of supergravity. Subsection \ref{SUBSEC::Parity} contains a discussion on the $Z_2$-parity ($B_{\mu \nu} \to - B_{\mu \nu}$) behavior of the different orders. Subsections \ref{SUBSEC:First} and \ref{SUBSEC::Second} are dedicated to discuss the state of the art regarding the first and second order corrections in the massless sector of HSZ theory. In Section \ref{SEC:Full} we present the main result of this paper, namely the full action to second order for the massless fields, and discuss some relevant interactions. We conclude in Section \ref{SEC::Conclusions} and include a number of Appendices to present some intermediate relevant results.

\section{Review of HSZ Double Field Theory} \label{SEC:HSZ}

In this section we introduce HSZ theory \cite{Hohm:2013jaa}, its field content, gauge transformations and action. We then show how to make contact with the generalized metric and how to relate its components with the standard fields in supergravity. We briefly review the state of the art on what is known about the first orders in the derivative expansion of the massless sector.

\subsection{Double metric, symmetries and action} \label{SUBSEC::DoubleMetric}

HSZ theory is defined on a double space of dimension $2D$, equipped with an $O(D,D,\mathbb{R})$ constant and symmetric invariant metric $\eta_{M N}$, which raises and lowers the duality indices $M,N,=1,\dots,2D$. The derivatives are denoted $\partial_M$, and all fields and gauge parameters are restricted by the strong constraint
\be
\partial_M \dots \partial^M \dots = \partial_M \partial^M \dots = 0 \ . \label{strongconstraint}
\ee

The fundamental fields in HSZ theory are the double metric ${\cal M}_{M N}$ and the generalized dilaton $d$. The double metric is symmetric, but otherwise unconstrained. Under infinitesimal generalized diffeomorphisms parameterized by a generalized vector $\xi^M$, the generalized dilaton transforms as usual
\be
\delta e^{-2d} = \partial_P \left( \xi^P e^{-2d}\right) \ \ \ \ \Leftrightarrow \ \ \ \ \delta d = \xi^P \partial_P d - \frac 1 2  \partial_P \xi^P \ , \label{gaugetransfdilaton}
\ee
but the double metric receives higher-derivative corrections
\be
\delta {\cal M}_{M N} = \widehat {\cal L}_\xi \, {\cal M}_{M N} + {\cal J}^{(1)}({\cal M})_{M N}  + {\cal J}^{(2)} ({\cal M})_{M N} \ . \label{gaugetransfdoublemet}
\ee
Here, the first term is the standard generalized Lie derivative
\be
\widehat {\cal L}_\xi \, {\cal M}_{M N} = \xi^{P} \partial_P {\cal M}_{M N} + K_{M}{}^P {\cal M}_{P N} + K_N{}^P {\cal M}_{M P} \ , \label{generalizedLiederivative}
\ee
where $K_{M N} = 2 \partial_{[M} \xi_{N]}$ and the other terms ${\cal J}^{(i)}$ constitute corrections with $2i + 1$ derivatives\footnote{We use the following convention for derivatives $\partial_{M_1 M_2 \dots M_k}  = \partial_{M_1} \partial_{M_2} \dots \partial_{M_k}$.}
\bea
{\cal J}^{(1)}({\cal M})_{M N} &=& - \frac 1 2 \partial_M {\cal M}^{P Q} \partial_P K_{Q N} - \partial_P {\cal M}_{Q M} \partial_N K^{Q P} + (M \leftrightarrow N) \\
{\cal J}^{(2)} ({\cal M})_{M N} &=& - \frac 1 4  \partial_{M K}  {\cal M}^{P Q} \partial_{N P} K_Q{}^K + (M \leftrightarrow N) \ .
\eea

Due to the strong constraint, the composition of gauge transformations closes
\be
\left[\delta_{\xi_1}, \, \delta_{\xi_2}\right] = \delta_{[\xi_1 ,\, \xi_2]_{\rm C'}} \ ,
\ee
with respect to a deformed C'-bracket
\be
[\xi_1 ,\, \xi_2]_{\rm C'}^M = [\xi_1 ,\, \xi_2]_{\rm C}^M + \partial_P \xi^Q_{[1} \partial^M \partial_Q\xi_{2]}^P \ ,
\ee
with the standard C-bracket defined as
\be
[\xi_1 ,\, \xi_2]_{\rm C}^M = 2 \xi_{[1}^P \partial_P \xi_{2]}^M - \xi_{[1}^P \partial^M \xi_{2]P}\ .
\ee

The deformed gauge transformations (\ref{gaugetransfdoublemet}) fail to satisfy the Leibnitz rule with respect to the standard product. Instead, one defines a star-product $\star$\footnote{Or for different rank-two tensors \bea
T_1 \star T_2 &=& \frac 1 2 (T_1 + T_2) \star (T_1 + T_2) - \frac 1 2\, T_1 \star T_1 - \frac 1 2\, T_2 \star T_2\ .
\eea}
\bea
(T\ \star\ T)_{M N} &=& {T}_{M P} {T}_{N}\,^{P} - \frac{1}{4}\, {\partial}_{M}{{T}_{P Q}}\,  {\partial}_{N}{{T}^{P Q}}\,  + 2\, {\partial}_{N}{{T}_{P}\,^{Q}}\,  {\partial}_{Q}{{T}_{M}\,^{P}}\,  - {\partial}_{P}{{T}_{M}\,^{Q}}\,  {\partial}_{Q}{{T}_{N}\,^{P}}\, \nn\\
 && +\,  \frac{1}{2}\, {\partial}_{M P}{{T}_{Q}\,^{I}}\,  {\partial}_{N I}{{T}^{Q P}}\,  - {\partial}_{N P}{{T}^{Q I}}\,  {\partial}_{Q I}{{T}_{M}\,^{P}}\,  - \frac{1}{8}\, {\partial}_{M P Q}{{T}^{I J}}\,  {\partial}_{N I J}{{T}^{P Q}}\, \nn \\
  &&  +\, \frac{1}{2}\, {T}^{P Q} {\partial}_{P Q}{{T}_{M N}}\,  + \frac{1}{2}\, {\partial}_{M}{{T}_{P}\,^{Q}}\,  {\partial}_{N Q}{{G}^{P}}\,  - \frac{1}{2}\, {\partial}_{N}{{T}^{P Q}}\,  {\partial}_{P Q}{{G}_{M}}\, \nn \\ && -\, \frac{1}{4}\, {\partial}_{M P}{{T}^{Q I}}\,  {\partial}_{N Q I}{{G}^{P}}\,  + \frac{1}{2}\, {G}^{P} {\partial}_{P}{{T}_{M N}}\, -\, {\partial}_{P}{{T}_{M}\,^{Q}}\,  {\partial}_{N Q}{{G}^{P}}\,\nn \\ &&+\, {T}_{M P} {\partial}_{N}{{G}^{P}}\,  - {T}_{M}\,^{P} {\partial}_{P}{{G}_{N}}\, + (M \leftrightarrow N) \vphantom{\frac 1 2} \ ,
\eea
where
\bea
G^M(T,d) &=& {\partial}_{N}{{T}^{M N}}\,  - 2\, {T}^{M N} {\partial}_{N}{d}\,  +\, T^{N P} \partial_{N P}{}^M d\label{G} \\
 &&- \frac 1 2 \partial^M \left[\partial_{N P} T^{N P} -4\, \partial_{N} T^{N P} \partial_P d - 2\, T^{N P} \left(\partial_{N P} d - 2 \partial_N d\, \partial_P d\right) \right] \ .\nn
\eea
The star product now does satisfy the Leibnitz rule with respect to deformed generalized diffeomorphisms
\be
\delta (T_1 \star T_2) = \delta T_1 \star T_2 + T_1 \star \delta T_2 \ ,
\ee
where $(T_1 \star T_2)_{M N}$, $T_{1 M N}$ and $T_{2 M N}$ transform in the same way as the double metric in (\ref{gaugetransfdoublemet}).

The standard inner product that follows from contractions with the duality invariant metric also turns out not to be covariant under the deformed gauge transformations. However, the following quantity
\bea
\langle T_1 | T_2 \rangle &=& \frac{1}{2}\, {T_1}_{P Q} {T_2}^{P Q} - {\partial}_{P}{{T_1}_{N}\,^{Q}}\,  {\partial}_{Q}{{T_2}^{N P}}\,  + \frac{1}{4}\, {\partial}_{M N}{{T_1}^{P Q}}\,  {\partial}_{P Q}{{T_2}^{M N}}\,  \\ && - \frac{3}{2}\, {G(T_1)}_{N} {G(T_2)}^{N} + \frac{3}{2}\, {\partial}_{M}{{G(T_1)}^{N}}\,  {\partial}_{N}{{G(T_2)}^{M}}\,  - \frac{3}{2}\, {T_2}_{P}\,^{N} {\partial}_{N}{{G(T_1)}^{P}}\, \nn \\ && - \frac{3}{2}\, {T_1}_{P}\,^{N} {\partial}_{N}{{G(T_2)}^{P}}\, + \frac{3}{4}\, {\partial}_{M}{{T_2}^{N P}}\,  {\partial}_{N P}{{G(T_1)}^{M}}\,  + \frac{3}{4}\, {\partial}_{M}{{T_1}^{N P}}\,  {\partial}_{N P}{{G(T_2)}^{M}}\ , \nn
\eea
transforms as a scalar under the deformed gauge transformations due to the strong constraint
\be
\delta \langle T_1 | T_2 \rangle = \langle \delta T_1 | T_2 \rangle + \langle T_1 | \delta T_2 \rangle = \xi^P \partial_P \langle T_1 | T_2 \rangle \ .
\ee

The action of HSZ theory is given by
\be
S = \int d^{2D}X \, e^{-2d}\, L \ , \ \ \ \ \ L = \bigg\langle {\cal M}\, \bigg| \, \eta - {\frac 1 6} {\cal M} \star {\cal M} \bigg\rangle \ .\label{ActionDoubleMetric}
 \ee
It is gauge invariant by construction, and was fixed to match the equations of motion that follow from the chiral CFT introduced in \cite{Hohm:2013jaa}
\be
({\cal M} \star {\cal M})_{M N} = 2 \eta_{M N}\ .
\ee
Since both the star and inner products are linear in each argument, the action turns out to be at most cubic in powers of the double metric. As a result, the strong constraint restricts the Lagrangian to contain no more than six derivatives. The full explicit form of the action is given in Appendix \ref{APP::DoubleMetricAction}.

\subsection{Generalized metric and massive fields} \label{SUBSEC::GeneralizedMetric}

Consider the following decomposition of the double metric \cite{Hohm:2015mka}
\be
{\cal M}_{M N} = {\cal H}_{M N} + F_{M N} \ ,\label{DecompMHF}
\ee
where ${\cal H}_{M N}$ is the generalized metric of double field theory, and $F_{M N}$ is a higher order contribution. The latter corresponds to massive fields \cite{Hohm:2016lim,Hohm:2015mka}, which can however be integrated out by replacing them with their equations of motion. We will be interested only in the massless spectrum of the theory, so we will assume that the integration has already been done, and consider the $F_{M N}$ as depending on the generalized metric and dilaton, i.e. $F = F({\cal H} , d)$.

It is sometimes convenient to use matrix notation, identifying the position of indices as follows
\be
\eta = \eta_{\bullet\bullet} \ , \ \ \ \eta^{-1} = \eta^{\bullet\bullet} \ , \ \ \ {\cal H} = {\cal H}_{\bullet\bullet} \ , \ \ \ {\cal H}^{-1} = {\cal H}^{\bullet\bullet}\ .
\ee
The generalized metric is symmetric and restricted by the constraint
\be
{\cal H} \eta^{-1} {\cal H} = \eta \ \ \ \Leftrightarrow \ \ \ \eta^{-1} {\cal H} \eta^{-1} = {\cal H}^{-1} \ \ \ \Leftrightarrow \ \ \ ({\cal H} \eta^{-1})^2 =  1\ .
\ee
As a consequence of this constraint, it is possible to define projectors
\be
P = \frac 1 2 (1 - {\cal H} \eta^{-1}) = P_{\bullet}{}^{\bullet} \ , \ \ \ \ \bar P = \frac 1 2 (1 + {\cal H} \eta^{-1}) = \bar P_{\bullet}{}^{\bullet} \ ,
\ee
which satisfy
\be
P + \bar P = 1 \ , \ \ \ \ P^2 = P \ , \ \ \ \ \bar P^2 = \bar P \ , \ \ \ \ P \bar P = \bar P P = 0 \ .
\ee

One can then define the following projections on a generic tensor $A = [A] + \{A\} = A_{\bullet \bullet}$
\be
[A] = P A \bar P^T + \bar P A P^T \ , \ \ \ \ \{A\} = P A P^T + \bar P A \bar P^T \ ,
\ee
such that
\be
[[A]] = [A] \ , \ \ \ \{\{A\}\} = \{A\} \ , \ \ \ [\{A\}] = \{[A]\} = 0 \ .
\ee
It is easy to check that the generalized metric and its variations satisfy
\be
{\cal H} = \{ {\cal H} \} \ , \ \ \ \delta {\cal H} = [\delta {\cal H}] \ .
\ee
On the other hand, it was shown in \cite{Hohm:2015mka} that the decomposition (\ref{DecompMHF}) is the most general ansatz for the double metric, provided the fields $F$ are constrained as
\be
F = \{F\} \ .
 \ee

We can now consider an expansion of the field $F = F_1 + F_2 + \dots$, where the subindex labels the order of the field. This means that when thought of as depending on the generalized metric and dilaton, $F_i$ carries $2i$ derivatives explicitly. Plugging the expansion in the gauge transformation (\ref{gaugetransfdoublemet}) one obtains on the one hand
\bea
\delta {\cal M} &=& \delta^{(0)} {\cal H} + \delta^{(1)} {\cal H} + \delta^{(2)} {\cal H} + \dots  \\
                   && + \delta^{(0)} F_1 + \delta^{(1)} F_1 + \delta^{(2)} F_1 + \dots \nn \\
                   && + \delta^{(0)} F_2 + \delta^{(1)} F_2 + \delta^{(2)} F_2 + \dots \ , \nn
\eea
 and so on, and on the other hand
 \bea
 \delta {\cal M} &=& \widehat {\cal L}_\xi\, {\cal M} + {\cal J}^{(1)}({\cal M}) + {\cal J}^{(2)}({\cal M}) \\
                 &=& \widehat {\cal L}_\xi\, {\cal H} + \widehat {\cal L}_\xi\, F_1 + \widehat {\cal L}_\xi\, F_2 + \dots \nn \\
                 && + {\cal J}^{(1)}({\cal H}) + {\cal J}^{(1)}(F_1) + {\cal J}^{(1)}(F_2) + \dots \nn \\
                 && + {\cal J}^{(2)}({\cal H}) + {\cal J}^{(2)}(F_1) + {\cal J}^{(2)}(F_2) + \dots \ . \nn
 \eea
 Equating order by order in this expansion one finds
 \bea
  \delta^{(0)} {\cal H} &=& \widehat {\cal L}_\xi\, {\cal H} \\
  \delta^{(1)} {\cal H} &=& - \Delta^{(0)} F_1 + {\cal J}^{(1)}({\cal H}) \\
  \delta^{(2)} {\cal H} &=& - \Delta^{(0)} F_2  + {\cal J}^{(2)} ({\cal H}) + {\cal J}^{(1)}(F_1) - \delta^{(1)} F_1  \\
   &\vdots& \nn
 \eea
where we have defined the operator $\Delta^{(0)} = \delta^{(0)} - \widehat {\cal L}_\xi$ that measures the failure of a tensor to transform covariantly to lowest order. Notice that since $\delta^{(i)} {\cal H} = [\delta^{(i)}{\cal H}]$, one has
 \bea
  \delta^{(0)} {\cal H} &=& \widehat {\cal L}_\xi\, {\cal H} \label{d0H}\\
  \delta^{(1)} {\cal H} &=& [{\cal J}^{(1)}({\cal H})] \label{d1H}\\
  \delta^{(2)} {\cal H} &=& [{\cal J}^{(2)} ({\cal H}) + {\cal J}^{(1)}(F_1) - \delta^{(1)} F_1] \label{d2H}\\
   &\vdots& \nn
 \eea
so it turns out that $\delta^{(n)}{\cal H}$ only depends on $F_{i}$ with $i < n$. Also, notice that since $\Delta^{(0)} P = \Delta^{(0)} \bar P = 0$, one has
 \bea
  \Delta^{(0)} F_1 &=& \{{\cal J}^{(1)}({\cal H})\} \label{AnomF1}\\
  \Delta^{(0)} F_2 &=& \{ {\cal J}^{(2)} ({\cal H}) + {\cal J}^{(1)}(F_1) - \delta^{(1)} F_1 \} \label{AnomF2}\\
   &\vdots& \ \ \ \ \ \ \ \ \ \ \ \ \ \ \ \ \ \ \ \ \ \ \ \ \ \ \ \ \ \ \ \ \ \ \ \ \ \  \ \ \ \  .\nn
 \eea

\subsection{Parameterizations and field redefinitions} \label{SUBSEC:Parameterizations}

Being symmetric and $O(D,D,\mathbb{R})$-valued, the generalized metric can always be parameterized in terms of a $D$-dimensional symmetric and invertible matrix $\widehat g_{\mu \nu}$ and an antisymmetric matrix $\widehat B_{\mu \nu}$
\be
{\cal H}_{M N} = \left(\begin{matrix} \widehat g^{\mu \nu} & - \widehat g^{\mu \rho} \widehat B_{\rho \nu} \\
\widehat B_{\mu \rho} \widehat g^{\rho \nu} & \widehat g_{\mu \nu} - \widehat B_{\mu \rho} \widehat g^{\rho \sigma} \widehat B_{\sigma \nu} \end{matrix} \right) \ .
\ee
This is consistent with the choice
\be
\eta_{M N} = \left(\begin{matrix} 0 & \delta^\mu_\nu \\
\delta^\nu_\mu & 0 \end{matrix} \right) \ , \ \ \  \partial_M = (\tilde \partial^\mu , \partial_\mu) = (0 , \partial_\mu) \ . \label{Eta}
\ee
The generalized dilaton is parameterized in terms of a dilaton $\widehat \phi$ and the determinant of $\widehat g$
\be
e^{- 2 d} = \sqrt{- \widehat g} e^{-2 \widehat \phi} \ ,
\ee
and the generalized parameters decompose as
\be
\xi^M = \left(\begin{matrix} \xi_\mu \\
\xi^\mu \end{matrix} \right) \ ,
\ee
where $\xi_\mu$ are the one-form parameters that generate the gauge transformations of the two-form, and $\xi^\mu$ are the vector parameters that generate infinitesimal diffeomorphisms.

To lowest order, the generalized metric transforms as usual with respect to the generalized Lie derivative. However, we have seen that in HSZ theory its gauge transformation receives in addition an infinite number of higher derivative corrections (\ref{d0H}-\ref{d2H})
\bea
\delta {\cal H} &=& \widehat {\cal L}_\xi\, {\cal H} + \delta^{(1)}{\cal H} + \delta^{(2)} {\cal H} + \dots \\
\delta d &=& \xi \cdot \partial d - \frac 1 2 \partial \cdot \xi \ .
\eea
 These in turn induce an infinite number of corrections to the gauge transformations of the components
\bea
\delta \widehat g_{\mu \nu} &=& L_\xi \widehat g_{\mu \nu} + \delta^{(1)} \widehat g_{\mu \nu} + \delta^{(2)} \widehat g_{\mu \nu} + \dots \label{transfG}\\
\delta \widehat B_{\mu \nu} &=& L_\xi \widehat B_{\mu \nu} + 2 \partial_{[\mu} \xi_{\nu]} + \delta^{(1)} \widehat B_{\mu \nu} + \delta^{(2)} \widehat B_{\mu \nu} + \dots \label{transfB}\\
\delta \widehat \phi &=& L_\xi \phi + \delta^{(1)} \widehat \phi + \delta^{(2)} \widehat \phi + \dots \ .
\eea
This signals an obstruction to identify the hatted fields $\widehat g_{\mu \nu}$, $\widehat B_{\mu \nu}$ and $\widehat \phi$ with the conventional metric $g_{\mu \nu}$, Kalb-Ramond $B_{\mu \nu}$ and dilaton field $\phi$ in the universal sector of supergravity, which transform as follows
\bea
\delta g_{\mu \nu} &=& L_\xi g_{\mu \nu} \label{TransfSugraG}\\
\delta B_{\mu \nu} &=& L_\xi B_{\mu \nu} + 2 \partial_{[\mu} \xi_{\nu]} + \widetilde \delta B_{\mu \nu} \label{TransfSugraB}\\
\delta \phi &=& L_\xi \phi \ .\label{TransfSugraP}
\eea
Here $\widetilde \delta B_{\mu \nu}$ gives us some freedom to consider higher order contributions, presumably of the Green-Schwarz type, which are present in some string theories (such as the heterotic string).

The link between both sets of fields -hatted and unhatted- must be given through an invertible field redefinition of the form
\bea
\widehat g_{\mu \nu} &=& g_{\mu \nu} + \Delta_{1} g_{\mu \nu} + \Delta_2 g_{\mu \nu} + \dots \label{DeltaG}\\
\widehat B_{\mu \nu} &=& B_{\mu \nu} + \Delta_{1} B_{\mu \nu} + \Delta_2 B_{\mu \nu} + \dots \label{DeltaB}\\
\widehat \phi &=& \phi + \Delta_{1} \phi + \Delta_2 \phi + \dots \ ,\label{DeltaP}
\eea
where $\Delta_i g_{\mu \nu}$, $\Delta_i B_{\mu \nu}$ and $\Delta_i g_{\mu \nu}$ are allowed to depend on $g_{\mu \nu}$, $B_{\mu \nu}$ and $\phi$, and contain $2i$ derivatives. Each of these terms must be non-covariant under the standard gauge transformations, and are of course defined only up to covariant contributions.

Note that since the transformation of $d$ is not deformed, it is possible to parameterize it in terms of the gauge covariant fields
\be
e^{- 2 d } = \sqrt{-g} e^{- 2 \phi} \ \ \  \Rightarrow \ \ \ \widehat \phi = \phi  + \frac 1 4 \log \frac {\widehat g} g \ ,
\ee
in such a way that the redefinitions $\Delta_i \phi$ are fixed by $\Delta_i g$.

We have then identified two sets of fields:
\begin{itemize}
\item Duality covariant fields $(\widehat g_{\mu \nu} ,\, \widehat B_{\mu \nu},\, \widehat \phi)$. They correspond to the components of the generalized metric and dilaton, and as such transform as usual under Buscher rules \cite{Buscher:1987qj} and more generally under the full $O(D,D,\mathbb{R})$. Instead, they transform non-covariantly under diffeomorphisms and gauge transformations.

\item Gauge and diffeomorphism covariant fields $(g_{\mu \nu} ,\, B_{\mu \nu},\, \phi)$. On the contrary, they transform as usual -covariantly- under diffeomorphisms and gauge transformations (possibly with a Green-Schwarz deformation), but the duality transformations of these fields receive higher derivative corrections.

\end{itemize}
The two sets of fields are related by field redefinitions that are neither covariant under diffeomorphisms and gauge transformations nor duality covariant. In order to find the final form of the action in terms of the standard supergravity fields, in a way that is manifestly gauge invariant (and where of course the duality symmetries are hidden) one has to perform the following chain of substitutions
\be
\left( {\cal M}, \, d \right) \to \left( {\cal H}, \, d \right) \to \left(\widehat g, \, \widehat B,\, \widehat \phi\right) \to \left(g, \, B,\, \phi\right) \ .
\ee

\subsection{$Z_2$-parity} \label{SUBSEC::Parity}

The two-derivative DFT features a discrete symmetry \cite{Hull:2009mi}, consisting in invariance under a $Z_2$-parity transformation generated by
\be
Z = Z_M{}^{N} = \left(\begin{matrix} - \delta^\mu_\nu & 0 \\ 0 & \delta^\nu_\mu\end{matrix}\right) \ , \ \ \  Z^2 = 1  \ , \ \ \ Z = Z^{-1} \ .
\ee
It transforms the generalized metric as
\be
Z_2({\cal H}) = Z {\cal H} Z^T \ ,
\ee
while leaving the generalized dilaton invariant. Adopting vectorial notation for derivatives $\partial = \partial_\bullet$ and gauge parameters $\xi = \xi^\bullet$, the parity transformation takes the form
\be
Z_{2}(\partial) = Z \partial = \partial Z^T \ , \ \ \ Z_2(\xi) = \xi Z = Z^T \xi \ .
\ee
At the level of components, these transformations imply
\bea
&& Z_2 (\widehat g_{\mu \nu}) = \widehat g_{\mu \nu} \ , \ \ \ Z_2 (\widehat B_{\mu \nu}) = - \widehat B_{\mu \nu} \ , \ \ \ Z_2(\widehat \phi) = \widehat \phi \\
&& Z_2 (\xi^\mu) = \xi^\mu \ , \ \ \ Z_2 (\xi_\mu) = - \xi_\mu \ , \ \ \ Z_2 (\partial_\mu) = \partial_\mu \ , \ \ \ Z_2 (\tilde \partial^\mu) = - \tilde \partial^\mu \ .
\eea
In particular, it exchanges the sign of the two-form while leaving the metric and dilaton invariant.

Note that the duality invariant metric (\ref{Eta}) changes sign when acted on by the parity matrices
\be
Z \eta Z^T = -  \eta \ , \ \ \ Z^T \eta^{-1} Z = -  \eta^{-1} \ .
\ee
The two-derivative DFT is $Z_2$-invariant because when the indices of the generalized metric and the derivatives are placed in canonical positions, namely ${\cal H}_{\bullet \bullet}$ and $\partial_\bullet$, the action contains an even number of $\eta^{-1}$. Since the effect of $Z_2$-parity is to change the sign of the duality invariant metric, the fact that it enters each term an even number of times renders the action invariant. This implies that the component action written in terms of standard fields must contain even powers of the two-form, as is.

Here, based on the $Z_2$-parity analysis of the HSZ action and gauge transformations, we wish to anticipate the powers with which the two-form enters the action order by order. Let us point out that (the following statements hold when the indices of the generalized fields and derivatives are placed downstairs ${\cal M}_{\bullet \bullet}$, ${\cal H}_{\bullet \bullet}$, $\partial_{\bullet}$ and all contractions are made with $\eta^{\bullet \bullet}$):
\begin{itemize}
\item In the action written in terms of the double metric (see Appendix \ref{APP::DoubleMetricAction}), terms with $2i$ derivatives depend explicitly on even/odd powers of $\eta^{-1}$ when $i$ is odd/even.

\item The fields $F_i$ contain (by definition) $2 i$ derivatives of the generalized metric and dilaton, and it can be inferred from their transformation behavior (\ref{AnomF1}-\ref{AnomF2}) that when expressed in terms of the generalized metric and dilaton they contain an even/odd number of $\eta^{-1}$ when $i$ is even/odd.

\item  As a consequence of the previous items, when the HSZ action is written purely in terms of the generalized metric and dilaton, terms containing explicitly $2i$ derivatives will involve an even/odd number of $\eta^{-1}$ when $i$ is odd/even. Then, terms with $2\ (mod\ 4)$ derivatives will contain even powers of $\widehat B_{\mu \nu}$ and terms with $4\ (mod\ 4)$ derivatives will contain odd powers of $\widehat B_{\mu \nu}$.

\item A close look into (\ref{d0H}-\ref{d2H}) shows that $\delta^{(i)} \widehat g_{\mu \nu}$ and $\delta^{(i)} \widehat \phi$ are even/odd under $Z_2$-parity when $i$ is even/odd, and $\delta^{(i)} \widehat B_{\mu \nu}$ is even/odd under $Z_2$-parity when $i$ is odd/even. As a result, $\Delta_i g_{\mu \nu}$ and $\Delta_i \phi$ must contain even/odd powers of $B_{\mu \nu}$ when $i$ is even/odd, and instead $\Delta_i B_{\mu \nu}$ contains even/odd powers of $B_{\mu \nu}$ when $i$ is odd/even.
\end{itemize}
Combining this information leads to the conclusion that when the final HSZ action is written in terms of the standard fields $g_{\mu \nu}$, $B_{\mu \nu}$ and $\phi$, terms with $2i$ derivatives contain even/odd powers of $B_{\mu \nu}$ when $i$ is odd/even. To be more precise, this statement is true only up to covariant field redefinitions that violate the parity behavior. Note that the  parity of $\delta^{(i)} \widehat g_{\mu \nu}$, $\delta^{(i)} \widehat B_{\mu \nu}$, $\delta^{(i)} \widehat \phi$ only fixes the parity of the non-covariant part of $\Delta_i g_{\mu \nu}$, $\Delta_i B_{\mu \nu}$ and $\Delta_i \phi$, but leaves the covariant part unrestricted. One could, for example, induce through covariant field redefinitions Ricci-squared terms in the action that seem to violate the claim that first order contributions are odd under $Z_2$-parity. But then of course by construction these terms are removable through covariant field redefinitions.

\subsection{First order corrections and the Green-Schwarz mechanism} \label{SUBSEC:First}

Following a preliminary discussion in \cite{Hohm:2013jaa}, the first order corrections to HSZ theory were worked out in detail in a series of papers \cite{Hohm:2015mka,Hohm:2014eba} by O. Hohm and B. Zwiebach. Here we briefly review the results that are relevant for our purpose of finding the second order corrections.

In \cite{Hohm:2016lim,Hohm:2015mka}, the fields $F_{M N}$ in (\ref{DecompMHF}) were shown to correspond to dynamical massive excitations. Once the action is expressed in terms of these fields  and of the the generalized metric and dilaton, one can integrate them out ending with an effective action for the massless fields. The result of such a procedure is a functional dependence of the fields $F_{i}$ in terms of the generalized metric and dilaton. For the first order, the solution was found to be
\bea
F_{1MN} &=& \frac{1}{4}\, {\cal H}_{M P}\, {\partial}_{Q}{{\cal H}_{N}\,^{I}}\,  {\partial}_{I}{{\cal H}^{P Q}}\,  - \frac{1}{4}\, {\cal H}_{M}\,^{P} {\partial}_{Q}{\cal H}_{N I}\,  {\partial}_{P}{{\cal H}^{Q I}}\,  - \frac{1}{4}\, {\cal H}_{M}\,^{P} {\partial}_{N}{{\cal H}^{Q I}}\,  {\partial}_{Q}{{\cal H}_{P I}}\, \nn \\ && + \frac{1}{16}\, {\cal H}_{M}\,^{P} {\partial}_{N}{{\cal H}^{Q I}}\,  {\partial}_{P}{{\cal H}_{Q I}}\,  + \frac{1}{8}\, {\cal H}^{P Q} {\partial}_{P}{{\cal H}_{M}\,^{I}}\,  {\partial}_{Q}{{\cal H}_{N I}} + (M \leftrightarrow N) \ . \label{F1HZ}
\eea
Since $F_1$ is already first order as it carries two derivatives, to first order the components of ${\cal H}_{M N}$ and $d$ can be taken to be the familiar unhatted gauge covariant $g_{\mu \nu}$, $B_{\mu \nu}$ and $\phi$.

In addition, the double metric (\ref{DecompMHF}) also receives a contribution to first order from the generalized metric. Indeed, the first order deformation of the gauge transformation of the generalized metric (\ref{d1H}) is given by
\bea
\delta^{(1)}{\cal H}_{M N} &=&  - \frac{1}{4}\, {\partial}_{M}{{\cal H}^{P Q}}\,  {\partial}_{P}{{K}_{Q N}}\,  + \frac{1}{4}\, {\cal H}_{M}\,^{P} {\cal H}_{N}\,^{Q} {\partial}_{P}{{\cal H}^{I J}}\,  {\partial}_{I}{{K}_{J Q}}\, \\ && - \frac{1}{2}\, {\partial}_{P}{{\cal H}_{Q M}}\,  {\partial}_{N}{{K}^{Q P}}\,  + \frac{1}{2}\, {\cal H}_{M}\,^{P} {\cal H}_{N}\,^{Q} {\partial}_{I}{{\cal H}_{J P}}\,  {\partial}_{Q}{{K}^{J I}} + (M \leftrightarrow N) \ .\nn
\eea
This, as explained above, induces a first order transformation of its components
\bea
\delta^{(1)} \widehat g_{\mu \nu} &=& \frac{1}{4}\, {\partial}_{\mu}{{\widehat B}_{\rho \sigma}}\,  {\partial}_{\nu \gamma}{{\xi}^{\rho}}\,  {\widehat g}^{\sigma \gamma} - \frac{1}{2}\, {\partial}_{\rho}{{\widehat B}_{\mu \sigma}}\,  {\partial}_{\nu \gamma}{{\xi}^{\rho}}\,  {\widehat g}^{\sigma \gamma} + \frac{1}{4}\, {\partial}_{\mu}{{\widehat g}_{\rho \sigma}}\,  {\partial}_{\gamma \epsilon}{{\xi}_{\nu}}\,  {\widehat g}^{\rho \gamma} {\widehat g}^{\sigma \epsilon} \nn \\ && - \frac{1}{4}\, {\partial}_{\mu}{{\widehat g}_{\rho \sigma}}\,  {\partial}_{\nu \gamma}{{\xi}_{\epsilon}}\,  {\widehat g}^{\rho \gamma} {\widehat g}^{\sigma \epsilon} - \frac{1}{4}\, {\widehat B}_{\mu \rho} {\partial}_{\nu}{{\widehat g}_{\sigma \gamma}}\,  {\partial}_{\epsilon \delta}{{\xi}^{\rho}}\,  {\widehat g}^{\sigma \epsilon} {\widehat g}^{\gamma \delta}  \\ && - \frac{1}{4}\, {\widehat B}_{\rho \sigma} {\partial}_{\mu}{{\widehat g}_{\gamma \epsilon}}\,  {\partial}_{\nu \delta}{{\xi}^{\rho}}\,  {\widehat g}^{\sigma \gamma} {\widehat g}^{\epsilon \delta} + (\mu \leftrightarrow \nu) \nn \\
\delta^{(1)} \widehat B_{\mu \nu} &=&  - \frac{1}{2}\, {\partial}_{\rho}{{\widehat g}_{\mu \sigma}}\,  {\partial}_{\nu \gamma}{{\xi}^{\rho}}\,  {\widehat g}^{\sigma \gamma} + \frac{1}{4}\, {\partial}_{\mu}{{\widehat g}_{\rho \sigma}}\,  {\partial}_{\gamma \epsilon}{{\xi}^{\delta}}\,  {\widehat g}_{\nu \delta} {\widehat g}^{\rho \gamma} {\widehat g}^{\sigma \epsilon} - (\mu \leftrightarrow \nu) \ .
 \eea

One can then try to perform a first order non-covariant field redefinition to make contact with the gauge covariant fields (\ref{TransfSugraG}-\ref{TransfSugraP}). These redefinitions were worked out in full detail in \cite{Hohm:2015mka}, and up to covariant terms are given by
\bea
\Delta_1 g_{\mu \nu} &=&  - \frac{1}{4}\, {\partial}_{\mu}{{B}_{\rho \sigma}}\,  {\partial}_{\gamma}{{g}_{\nu \epsilon}}\,  {g}^{\rho \gamma} {g}^{\sigma \epsilon} - \frac{1}{4}\, {\partial}_{\rho}{{B}_{\mu \sigma}}\,  {\partial}_{\gamma}{{g}_{\nu \epsilon}}\,  {g}^{\rho \epsilon} {g}^{\sigma \gamma} \nn \\ && + \frac{1}{4}\, {\partial}_{\rho}{{B}_{\mu \sigma}}\,  {\partial}_{\gamma}{{g}_{\nu \epsilon}}\,  {g}^{\rho \gamma} {g}^{\sigma \epsilon} - \frac{1}{4}\, {\partial}_{\rho}{{B}_{\mu \sigma}}\,  {\partial}_{\nu}{{g}_{\gamma \epsilon}}\,  {g}^{\rho \gamma} {g}^{\sigma \epsilon} + (\mu \leftrightarrow \nu) \nn\\
\Delta_1 B_{\mu \nu} &=& \frac{1}{4}\, {\partial}_{\mu}{{g}_{\rho \sigma}}\,  {\partial}_{\gamma}{{g}_{\nu \epsilon}}\,  {g}^{\rho \gamma} {g}^{\sigma \epsilon} - (\mu \leftrightarrow \nu) \ .
\eea

It turns out to be impossible to trivialize the transformation of the Kalb-Ramond field to first order \cite{Hohm:2014eba}, and one is forced to end with a two-form that transforms to first order under a Green-Schwarz-like transformation
\be
\widetilde \delta B_{\mu \nu} = \partial_{[\mu|} \partial_\rho \xi^\sigma \Gamma_{|\nu]\sigma}^\rho \ , \label{TildedeltaB}
\ee
where $\widetilde \delta B_{\mu \nu}$ was defined in (\ref{TransfSugraB}). Due to this transformation, the standard supergravity action fails to be diffeomorphism invariant to first order, so the three-form field strength must be corrected with a Chern-Simons term. This is reminiscent to what happens in the heterotic string, where anomaly cancellation requires a deformation of the three-form field strength with a Lorentz Chern-Simons term constructed from a spin connection with torsion. In HSZ theory the connection in the Chern-Simons term is the Levi-Civita (or Christoffel) connection. The minimal action consistent with the deformed gauge transformations and the standard zeroth order action is given by
\be
S = \int d^Dx \sqrt{-g} e^{-2 \phi}\left(R + 4 \nabla_\mu \nabla^\mu \phi - 4 \nabla_\mu \phi \nabla^\mu \phi - \frac 1 {12} \widehat H_{\mu \nu \rho} \widehat H^{\mu \nu \rho} \right) \ , \label{Action1stOrder}
\ee
where
\be
\widehat H_{\mu \nu \rho} = H_{\mu \nu \rho} + 3 \Omega_{\mu \nu \rho} \ .\label{WidehatH}
\ee
All these quantities were defined in Appendix \ref{APP::Conventions}. The first order contribution to the Lagrangian is given by $- \frac 1 2\, H_{\mu \nu \rho}\, \Omega^{\mu \nu \rho}$, and consistently satisfies the expectation that first order corrections involve odd powers of the two-form field. The bosonic and heterotic strings contain a Riemann-squared term to first order, but in HSZ theory such a term is forbidden by $Z_2$-parity.

\subsection{Second order corrections and the Riemann-cubed terms} \label{SUBSEC::Second}

Already from the results of the previous subsection one can anticipate a second order contribution to HSZ theory given by the Chern-Simons squared term in (\ref{Action1stOrder}-\ref{WidehatH}), namely $- \frac 3 4\, \Omega_{\mu \nu \rho}\, \Omega^{\mu \nu \rho}$. In \cite{Hohm:2015doa}, O. Hohm and B. Zwiebach explored to what extent higher derivative corrections are constrained by T-duality. Building on the seminal work by K. Meissner \cite{Meissner:1996sa} (see also \cite{godazgar}), they refined a procedure to test the T-duality invariance of a generic gauge invariant action, offering necessary conditions. By applying the method to the action (\ref{Action1stOrder}), they found that the theory is consistent with T-duality to first order, but fails to be duality invariant to second order. As a result, the action (\ref{Action1stOrder}) faithfully reproduces the interactions in HSZ theory to first order, but is expected to contain additional second order interactions beyond the Chern-Simons squared terms.

Analyzing on-shell three point functions to all orders in the derivative expansion, U. Naseer and B. Zwiebach concluded in \cite{Naseer:2016izx} that HSZ theory must contain, in addition to the quadratic Chern-Simons terms, a Riemann-cubed interaction to second order. Their work shows that to second order, the following is the gauge invariant action that reproduces the on-shell cubic amplitudes of HSZ theory
\bea
S &=& \int d^Dx \sqrt{-g} e^{-2 \phi}\left(R + 4 \nabla_\mu \nabla^\mu \phi - 4 \nabla_\mu \phi \nabla^\mu \phi - \frac 1 {12} \widehat H_{\mu \nu \rho} \widehat H^{\mu \nu \rho} \right.\nn \\ && \ \ \ \ \ \ \ \ \ \ \ \ \ \ \ \ \ \ \ \  \ \ \ \ \    \left. - \frac 1 {48}\, R_{\mu \nu}{}^{\alpha \beta} R_{\alpha \beta}{}^{\rho \sigma} R_{\rho \sigma}{}^{\mu \nu}\right)\ . \label{Action2ndOrderNZ}
\eea
The coefficient of the Riemann-cubed term is minus the coefficient of the same term in bosonic string theory. No such term appears in the heterotic string.  It was shown in \cite{Camanho:2014apa} that interactions that contribute to three-point amplitudes with three external gravitons lead to causality violations that require the existence of an infinite tower of new particles with spin higher than two. Such violations are avoided when the full structure of string theory is taken into account \cite{D'Appollonio:2015gpa}.

In bosonic string theory there is also a non-zero Gauss-Bonnet Riemann-cubed term $G_3$
\be
G_3 = R_{\mu \nu}{}^{\alpha \beta} R_{\alpha \beta}{}^{\rho \sigma} R_{\rho \sigma}{}^{\mu \nu} - 2\, R^{\mu \nu \alpha \beta} R_{\nu \lambda \beta \gamma} R^\lambda{}_\mu{}^\gamma{}_\alpha \ ,
\ee
but its presence can only be seen from four point amplitudes \cite{Metsaev:1986yb}. Note again that these corrections satisfy the expectation that the second order corrections must involve even powers of the two-form.

\section{The full massless action to second order} \label{SEC:Full}

In this section we present the original results of this paper. Since many computations produce large outputs, we separate some intermediate results and show them in Appendix \ref{APP::Results}. Most computations are conceptually simple but computationally challenging. Without the aid of Cadabra software \cite{Peeters:2007wn}, some of the computations would have been for all practical purposes impossible to carry on.

The main result is the computation of the full massless second order action of HSZ theory. We begin explaining the procedure, then display the result and conclude by analyzing some relevant interactions.

\subsection{Second order gauge transformations and field redefinitions} \label{SUBSEC::Procedure}
Let us schematically illustrate the procedure and some intermediate relevant results:
\begin{itemize}
\item {\bf \it First.} We compute explicitly how the generalized metric transforms to second order. The formal expression is given in (\ref{d2H})
\be
\delta^{(2)} {\cal H} = [{\cal J}^{(2)} ({\cal H}) + {\cal J}^{(1)}(F_1) - \delta^{(1)} F_1] \ ,
\ee
and we already have all the ingredients to perform the computation, namely $F_1[{\cal H}]$ in (\ref{F1HZ}). Note that $\delta^{(2)}{\cal H}$ is independent of the generalized dilaton, because so is $F_1$.

\item {\bf \it Second.} We compute the second order transformation induced on the set of fields that parameterize the generalized metric. The results can be found in Appendix \ref{APP::Results}
    \bea
        \delta^{(2)} \widehat g_{\mu \nu}\ \ \  &\to& \ \ \ {\rm eq. \ (\ref{d2tildeGa})} \\
         \delta^{(2)} \widehat B_{\mu \nu}\ \ \ &\to& \ \ \ {\rm eq. \ (\ref{d2tildeBa})} \ .
    \eea
\item {\bf \it Third.} We find the field redefinitions that connect the duality covariant fields $(\widehat g_{\mu \nu},\, \widehat B_{\mu \nu})$ to the gauge covariant fields $(g_{\mu \nu}, \, B_{\mu \nu})$ by demanding that the gauge transformations trivialize to (\ref{TransfSugraG}-\ref{TransfSugraB}). The results for the metric (\ref{DeltaG}) and two-form (\ref{DeltaB}) are given in Appendix \ref{APP::Results}
    \bea
        \Delta_2 \widehat g_{\mu \nu}\ \ \  &\to& \ \ \ {\rm eq. \ (\ref{Delta2tildeGa})} \\
         \Delta_2 \widehat B_{\mu \nu}\ \ \ &\to& \ \ \ {\rm eq. \ (\ref{Delta2tildeBa})} \ .
    \eea
    Interestingly, under these redefinitions, $\widetilde \delta B_{\mu \nu}$ in (\ref{TransfSugraB}) does not receive a second order deformation, and remains the first order Green-Schwarz transformation (\ref{TildedeltaB}).

\item {\bf {\it Fourth.}} The last piece of information we need is the functional dependence of $F_{2}[{\cal H}, \, d]$. With this, and the information obtained above on how the generalized metric and dilaton are parameterized in terms of the gauge covariant fields, we are ready to compute the  action in terms of the standard supergravity fields. We then seek the most general $F_2$ that satisfies equation (\ref{AnomF2}). This of course determines $F_2$ up to terms that are covariant with respect to the lowest order transformations. One expects a splitting of the form
    \be
    F_2 = F_2^{\rm non-cov} + F_2^{\rm cov} \ , \ \ \ \  \ \Delta^{(0)} F_2^{\rm cov} = 0 \ , \label{f2covnoncov}
    \ee
    where $F_2^{\rm non-cov}$ is unique. We show in Appendix \ref{APP::F2covariant} that the covariant piece can be simply ignored to second order, because it leads to covariant boundary terms in the action. Since the right hand side in (\ref{AnomF2}) does not depend on the generalized dilaton (because $F_1$ in (\ref{F1HZ}) is independent) it follows that $F_2^{\rm non-cov}$ is independent as well. Again, the result is shown in Appendix \ref{APP::Results}
    \be
       F_2^{\rm non-cov} \ \ \ \to \ \ \ {\rm eq. \ (\ref{F2Resulta})} \ .
    \ee
\end{itemize}

\subsection{Full second order action in covariant form} \label{SUBSEC::FullAction}

We now have all the necessary ingredients to compute the full second order correction to the HSZ action. The procedure has been outlined before, but let us sketch it once again. The starting point is the action in terms of the double metric and generalized dilaton (\ref{ActionDoubleMetric}) (see also Appendix \ref{APP::DoubleMetricAction}). Since we are interested in the second order terms, we need to consider the full action with all up to six derivatives. Then, one needs to find the parameterization of the double metric ${\cal M}$ and dilaton $d$ in terms of the gauge covariant fields $g_{\mu \nu}$, $B_{\mu \nu}$ and $\phi$. This is accomplished through the substitution chain
\be
\left( {\cal M}, \, d \right) \to \left( {\cal H}, \, d \right) \to \left(\widehat g, \, \widehat B,\, \widehat \phi\right)\to \left(g, \, B,\, \phi\right) \ ,
\ee
truncated to second order in the action. After this sequence is implemented, the output is a huge expression in terms of fields and their derivatives, which is covariant under diffeomorphisms and gauge transformations by construction, but the symmetries are far from being manifest. One then has to re-write the final output in terms of contractions of covariant quantities, such as the Riemann tensor, three-form field strength, dilaton and covariant derivatives acting on them.

The result is the following action
\bea
S = \int d^D x \sqrt{-g} e^{-2 \phi} \left(L^{(0)} + L^{(1)} + L^{(2)}\right) \ . \label{OrdersAction}
\eea
Here, $L^{(0)}$ is of course the usual two-derivative action
\be
L^{(0)} = R - 4 \nabla_\mu \phi \nabla^\mu \phi + 4 \nabla_\mu \nabla^\mu \phi - \frac 1 {12} H_{\mu \nu \rho} H^{\mu \nu \rho} \ . \label{L0}
\ee
The first order contribution is exactly the expression anticipated in \cite{Hohm:2014eba}
\be
L^{(1)} = - \frac 1 2 H_{\mu \nu \rho} \Omega^{\mu \nu \rho} \ , \label{L1}
\ee
and contains neither extra boundary terms, nor other extra covariant terms that could have been required by duality. This is consistent with the findings in \cite{Marques:2015vua}. The second order contribution is rather lengthy, and it helps to separate it in powers of the two-form and dilaton as follows
\bea
L^{(2)} &=& L^{[0,0]} + L^{[0,1]} + L^{[0,2]} + L^{[0,3]} + L^{[0,4]} \nn \\
    && L^{[2,0]} + L^{[2,1]} + L^{[2,2]} + L^{[2,3]} \label{L2} \\
    && L^{[4,0]} + L^{[4,1]} + L^{[4,2]} + L^{[6,0]} \ .\nn
\eea
The notation $L^{[n_B , n_\phi]}$ indicates that such terms contain $n_B$ two-forms and $n_\phi$ dilatons.

Concretely, each contribution is
\begin{dgroup*}
\begin{dmath*}[compact, spread=2pt]
L^{[0,0]} =  - \frac{3}{4}\, {\Omega}^{\mu \nu \rho} {\Omega}_{\mu \nu \rho} - \frac{1}{8}\, {\nabla}^{\mu}{{R}^{\nu \rho \sigma \gamma}}\,  {\nabla}_{\nu}{{R}_{\mu \rho \sigma \gamma}}\,  + \frac{1}{3}\, {R}^{\mu \nu \rho \sigma} {R}_{\mu}\,^{\gamma}\,_{\rho}\,^{\epsilon} {R}_{\nu \gamma \sigma \epsilon} - \frac{1}{8}\, {\nabla}^{\mu}{{\nabla}_{\mu}{{\nabla}^{\nu}{{\nabla}_{\nu}{R}\, }\, }\, }\, + \frac{1}{4}\, {R}^{\mu \nu} {\nabla}^{\rho}{{\nabla}_{\rho}{{R}_{\mu \nu}}\, }\,  + {R}^{\mu \nu} {\nabla}_{\mu}{{\nabla}^{\rho}{{R}_{\nu \rho}}\, }\, + \frac{1}{2}\, {\nabla}^{\mu}{{R}^{\nu \rho}}\,  {\nabla}_{\mu}{{R}_{\nu \rho}}\,  + \frac{1}{2}\, {\nabla}^{\mu}{{R}_{\mu}\,^{\nu}}\,  {\nabla}^{\rho}{{R}_{\nu \rho}}\,  - \frac{1}{2}\, {R}^{\mu \nu} {R}_{\mu}\,^{\rho}\,_{\nu}\,^{\sigma} {R}_{\rho \sigma} \ ,
\end{dmath*}
\begin{dmath*}[compact, spread=2pt]
L^{[0,1]}=  - {\nabla}^{\mu}{{\nabla}^{\nu}{R}\, }\,  {\nabla}_{\mu}{{\nabla}_{\nu}{\phi}\, }\,  - \frac{3}{4}\, {\nabla}^{\mu}{{\nabla}^{\nu}{{\nabla}_{\nu}{R}\, }\, }\,  {\nabla}_{\mu}{\phi}\,  + \frac{5}{2}\, {\nabla}^{\mu}{{\nabla}_{\mu}{{\nabla}^{\nu}{{\nabla}^{\rho}{{\nabla}_{\nu}{{\nabla}_{\rho}{\phi}\, }\, }\, }\, }\, }\,  - 3\, {\nabla}^{\mu}{{\nabla}_{\mu}{{\nabla}^{\nu}{{\nabla}_{\nu}{{\nabla}^{\rho}{{\nabla}_{\rho}{\phi}\, }\, }\, }\, }\, }\,  - 2\, {R}^{\mu \nu} {\nabla}_{\mu}{R}\,  {\nabla}_{\nu}{\phi}\,  - 3\, {\nabla}^{\mu}{{R}^{\nu \rho}}\,  {\nabla}_{\mu}{{\nabla}_{\nu}{{\nabla}_{\rho}{\phi}\, }\, }\,  - 2\, {\nabla}^{\mu}{{\nabla}_{\mu}{{R}^{\nu \rho}}\, }\,  {\nabla}_{\nu}{{\nabla}_{\rho}{\phi}\, }\,  - 4\, {R}^{\mu \nu} {R}_{\mu}\,^{\rho} {\nabla}_{\nu}{{\nabla}_{\rho}{\phi}\, }\,  - \frac{5}{2}\, {R}^{\mu \nu} {\nabla}_{\mu}{{R}_{\nu}\,^{\rho}}\,  {\nabla}_{\rho}{\phi}\,  + 2\, {R}^{\mu \nu} {R}_{\mu}\,^{\rho}\,_{\nu}\,^{\sigma} {\nabla}_{\rho}{{\nabla}_{\sigma}{\phi}\, }\,  + \frac{3}{2}\, {R}^{\mu \nu} {\nabla}^{\rho}{{R}_{\mu \rho \nu}\,^{\sigma}}\,  {\nabla}_{\sigma}{\phi}\ ,
\end{dmath*}
\begin{dmath*}[compact, spread=2pt]
L^{[0,2]}=  - \frac{1}{2}\, {\nabla}^{\mu}{{\nabla}^{\nu}{R}\, }\,  {\nabla}_{\mu}{\phi}\,  {\nabla}_{\nu}{\phi}\,  + 3\, {\nabla}^{\mu}{{\nabla}^{\nu}{{\nabla}_{\nu}{{\nabla}^{\rho}{{\nabla}_{\rho}{\phi}\, }\, }\, }\, }\,  {\nabla}_{\mu}{\phi}\,  + 11\, {\nabla}^{\mu}{{\nabla}^{\nu}{{\nabla}^{\rho}{{\nabla}_{\rho}{\phi}\, }\, }\, }\,  {\nabla}_{\mu}{{\nabla}_{\nu}{\phi}\, }\,  + 4\, {\nabla}^{\mu}{{\nabla}_{\mu}{{\nabla}^{\nu}{\phi}\, }\, }\,  {\nabla}_{\nu}{{\nabla}^{\rho}{{\nabla}_{\rho}{\phi}\, }\, }\,  + 4\, {\nabla}^{\mu}{{\nabla}^{\nu}{{\nabla}^{\rho}{\phi}\, }\, }\,  {\nabla}_{\mu}{{\nabla}_{\nu}{{\nabla}_{\rho}{\phi}\, }\, }\,  + 2\, {R}^{\mu \nu} {\nabla}^{\rho}{{\nabla}_{\rho}{{\nabla}_{\mu}{\phi}\, }\, }\,  {\nabla}_{\nu}{\phi}\,  - 5\, {R}^{\mu \nu} {\nabla}_{\mu}{{\nabla}_{\nu}{{\nabla}^{\rho}{\phi}\, }\, }\,  {\nabla}_{\rho}{\phi}\,  + 4\, {R}^{\mu \nu} {\nabla}^{\rho}{{\nabla}_{\mu}{\phi}\, }\,  {\nabla}_{\rho}{{\nabla}_{\nu}{\phi}\, }\,  - 4\, {\nabla}^{\mu}{{R}^{\nu \rho}}\,  {\nabla}_{\nu}{{\nabla}_{\rho}{\phi}\, }\,  {\nabla}_{\mu}{\phi}\,  + 10\, {\nabla}^{\mu}{{R}^{\nu \rho}}\,  {\nabla}_{\nu}{{\nabla}_{\mu}{\phi}\, }\,  {\nabla}_{\rho}{\phi}\,  - 5\, {\nabla}^{\mu}{{R}_{\mu}\,^{\nu}}\,  {\nabla}^{\rho}{{\nabla}_{\nu}{\phi}\, }\,  {\nabla}_{\rho}{\phi}\,  + {\nabla}^{\mu}{{\nabla}_{\mu}{{R}^{\nu \rho}}\, }\,  {\nabla}_{\nu}{\phi}\,  {\nabla}_{\rho}{\phi}\,  - 8\, {R}^{\mu \nu \rho \sigma} {\nabla}_{\mu}{{\nabla}_{\rho}{\phi}\, }\,  {\nabla}_{\nu}{{\nabla}_{\sigma}{\phi}\, }\,  + {R}^{\mu \nu} {R}_{\mu}\,^{\rho}\,_{\nu}\,^{\sigma} {\nabla}_{\rho}{\phi}\,  {\nabla}_{\sigma}{\phi}\ ,
\end{dmath*}
\begin{dmath*}[compact, spread=2pt]
L^{[0,3]}=  - 2\, {\nabla}^{\mu}{{\nabla}^{\nu}{{\nabla}^{\rho}{{\nabla}_{\rho}{\phi}\, }\, }\, }\,  {\nabla}_{\mu}{\phi}\,  {\nabla}_{\nu}{\phi}\,  + 4\, {\nabla}^{\mu}{{\nabla}^{\nu}{{\nabla}^{\rho}{{\nabla}_{\nu}{\phi}\, }\, }\, }\,  {\nabla}_{\mu}{\phi}\,  {\nabla}_{\rho}{\phi}\,  - 8\, {\nabla}^{\mu}{{\nabla}^{\nu}{{\nabla}^{\rho}{{\nabla}_{\mu}{\phi}\, }\, }\, }\,  {\nabla}_{\nu}{\phi}\,  {\nabla}_{\rho}{\phi}\,  - 2\, {\nabla}^{\mu}{{\nabla}^{\nu}{{\nabla}^{\rho}{\phi}\, }\, }\,  {\nabla}_{\nu}{{\nabla}_{\rho}{\phi}\, }\,  {\nabla}_{\mu}{\phi}\,  - 16\, {\nabla}^{\mu}{{\nabla}^{\nu}{{\nabla}_{\nu}{\phi}\, }\, }\,  {\nabla}^{\rho}{{\nabla}_{\mu}{\phi}\, }\,  {\nabla}_{\rho}{\phi}\,  - 20\, {\nabla}^{\mu}{{\nabla}^{\nu}{{\nabla}^{\rho}{\phi}\, }\, }\,  {\nabla}_{\mu}{{\nabla}_{\rho}{\phi}\, }\,  {\nabla}_{\nu}{\phi}\,  - 12\, {\nabla}^{\mu}{{\nabla}^{\nu}{\phi}\, }\,  {\nabla}_{\mu}{{\nabla}^{\rho}{\phi}\, }\,  {\nabla}_{\nu}{{\nabla}_{\rho}{\phi}\, }\ ,
\end{dmath*}
\begin{dmath*}[compact, spread=2pt]
L^{[0,4]}= 4\, {\nabla}^{\mu}{{\nabla}^{\nu}{{\nabla}^{\rho}{\phi}\, }\, }\,  {\nabla}_{\mu}{\phi}\,  {\nabla}_{\nu}{\phi}\,  {\nabla}_{\rho}{\phi}\,  + 16\, {\nabla}^{\mu}{{\nabla}^{\nu}{\phi}\, }\,  {\nabla}^{\rho}{{\nabla}_{\nu}{\phi}\, }\,  {\nabla}_{\mu}{\phi}\,  {\nabla}_{\rho}{\phi}\ ,
\end{dmath*}
\begin{dmath*}[compact, spread=2pt]
L^{[2,0]}= \frac{1}{32}\, {\nabla}^{\mu}{{\nabla}_{\mu}{{\nabla}^{\nu}{{\nabla}^{\rho}{({H}^{\sigma}\,_{\nu}\,^{\gamma} {H}_{\rho \gamma \sigma})}\, }\, }\, }\,  - \frac{1}{8}\, {H}^{\mu \nu \rho} {H}_{\mu \nu}\,^{\sigma} {\nabla}_{\rho}{{\nabla}_{\sigma}{R}\, }\,  - \frac{1}{24}\, {H}^{\mu \nu \rho} {\nabla}^{\sigma}{{H}_{\mu \nu \rho}}\,  {\nabla}_{\sigma}{R}\,  - \frac{1}{8}\, {H}^{\mu \nu \rho} {\nabla}^{\sigma}{{H}_{\mu \nu \sigma}}\,  {\nabla}_{\rho}{R}\,  + \frac{1}{48}\, {\nabla}^{\mu}{{H}^{\nu \rho \sigma}}\,  {\nabla}^{\gamma}{{\nabla}_{\gamma}{{\nabla}_{\mu}{{H}_{\nu \rho \sigma}}\, }\, }\,  + \frac{1}{16}\, {\nabla}^{\mu}{{\nabla}^{\nu}{{H}_{\mu}\,^{\rho \sigma}}\, }\,  {\nabla}^{\gamma}{{\nabla}_{\nu}{{H}_{\rho \sigma \gamma}}\, }\,  - \frac{1}{32}\, {H}^{\mu \nu \rho} {H}_{\mu \nu}\,^{\sigma} {\nabla}^{\gamma}{{\nabla}_{\gamma}{{R}_{\rho \sigma}}\, }\,  - \frac{1}{4}\, {H}^{\mu \nu \rho} {R}^{\sigma \gamma} {\nabla}_{\mu}{{\nabla}_{\sigma}{{H}_{\nu \rho \gamma}}\, }\,  + \frac{1}{8}\, {H}^{\mu \nu \rho} {R}_{\mu}\,^{\sigma} {\nabla}_{\nu}{{\nabla}^{\gamma}{{H}_{\rho \sigma \gamma}}\, }\,  - \frac{3}{16}\, {H}^{\mu \nu \rho} {R}_{\mu}\,^{\sigma} {\nabla}_{\sigma}{{\nabla}^{\gamma}{{H}_{\nu \rho \gamma}}\, }\,  - \frac{3}{8}\, {H}^{\mu \nu \rho} {\nabla}^{\sigma}{{H}_{\mu \nu}\,^{\gamma}}\,  {\nabla}_{\sigma}{{R}_{\rho \gamma}}\,  - \frac{1}{4}\, {H}^{\mu \nu \rho} {\nabla}^{\sigma}{{H}_{\mu \sigma}\,^{\gamma}}\,  {\nabla}_{\nu}{{R}_{\rho \gamma}}\,  - \frac{1}{4}\, {R}^{\mu \nu} {\nabla}_{\mu}{{H}_{\nu}\,^{\rho \sigma}}\,  {\nabla}^{\gamma}{{H}_{\rho \sigma \gamma}}\,  + \frac{1}{2}\, {R}^{\mu \nu} {\nabla}^{\rho}{{H}_{\mu}\,^{\sigma \gamma}}\,  {\nabla}_{\sigma}{{H}_{\nu \rho \gamma}}\,  - \frac{5}{16}\, {R}^{\mu \nu} {\nabla}^{\rho}{{H}_{\mu}\,^{\sigma \gamma}}\,  {\nabla}_{\rho}{{H}_{\nu \sigma \gamma}}\,  + \frac{1}{8}\, {H}^{\mu \nu \rho} {H}_{\mu}\,^{\sigma \gamma} {R}_{\nu \sigma} {R}_{\rho \gamma} - \frac{1}{32}\, {H}^{\mu \nu \rho} {H}_{\mu}\,^{\sigma \gamma} {\nabla}^{\epsilon}{{\nabla}_{\epsilon}{{R}_{\nu \rho \sigma \gamma}}\, }\,  - \frac{1}{8}\, {H}^{\mu \nu \rho} {R}_{\mu}\,^{\sigma}\,_{\nu}\,^{\gamma} {\nabla}_{\rho}{{\nabla}^{\epsilon}{{H}_{\sigma \gamma \epsilon}}\, }\,  - \frac{1}{4}\, {H}^{\mu \nu \rho} {R}_{\mu}\,^{\sigma \gamma \epsilon} {\nabla}_{\nu}{{\nabla}_{\sigma}{{H}_{\rho \gamma \epsilon}}\, }\, %
 - \frac{1}{8}\, {H}^{\mu \nu \rho} {\nabla}^{\sigma}{{H}_{\mu}\,^{\gamma \epsilon}}\,  {\nabla}_{\nu}{{R}_{\rho \sigma \gamma \epsilon}}\,  + \frac{1}{16}\, {R}^{\mu \nu \rho \sigma} {\nabla}^{\gamma}{{H}_{\mu \nu}\,^{\epsilon}}\,  {\nabla}_{\gamma}{{H}_{\rho \sigma \epsilon}}\,  - \frac{1}{16}\, {R}^{\mu \nu \rho \sigma} {\nabla}^{\gamma}{{H}_{\mu \nu \gamma}}\,  {\nabla}^{\epsilon}{{H}_{\rho \sigma \epsilon}}\,  + \frac{1}{4}\, {R}^{\mu \nu \rho \sigma} {\nabla}_{\mu}{{H}_{\rho}\,^{\gamma \epsilon}}\,  {\nabla}_{\sigma}{{H}_{\nu \gamma \epsilon}}\,  + \frac{3}{4}\, {H}^{\mu \nu \rho} {H}_{\mu}\,^{\sigma \gamma} {R}_{\nu}\,^{\epsilon} {R}_{\rho \sigma \gamma \epsilon} + \frac{3}{16}\, {H}^{\mu \nu \rho} {H}^{\sigma \gamma \epsilon} {R}_{\mu \nu \sigma}\,^{\delta} {R}_{\rho \gamma \epsilon \delta} - \frac{1}{4}\, {H}^{\mu \nu \rho} {H}_{\mu}\,^{\sigma \gamma} {R}_{\nu}\,^{\epsilon}\,_{\rho}\,^{\delta} {R}_{\sigma \epsilon \gamma \delta} - \frac{5}{8}\, {H}^{\mu \nu \rho} {H}_{\mu}\,^{\sigma \gamma} {R}_{\nu}\,^{\epsilon}\,_{\sigma}\,^{\delta} {R}_{\rho \epsilon \gamma \delta} + \frac{1}{8}\, {H}^{\mu \nu \rho} {H}_{\mu}\,^{\sigma \gamma} {R}_{\nu \sigma}\,^{\epsilon \delta} {R}_{\rho \gamma \epsilon \delta} - \frac{1}{16}\, {H}^{\mu \nu \rho} {H}_{\mu \nu}\,^{\sigma} {R}_{\rho}\,^{\gamma \epsilon \delta} {R}_{\sigma \gamma \epsilon \delta} \ ,
\end{dmath*}
\begin{dmath*}[compact, spread=2pt]
L^{[2,1]}= \frac{1}{4}\, {H}^{\mu \nu \rho} {H}_{\mu \nu}\,^{\sigma} {\nabla}_{\rho}{R}\,  {\nabla}_{\sigma}{\phi}\,  - \frac{5}{8}\, {H}^{\mu \nu \rho} {H}_{\mu \nu}\,^{\sigma} {\nabla}^{\gamma}{{\nabla}_{\gamma}{{\nabla}_{\rho}{{\nabla}_{\sigma}{\phi}\, }\, }\, }\,  - \frac{5}{8}\, {H}^{\mu \nu \rho} {\nabla}^{\sigma}{{H}_{\mu \nu \sigma}}\,  {\nabla}^{\gamma}{{\nabla}_{\gamma}{{\nabla}_{\rho}{\phi}\, }\, }\,  - \frac{5}{24}\, {H}^{\mu \nu \rho} {\nabla}^{\sigma}{{H}_{\mu \nu \rho}}\,  {\nabla}^{\gamma}{{\nabla}_{\gamma}{{\nabla}_{\sigma}{\phi}\, }\, }\,  - {H}^{\mu \nu \rho} {\nabla}^{\sigma}{{H}_{\mu \nu}\,^{\gamma}}\,  {\nabla}_{\rho}{{\nabla}_{\sigma}{{\nabla}_{\gamma}{\phi}\, }\, }\,  - \frac{3}{4}\, {H}^{\mu \nu \rho} {\nabla}^{\sigma}{{\nabla}^{\gamma}{{H}_{\mu \nu \gamma}}\, }\,  {\nabla}_{\rho}{{\nabla}_{\sigma}{\phi}\, }\,  - \frac{1}{4}\, {H}^{\mu \nu \rho} {\nabla}^{\sigma}{{\nabla}^{\gamma}{{H}_{\mu \nu \rho}}\, }\,  {\nabla}_{\sigma}{{\nabla}_{\gamma}{\phi}\, }\,  - \frac{1}{8}\, {H}^{\mu \nu \rho} {\nabla}^{\sigma}{{\nabla}_{\sigma}{{\nabla}^{\gamma}{{H}_{\mu \nu \gamma}}\, }\, }\,  {\nabla}_{\rho}{\phi}\,  - \frac{1}{4}\, {H}^{\mu \nu \rho} {\nabla}^{\sigma}{{\nabla}^{\gamma}{{\nabla}_{\mu}{{H}_{\nu \rho \gamma}}\, }\, }\,  {\nabla}_{\sigma}{\phi}\,  - \frac{1}{4}\, {\nabla}^{\mu}{{H}^{\nu \rho \sigma}}\,  {\nabla}^{\gamma}{{H}_{\nu \rho \sigma}}\,  {\nabla}_{\mu}{{\nabla}_{\gamma}{\phi}\, }\,  - \frac{1}{4}\, {\nabla}^{\mu}{{H}^{\nu \rho \sigma}}\,  {\nabla}_{\mu}{{H}_{\nu \rho}\,^{\gamma}}\,  {\nabla}_{\sigma}{{\nabla}_{\gamma}{\phi}\, }\,  - \frac{3}{4}\, {\nabla}^{\mu}{{H}_{\mu}\,^{\nu \rho}}\,  {\nabla}^{\sigma}{{H}_{\nu \rho}\,^{\gamma}}\,  {\nabla}_{\sigma}{{\nabla}_{\gamma}{\phi}\, }\,  - \frac{1}{4}\, {\nabla}^{\mu}{{H}^{\nu \rho \sigma}}\,  {\nabla}^{\gamma}{{\nabla}_{\gamma}{{H}_{\mu \nu \rho}}\, }\,  {\nabla}_{\sigma}{\phi}\,  - \frac{1}{2}\, {\nabla}^{\mu}{{H}^{\nu \rho \sigma}}\,  {\nabla}_{\mu}{{\nabla}^{\gamma}{{H}_{\nu \rho \gamma}}\, }\,  {\nabla}_{\sigma}{\phi}\,  - \frac{1}{2}\, {\nabla}^{\mu}{{H}^{\nu \rho \sigma}}\,  {\nabla}_{\mu}{{\nabla}_{\nu}{{H}_{\rho \sigma}\,^{\gamma}}\, }\,  {\nabla}_{\gamma}{\phi}\,  - \frac{1}{4}\, {\nabla}^{\mu}{{H}_{\mu}\,^{\nu \rho}}\,  {\nabla}^{\sigma}{{\nabla}_{\sigma}{{H}_{\nu \rho}\,^{\gamma}}\, }\,  {\nabla}_{\gamma}{\phi}\,  - \frac{1}{4}\, {\nabla}^{\mu}{{H}_{\mu}\,^{\nu \rho}}\,  {\nabla}_{\nu}{{\nabla}^{\sigma}{{H}_{\rho \sigma}\,^{\gamma}}\, }\,  {\nabla}_{\gamma}{\phi}\,  + \frac{5}{4}\, {H}^{\mu \nu \rho} {H}_{\mu \nu}\,^{\sigma} {R}_{\rho}\,^{\gamma} {\nabla}_{\sigma}{{\nabla}_{\gamma}{\phi}\, }\,  + {H}^{\mu \nu \rho} {H}_{\mu \nu}\,^{\sigma} {\nabla}_{\rho}{{R}_{\sigma}\,^{\gamma}}\,  {\nabla}_{\gamma}{\phi}\, %
 - \frac{3}{8}\, {H}^{\mu \nu \rho} {H}_{\mu \nu}\,^{\sigma} {\nabla}^{\gamma}{{R}_{\rho \sigma}}\,  {\nabla}_{\gamma}{\phi}\,  + \frac{1}{8}\, {H}^{\mu \nu \rho} {R}^{\sigma \gamma} {\nabla}_{\sigma}{{H}_{\mu \nu \rho}}\,  {\nabla}_{\gamma}{\phi}\,  + \frac{1}{2}\, {H}^{\mu \nu \rho} {R}^{\sigma \gamma} {\nabla}_{\sigma}{{H}_{\mu \nu \gamma}}\,  {\nabla}_{\rho}{\phi}\,  - \frac{1}{4}\, {H}^{\mu \nu \rho} {R}_{\mu}\,^{\sigma} {\nabla}^{\gamma}{{H}_{\nu \sigma \gamma}}\,  {\nabla}_{\rho}{\phi}\,  + \frac{5}{8}\, {H}^{\mu \nu \rho} {R}_{\mu}\,^{\sigma} {\nabla}^{\gamma}{{H}_{\nu \rho \gamma}}\,  {\nabla}_{\sigma}{\phi}\,  + \frac{1}{4}\, {H}^{\mu \nu \rho} {R}_{\mu}\,^{\sigma} {\nabla}_{\sigma}{{H}_{\nu \rho}\,^{\gamma}}\,  {\nabla}_{\gamma}{\phi}\,  + \frac{1}{4}\, {H}^{\mu \nu \rho} {H}^{\sigma \gamma \epsilon} {R}_{\mu \nu \sigma \gamma} {\nabla}_{\rho}{{\nabla}_{\epsilon}{\phi}\, }\,  + \frac{1}{2}\, {H}^{\mu \nu \rho} {H}_{\mu}\,^{\sigma \gamma} {R}_{\nu \rho \sigma}\,^{\epsilon} {\nabla}_{\gamma}{{\nabla}_{\epsilon}{\phi}\, }\,  - \frac{1}{2}\, {H}^{\mu \nu \rho} {H}_{\mu \nu}\,^{\sigma} {R}_{\rho}\,^{\gamma}\,_{\sigma}\,^{\epsilon} {\nabla}_{\gamma}{{\nabla}_{\epsilon}{\phi}\, }\,  + \frac{1}{4}\, {H}^{\mu \nu \rho} {H}_{\mu}\,^{\sigma \gamma} {\nabla}^{\epsilon}{{R}_{\nu \rho \sigma \epsilon}}\,  {\nabla}_{\gamma}{\phi}\,  + \frac{1}{2}\, {H}^{\mu \nu \rho} {R}_{\mu}\,^{\sigma \gamma \epsilon} {\nabla}_{\sigma}{{H}_{\nu \rho \gamma}}\,  {\nabla}_{\epsilon}{\phi}\,  + \frac{1}{8}\, {H}^{\mu \nu \rho} {R}_{\mu \nu}\,^{\sigma \gamma} {\nabla}^{\epsilon}{{H}_{\sigma \gamma \epsilon}}\,  {\nabla}_{\rho}{\phi}\,  + \frac{1}{2}\, {H}^{\mu \nu \rho} {R}_{\mu}\,^{\sigma \gamma \epsilon} {\nabla}_{\nu}{{H}_{\rho \gamma \epsilon}}\,  {\nabla}_{\sigma}{\phi}\ ,
\end{dmath*}
\begin{dmath*}[compact, spread=2pt]
L^{[2,2]}= \frac{5}{4}\, {H}^{\mu \nu \rho} {H}_{\mu \nu}\,^{\sigma} {\nabla}^{\gamma}{{\nabla}_{\gamma}{{\nabla}_{\rho}{\phi}\, }\, }\,  {\nabla}_{\sigma}{\phi}\,  + \frac{5}{4}\, {H}^{\mu \nu \rho} {H}_{\mu \nu}\,^{\sigma} {\nabla}_{\rho}{{\nabla}_{\sigma}{{\nabla}^{\gamma}{\phi}\, }\, }\,  {\nabla}_{\gamma}{\phi}\,  + 2\, {H}^{\mu \nu \rho} {H}_{\mu \nu}\,^{\sigma} {\nabla}_{\rho}{{\nabla}^{\gamma}{\phi}\, }\,  {\nabla}_{\sigma}{{\nabla}_{\gamma}{\phi}\, }\,  + \frac{3}{2}\, {H}^{\mu \nu \rho} {\nabla}^{\sigma}{{H}_{\mu \nu}\,^{\gamma}}\,  {\nabla}_{\gamma}{{\nabla}_{\rho}{\phi}\, }\,  {\nabla}_{\sigma}{\phi}\,  + \frac{5}{4}\, {H}^{\mu \nu \rho} {\nabla}^{\sigma}{{H}_{\mu \nu \sigma}}\,  {\nabla}^{\gamma}{{\nabla}_{\rho}{\phi}\, }\,  {\nabla}_{\gamma}{\phi}\,  + \frac{3}{2}\, {H}^{\mu \nu \rho} {\nabla}^{\sigma}{{H}_{\mu \nu}\,^{\gamma}}\,  {\nabla}_{\sigma}{{\nabla}_{\gamma}{\phi}\, }\,  {\nabla}_{\rho}{\phi}\,  + \frac{5}{12}\, {H}^{\mu \nu \rho} {\nabla}^{\sigma}{{H}_{\mu \nu \rho}}\,  {\nabla}^{\gamma}{{\nabla}_{\sigma}{\phi}\, }\,  {\nabla}_{\gamma}{\phi}\,  - 3\, {H}^{\mu \nu \rho} {\nabla}_{\mu}{{H}_{\nu}\,^{\sigma \gamma}}\,  {\nabla}_{\sigma}{{\nabla}_{\rho}{\phi}\, }\,  {\nabla}_{\gamma}{\phi}\,  + \frac{1}{2}\, {H}^{\mu \nu \rho} {\nabla}^{\sigma}{{\nabla}^{\gamma}{{H}_{\mu \nu \gamma}}\, }\,  {\nabla}_{\rho}{\phi}\,  {\nabla}_{\sigma}{\phi}\,  + \frac{1}{12}\, {H}^{\mu \nu \rho} {\nabla}^{\sigma}{{\nabla}^{\gamma}{{H}_{\mu \nu \rho}}\, }\,  {\nabla}_{\sigma}{\phi}\,  {\nabla}_{\gamma}{\phi}\,  - \frac{1}{2}\, {H}^{\mu \nu \rho} {\nabla}^{\sigma}{{\nabla}_{\mu}{{H}_{\nu \sigma}\,^{\gamma}}\, }\,  {\nabla}_{\rho}{\phi}\,  {\nabla}_{\gamma}{\phi}\,  - \frac{1}{2}\, {\nabla}^{\mu}{{H}^{\nu \rho \sigma}}\,  {\nabla}_{\nu}{{H}_{\mu \rho}\,^{\gamma}}\,  {\nabla}_{\sigma}{\phi}\,  {\nabla}_{\gamma}{\phi}\,  + \frac{3}{4}\, {\nabla}^{\mu}{{H}^{\nu \rho \sigma}}\,  {\nabla}_{\mu}{{H}_{\nu \rho}\,^{\gamma}}\,  {\nabla}_{\sigma}{\phi}\,  {\nabla}_{\gamma}{\phi}\,  + \frac{1}{4}\, {\nabla}^{\mu}{{H}_{\mu}\,^{\nu \rho}}\,  {\nabla}^{\sigma}{{H}_{\nu \rho}\,^{\gamma}}\,  {\nabla}_{\sigma}{\phi}\,  {\nabla}_{\gamma}{\phi}\,  - \frac{3}{4}\, {H}^{\mu \nu \rho} {H}_{\mu \nu}\,^{\sigma} {R}_{\rho}\,^{\gamma} {\nabla}_{\sigma}{\phi}\,  {\nabla}_{\gamma}{\phi}\,  - \frac{1}{2}\, {H}^{\mu \nu \rho} {H}^{\sigma \gamma \epsilon} {R}_{\mu \sigma \nu \gamma} {\nabla}_{\rho}{\phi}\,  {\nabla}_{\epsilon}{\phi}\,  - \frac{3}{2}\, {H}^{\mu \nu \rho} {H}_{\mu}\,^{\sigma \gamma} {R}_{\nu \sigma \rho}\,^{\epsilon} {\nabla}_{\gamma}{\phi}\,  {\nabla}_{\epsilon}{\phi}\,  - \frac{1}{4}\, {H}^{\mu \nu \rho} {H}_{\mu \nu}\,^{\sigma} {R}_{\rho}\,^{\gamma}\,_{\sigma}\,^{\epsilon} {\nabla}_{\gamma}{\phi}\,  {\nabla}_{\epsilon}{\phi}\ ,
\end{dmath*}
\begin{dmath*}[compact, spread=2pt]
L^{[2,3]}=  - \frac{5}{2}\, {H}^{\mu \nu \rho} {H}_{\mu \nu}\,^{\sigma} {\nabla}_{\rho}{{\nabla}^{\gamma}{\phi}\, }\,  {\nabla}_{\sigma}{\phi}\,  {\nabla}_{\gamma}{\phi}\,  - \frac{1}{2}\, {H}^{\mu \nu \rho} {\nabla}^{\sigma}{{H}_{\mu \nu}\,^{\gamma}}\,  {\nabla}_{\rho}{\phi}\,  {\nabla}_{\sigma}{\phi}\,  {\nabla}_{\gamma}{\phi}\ ,
\end{dmath*}
\begin{dmath*}[compact, spread=2pt]
L^{[4,0]}= \frac{1}{32}\, {H}_{\mu \nu}\,^{\rho} {H}_{\sigma}\,^{\gamma \epsilon} {H}^{\mu \nu \sigma} {\nabla}_{\rho}{{\nabla}^{\delta}{{H}_{\gamma \epsilon \delta}}\, }\,  - \frac{1}{16}\, {H}_{\mu}\,^{\nu \rho} {H}_{\sigma}\,^{\gamma \epsilon} {H}^{\mu \sigma \delta} {\nabla}_{\delta}{{\nabla}_{\nu}{{H}_{\rho \gamma \epsilon}}\, }\,  + \frac{1}{48}\, {H}_{\mu \nu}\,^{\rho} {H}^{\sigma \gamma \epsilon} {H}^{\mu \nu \delta} {\nabla}_{\delta}{{\nabla}_{\rho}{{H}_{\sigma \gamma \epsilon}}\, }\,  + \frac{1}{16}\, {H}^{\mu \nu \rho} {H}_{\mu \nu}\,^{\sigma} {\nabla}_{\rho}{{H}_{\sigma}\,^{\gamma \epsilon}}\,  {\nabla}^{\delta}{{H}_{\gamma \epsilon \delta}}\,  + \frac{3}{64}\, {H}^{\mu \nu \rho} {H}_{\mu}\,^{\sigma \gamma} {\nabla}_{\nu}{{H}_{\rho}\,^{\epsilon \delta}}\,  {\nabla}_{\sigma}{{H}_{\gamma \epsilon \delta}}\,  - \frac{1}{32}\, {H}^{\mu \nu \rho} {H}_{\mu}\,^{\sigma \gamma} {\nabla}_{\nu}{{H}_{\sigma}\,^{\epsilon \delta}}\,  {\nabla}_{\gamma}{{H}_{\rho \epsilon \delta}}\,  + \frac{1}{64}\, {H}^{\mu \nu \rho} {H}_{\mu \nu}\,^{\sigma} {\nabla}_{\rho}{{H}^{\gamma \epsilon \delta}}\,  {\nabla}_{\sigma}{{H}_{\gamma \epsilon \delta}}\,  + \frac{1}{16}\, {H}^{\mu \nu \rho} {H}^{\sigma \gamma \epsilon} {\nabla}_{\mu}{{H}_{\nu \sigma}\,^{\delta}}\,  {\nabla}_{\rho}{{H}_{\gamma \epsilon \delta}}\,  - \frac{1}{32}\, {H}^{\mu \nu \rho} {H}^{\sigma \gamma \epsilon} {\nabla}_{\mu}{{H}_{\nu \sigma}\,^{\delta}}\,  {\nabla}_{\gamma}{{H}_{\rho \epsilon \delta}}\,  + \frac{3}{64}\, {H}^{\mu \nu \rho} {H}^{\sigma \gamma \epsilon} {\nabla}_{\mu}{{H}_{\sigma \gamma}\,^{\delta}}\,  {\nabla}_{\epsilon}{{H}_{\nu \rho \delta}}\,  - \frac{1}{32}\, {H}^{\mu \nu \rho} {H}_{\mu \nu}\,^{\sigma} {\nabla}^{\gamma}{{H}_{\rho \gamma}\,^{\epsilon}}\,  {\nabla}^{\delta}{{H}_{\sigma \epsilon \delta}}\,  + \frac{1}{16}\, {H}^{\mu \nu \rho} {H}_{\mu}\,^{\sigma \gamma} {\nabla}_{\nu}{{H}_{\sigma \gamma}\,^{\epsilon}}\,  {\nabla}^{\delta}{{H}_{\rho \epsilon \delta}}\,  + \frac{1}{32}\, {H}^{\mu \nu \rho} {H}_{\mu}\,^{\sigma \gamma} {\nabla}^{\epsilon}{{H}_{\nu \rho \epsilon}}\,  {\nabla}^{\delta}{{H}_{\sigma \gamma \delta}}\,  + \frac{1}{288}\, {H}^{\mu \nu \rho} {H}^{\sigma \gamma \epsilon} {\nabla}^{\delta}{{H}_{\mu \nu \rho}}\,  {\nabla}_{\delta}{{H}_{\sigma \gamma \epsilon}}\,  + \frac{1}{48}\, {H}^{\mu \nu \rho} {H}^{\sigma \gamma \epsilon} {\nabla}_{\mu}{{H}_{\sigma \gamma \epsilon}}\,  {\nabla}^{\delta}{{H}_{\nu \rho \delta}}\,  - \frac{1}{16}\, {H}^{\mu \nu \rho} {H}^{\sigma \gamma \epsilon} {\nabla}_{\mu}{{H}_{\nu \sigma \gamma}}\,  {\nabla}^{\delta}{{H}_{\rho \epsilon \delta}}\,  - \frac{1}{16}\, {H}_{\mu \nu \rho} {H}_{\sigma}\,^{\nu \rho} {H}_{\gamma}\,^{\epsilon \sigma} {H}_{\delta \epsilon}\,^{\mu} {R}^{\gamma \delta} - \frac{1}{32}\, {H}_{\mu \nu \rho} {H}_{\sigma \gamma}\,^{\rho} {H}_{\epsilon}\,^{\sigma \gamma} {H}_{\delta}\,^{\mu \nu} {R}^{\epsilon \delta} - \frac{1}{32}\, {H}_{\mu \nu \rho} {H}_{\sigma}\,^{\nu \rho} {H}_{\gamma \epsilon}\,^{\delta} {H}_{\lambda \delta}\,^{\mu} {R}^{\gamma \epsilon \lambda \sigma}%
 - \frac{1}{16}\, {H}_{\mu \nu \rho} {H}_{\sigma \gamma}\,^{\rho} {H}_{\epsilon}\,^{\delta \gamma} {H}_{\lambda \delta}\,^{\nu} {R}^{\epsilon \lambda \sigma \mu} - \frac{1}{32}\, {H}_{\mu \nu \rho} {H}_{\sigma \gamma \epsilon} {H}_{\delta}\,^{\gamma \epsilon} {H}_{\lambda}\,^{\nu \rho} {R}^{\delta \lambda \sigma \mu} \ ,
\end{dmath*}
\begin{dmath*}[compact, spread=2pt]
L^{[4,1]}=  - \frac{1}{8}\, {H}_{\mu}\,^{\nu \rho} {H}^{\sigma \gamma \mu} {H}_{\epsilon \nu}\,^{\delta} {H}_{\sigma \gamma}\,^{\epsilon} {\nabla}_{\rho}{{\nabla}_{\delta}{\phi}\, }\,  - \frac{1}{8}\, {H}_{\mu}\,^{\nu \rho} {H}^{\sigma \gamma \mu} {H}_{\nu \rho}\,^{\epsilon} {H}_{\sigma \gamma}\,^{\delta} {\nabla}_{\delta}{{\nabla}_{\epsilon}{\phi}\, }\,  + \frac{1}{8}\, {H}_{\mu}\,^{\nu \rho} {H}^{\sigma \mu \gamma} {H}_{\sigma}\,^{\epsilon \delta} {\nabla}_{\gamma}{{H}_{\epsilon \delta \nu}}\,  {\nabla}_{\rho}{\phi}\,  - \frac{1}{8}\, {H}_{\mu}\,^{\nu \rho} {H}^{\sigma \gamma \mu} {H}_{\sigma \gamma}\,^{\epsilon} {\nabla}^{\delta}{{H}_{\epsilon \nu \delta}}\,  {\nabla}_{\rho}{\phi}\,  - \frac{1}{8}\, {H}_{\mu}\,^{\nu \rho} {H}^{\sigma \gamma \mu} {H}_{\sigma \gamma}\,^{\epsilon} {\nabla}_{\nu}{{H}_{\epsilon \rho}\,^{\delta}}\,  {\nabla}_{\delta}{\phi}\,  + \frac{1}{8}\, {H}_{\mu}\,^{\nu \rho} {H}^{\sigma \mu \gamma} {H}_{\sigma}\,^{\epsilon \delta} {\nabla}_{\epsilon}{{H}_{\delta \nu \rho}}\,  {\nabla}_{\gamma}{\phi}\,  - \frac{1}{8}\, {H}^{\mu \nu \rho} {H}^{\sigma \gamma \epsilon} {H}_{\sigma \gamma}\,^{\delta} {\nabla}_{\mu}{{H}_{\epsilon \nu \rho}}\,  {\nabla}_{\delta}{\phi}\,  - \frac{1}{8}\, {H}^{\mu \nu \rho} {H}^{\sigma \gamma \epsilon} {H}_{\sigma \gamma}\,^{\delta} {\nabla}_{\epsilon}{{H}_{\delta \mu \nu}}\,  {\nabla}_{\rho}{\phi}\,  - \frac{1}{16}\, {H}_{\mu}\,^{\nu \rho} {H}^{\sigma \gamma \mu} {H}_{\sigma \gamma}\,^{\epsilon} {\nabla}^{\delta}{{H}_{\epsilon \nu \rho}}\,  {\nabla}_{\delta}{\phi}\,  - \frac{1}{8}\, {H}_{\mu}\,^{\nu \rho} {H}^{\sigma \gamma \mu} {H}_{\sigma \gamma}\,^{\epsilon} {\nabla}^{\delta}{{H}_{\nu \rho \delta}}\,  {\nabla}_{\epsilon}{\phi}\ ,
\end{dmath*}
\begin{dmath*}[compact, spread=2pt]
L^{[4,2]}=  - \frac{1}{8}\, {H}^{\mu}\,_{\nu}\,^{\rho} {H}^{\sigma \gamma \nu} {H}_{\rho \mu}\,^{\epsilon} {H}_{\gamma \sigma}\,^{\delta} {\nabla}_{\delta}{\phi}\,  {\nabla}_{\epsilon}{\phi}\,  - \frac{1}{8}\, {H}^{\mu}\,_{\nu}\,^{\rho} {H}^{\sigma \gamma \nu} {H}^{\epsilon}\,_{\delta \rho} {H}_{\gamma \sigma}\,^{\delta} {\nabla}_{\mu}{\phi}\,  {\nabla}_{\epsilon}{\phi}\ ,
\end{dmath*}
\begin{dmath*}[compact, spread=2pt]
L^{[6,0]}=  - \frac{1}{384}\, {H}^{\mu}\,_{\nu \rho} {H}^{\sigma \nu \rho} {H}^{\gamma}\,_{\epsilon \delta} {H}^{\lambda}\,_{\mu}\,^{\alpha} {H}_{\alpha \gamma \lambda} {H}_{\sigma}\,^{\epsilon \delta} + \frac{1}{128}\, {H}^{\mu \nu \rho} {H}_{\sigma \gamma}\,^{\epsilon} {H}^{\delta}\,_{\lambda \alpha} {H}_{\delta \epsilon}\,^{\alpha} {H}_{\mu \nu}\,^{\gamma} {H}^{\sigma}\,_{\rho}\,^{\lambda} \ .
\end{dmath*}
\end{dgroup*}

This action can of course be simplified combining Bianchi identities, integration by parts and field redefinitions. Finding the minimal form of the full action is an ambitious program that we will not pursue here. In the case of the four-derivative theory, it was shown in \cite{Metsaev:1987zx} that any gauge invariant action can be reduced to contain only eight terms. Some of these terms were dubbed unambiguous because their coefficients  can not be modified through field redefinitions nor integrations by parts. An example of an unambiguous term is the Riemann-squared contained in the bosonic string. The six-derivative case is far more complicated, and was discussed in \cite{Jones:1988hk} for the case of vanishing dilaton. There, it was shown that the most general theory can be reduced to
only twenty-one terms, ten of which are unambiguous. In the next section we will briefly discuss some limit cases that will help in identifying some relevant six-derivative interactions in the theory. In particular, we will provide the minimal form this action when $\phi = 0$.

Let us finally note here that gauge invariance (more precisely, the Green-Schwarz transformation of the two-form (\ref{TildedeltaB})) requires putting a hat over all the three-form field strengths $H_{\mu \nu \rho} \to \widehat H_{\mu \nu \rho}$, in order to restore the Chern-Simons terms (\ref{WidehatH}). This replacement leaves the second order terms untouched, but would be required by gauge invariance when analyzing the next orders in $\alpha'$.

\subsection{Riemann-cubed, Gauss-Bonnet and more} \label{SUBSEC::Comparison}

Here we wish to study the action in some limit cases, in order to understand what the most relevant terms are.

As a first check, we have verified explicitly full consistency with the results in \cite{Naseer:2016izx}. Setting the dilaton to zero $\phi = 0$, after repeated use of integrations by part and the lowest order equations of motion (which as we show in Appendix \ref{APP::FieldRedefs} corresponds to performing second order field redefinitions), we expanded the metric around a flat background $g_{\mu \nu} = \eta_{\mu \nu} + h_{\mu \nu}$ keeping terms up to cubic order in fluctuations\footnote{For a discussion on the cubic-order duality covariant expansion of DFT see \cite{Hohm:2015ugy}.}. We then imposed on-shell and polarization conditions, namely $\eta^{\mu \nu} \partial_{\mu} \dots \partial_{\nu} \dots = 0$ and $\partial^{\mu} h_{\mu \nu} = \partial^\mu B_{\mu \nu} = 0$, and verified that the only contribution to the cubic action is given by
\be
- \frac 1 {48} \partial^{\mu \nu} h_{\rho \sigma} \partial^{\delta \sigma} h_{\lambda \mu} \partial^{\lambda \rho} h_{\delta \nu} \ ,
\ee
in agreement with \cite{Naseer:2016izx}.

Let us now take both the dilaton and two-form to zero $B_{\mu \nu} = \phi = 0$, and keep the full (unexpanded) metric field. In this case, the only terms that survive are those in $L^{[0,0]}$, and moreover integrating by parts, using Bianchi identities (\ref{Bi1}-\ref{Bi2}) and taking $R_{\mu \nu} = 0$ and $R = 0$ (which corresponds to performing field redefinitions) one is left with
\be
L^{[0,0]} =  - \frac{3}{4}\, {\Omega}^{\mu \nu \rho} {\Omega}_{\mu \nu \rho} - \frac 1 {48} \, R_{\mu \nu}{}^{\alpha \beta} R_{\alpha \beta}{}^{\rho \sigma} R_{\rho \sigma}{}^{\mu \nu} - \frac 1 {24} G_3  \ ,
\ee
where $G_3$ is the cubic Gauss-Bonnet interaction term
\be
G_3 = R_{\mu \nu}{}^{\alpha \beta} R_{\alpha \beta}{}^{\rho \sigma} R_{\rho \sigma}{}^{\mu \nu} - 2\, R^{\mu \nu \alpha \beta} R_{\nu \lambda \beta \gamma} R^\lambda{}_\mu{}^\gamma{}_\alpha \ .
\ee
This term vanishes on-shell to cubic order and is then transparent to three-point amplitudes.

Let us finally discuss another limit case with vanishing dilaton $\phi = 0$, but non-vanishing metric and two-form. After some redefinitions and integrations by parts the Lagrangian can be drastically simplified to
\bea
L^{(2)} \!\!\!&=& \!\!\! -\frac{3}{4} \Omega^{\mu \nu \rho} \Omega_{\mu \nu \rho} - \frac{1}{16} R_{\mu \nu}{}^{\alpha \beta} R_{\alpha \beta}{}^{\rho \sigma} R_{\rho \sigma}{}^{\mu \nu} + \frac{1}{12} R_{\rho \sigma \alpha \beta} R^{\sigma \mu \beta \nu} R_{\mu}{}^{\rho}{}_{\nu}{}^{\alpha}
+ \frac{1}{32} {\widehat{H}}^2_{\rho \sigma} R^{\rho \alpha \beta \gamma} R^{\sigma}{}_{\alpha \beta \gamma} \nn \\
&& - \frac{1}{64} R_{\alpha \beta \mu \rho} R_{\nu}{}^{\rho} {\widehat{H}}_{\gamma}{}^{\alpha \beta} {\widehat{H}}^{\gamma \mu \nu}
+ \frac{1}{128} R_{\alpha \beta \sigma \mu} R_{\gamma \nu} {\widehat{H}}^{\alpha \beta \gamma} {\widehat{H}}^{\sigma \mu \nu} - \frac{1}{128} R_{\alpha \beta \gamma \sigma} {\widehat{H}}^{2 \alpha \gamma} {\widehat{H}}^{2 \beta \sigma}  \\ && - \frac{1}{768} R_{\alpha \beta} {\widehat{H}}^{2 \alpha}{}_{\mu}  {\widehat{H}}^{2 \beta \mu} - \frac{1}{256} R_{\alpha \beta} {\widehat{H}}^{\alpha}{}_{\rho \gamma} {\widehat{H}}^{\beta \rho}{}_{\nu} {\widehat{H}}^{2 \gamma \nu}
- \frac{1}{128} \nabla_{\sigma} {\widehat{H}}_{\alpha \mu \nu} \nabla^{\sigma} {\widehat{H}}_{\beta}{}^{\mu \nu} {\widehat{H}}^{2 \alpha \beta}\ , \nn
\eea
where we have defined $\widehat H^2_{\mu \nu} = \widehat H_{\mu \rho \sigma} \widehat H_\nu{}^{\rho \sigma}$.
Written in this way, except for the Chern-Simons-squared term, it fits into the minimal form of the action introduced in \cite{Jones:1988hk}, implying that the number of terms can not be reduced further\footnote{Comparing with the minimal action (5.1) in \cite{Jones:1988hk} we see that the only non-vanishing coefficients are $B_1 = - \frac{1}{16}$, $B_2 = \frac{1}{12}$, $A_6 = \frac{1}{32}$, $A_9 = - \frac{1}{64}$, $A_{14} = \frac{1}{128}$, $A_{20} = - \frac{1}{128}$, $A_{22} = - \frac{1}{768}$, $A_{24} = - \frac{1}{256}$, $A_{38} = - \frac{1}{128}$.}. One can however ask if the dependence on $\widehat H_{\mu \nu \rho}$ can be eliminated by field redefinitions. Using the equation of motion with vanishing dilaton $\widehat H^2_{\mu \nu} = 4 R_{\mu \nu}$, the Lagrangian can be rewritten as
\bea
L^{(2)} &=&  - \frac{3}{4}\, {\Omega}^{\mu \nu \rho} {\Omega}_{\mu \nu \rho} - \frac 1 {48}\, R_{\mu \nu}{}^{\alpha \beta} R_{\alpha \beta}{}^{\rho \sigma} R_{\rho \sigma}{}^{\mu \nu} - \frac 1 {24} G_3
+ \frac{1}{8} R_{\rho \sigma} R^{\rho \alpha \beta \gamma} R^{\sigma}{}_{\alpha \beta \gamma} - \frac{1}{8} R_{\alpha \beta \gamma \sigma} R^{\alpha \gamma} R^{\beta \sigma} \nn \\
 &&
 - \frac{1}{48} R_{\alpha \beta} R^{\alpha}{}_{\mu} R^{\beta \mu}
- \frac{1}{64} R_{\alpha \beta \mu \rho} R_{\nu}{}^{\rho} {\widehat{H}}_{\gamma}{}^{\alpha \beta} {\widehat{H}}^{\gamma \mu \nu}
+ \frac{1}{128} R_{\alpha \beta \sigma \mu} R_{\gamma \nu} {\widehat{H}}^{\alpha \beta \gamma} {\widehat{H}}^{\sigma \mu \nu} \label{minimalaction}  \\
 && - \frac{1}{64} R_{\alpha \beta} {\widehat{H}}^{\alpha}{}_{\rho \gamma} {\widehat{H}}^{\beta \rho}{}_{\nu} R^{\gamma \nu}
- \frac{1}{64} \nabla_{\sigma} {\widehat{H}}_{\alpha \mu \nu} \nabla^{\sigma} {\widehat{H}}_{\beta}{}^{\mu \nu} R^{\alpha \beta}\ . \nn
\eea
We conclude that $\widehat H_{\mu \nu \rho}$ can not be totally eliminated from the action, and taking into account that it hides a first order correction, the fact that it appears quadratically announces that the full action must at least go up to ${\cal O}(\alpha'{}^{4})$. This confirms the expectation that the second order action is not exactly duality invariant, and that duality invariance will require an infinite expansion in powers of $\alpha'$.

Let us finally note that the lowest order equations of motion combined with integration by parts always allow to eliminate from the Lagrangian all terms with covariant derivatives of the dilaton, even without assuming that some particular field vanishes. In fact, whenever the dilaton appears covariantly derived once, it can be eliminated through an integration by parts $e^{- 2 \phi} V^\mu \nabla_\mu \phi = - \frac 1 2 V^\mu \nabla_{\mu} e^{-2 \phi} = \frac 1 2 \nabla_\mu V^\mu e^{-2 \phi}$. If instead it appears derived more that once, the equations of motion can be used to eliminate the dependence on it $\nabla_{\mu \nu} \phi = - \frac 1 2 R_{\mu \nu} + \frac 1 8 H_{\mu \rho \sigma} H_{\nu}{}^{\rho \sigma}$.

\section{Outlook and concluding remarks} \label{SEC::Conclusions}

We have computed the full second order action for the massless sector of HSZ Double Field Theory \cite{Hohm:2013jaa}. The theory features an exact duality invariant extension of the Green-Schwarz mechanism \cite{Hohm:2014eba}. The first order $\alpha'$-corrections are given by Chern-Simons deformations of the three-form field strength. To second order in $\alpha'$ the theory was known to contain a Riemann-cubed interaction, with minus the coefficient of the same term in the bosonic string \cite{Naseer:2016izx}.  Here we provide the full second order action in a manifestly covariant form (the result can be found in Subsection \ref{SUBSEC::FullAction}). The gravitational sector includes a cubic Gauss-Bonnet interaction, also with opposite coefficient to that in the bosonic string. The second order contributions contain even powers of the three-form field strength, and the fact its dependence can not be eliminated from the action anticipates that exact duality invariance requires further higher-derivative terms.

It would be desirable to take the full second order action to a minimal form, combining field redefinitions, Bianchi identities and integrations by part. We have done this in the particular case of vanishing dilaton, obtaining a very simple expression for the action (\ref{minimalaction}). It would be interesting to compare our results with the second order corrections to the bosonic string, and explore to what extent the coincidences that occur in the gravitational sector apply to the general action.

HSZ theory is presumably the simplest higher-derivative theory of gravity with exact duality invariance. Although it is not a conventional string theory, understanding how duality and gauge invariance organize the action can shed light on how duality constrains higher-derivative interactions in more general set-ups. In \cite{Marques:2015vua} a deformation of DFT was proposed in the frame-formalism, which captures the first order corrections to the bosonic, heterotic and HSZ theory. It would be interesting to see how to extend these deformations to higher-orders, so understanding them in the particular case of HSZ theory is a convenient starting point.

~

\noindent {\bf \underline{Acknowledgments:}} We are indebted with J. Edelstein, O. Hohm, U. Naseer, C. Nu\~nez and B. Zwiebach for enlightening discussions.
Support by CONICET, UBA and ANPCyT is also gratefully acknowledged.

\begin{appendix}
\section{Conventions and definitions}\label{APP::Conventions}
In this Appendix we introduce some notation used throughout the paper.

Space-time indices are denoted with greek letters $\mu, \nu, \dots$. The Lie derivative of a tensor is given by
\be
L_{\xi} V_{\mu}{}^{\nu}= \xi^{\rho} \partial_{\rho}V_{\mu}{}^{\nu} + \partial_{\mu} \xi^{\rho} V_{\rho}{}^{\nu} - \partial_{\rho} \xi^{\nu} V_{\mu}{}^{\rho} \ .
\ee
The Christoffel connection is defined in terms of the metric as
\be
\Gamma_{\mu \nu}^{\rho} = \frac 1 2 g^{\rho \sigma} \left( \partial_{\mu} g_{\nu \sigma} + \partial_{\nu} g_{\mu \sigma} - \partial_{\sigma} g_{\mu \nu}\right) \ , \ \ \ \ \ \ \Gamma_{[\mu \nu]}^{\rho} = 0 \ ,
\ee
and transforms anomalously under infinitesimal diffeomorphisms (whenever the Lie derivative acts on a
non-tensorial object, we use the convention that it acts as if it
were covariant)
\be
\delta_{\xi} \Gamma_{\mu \nu}^\rho = L_{\xi}  \Gamma_{\mu \nu}^\rho + \partial_{\mu} \partial_{\nu} \xi^{\rho} \ ,
\ee
so it allows to define a covariant derivative, given by
\be
\nabla_{\rho} V_{\mu}{}^{\nu} = \partial_{\rho} V_{\mu}{}^{\nu} - \Gamma_{\rho \mu}^{\sigma} V_{\sigma}{}^{\nu} + \Gamma_{\rho \sigma}^{\nu} V_{\mu}{}^{\sigma} \ .
\ee
The commutator of two covariant derivatives
\be
\left[ \nabla_{\mu} ,\ \nabla_{\nu}\right] V_{\rho}{}^{\sigma} = - R^\delta{}_{\rho \mu \nu} \, V_\delta{}^\sigma + R^\sigma{}_{\delta \mu \nu} V_\rho{}^\delta \ , \label{Bi1}
\ee
is expressed in terms of the Riemann tensor, which can be expressed as
\be
R^{\rho}{}_{\sigma \mu \nu} = \partial_{\mu} \Gamma_{\nu \sigma}^{\rho} - \partial_{\nu} \Gamma_{\mu \sigma}^\rho + \Gamma_{\mu \delta}^\rho \Gamma_{\nu \sigma}^\delta - \Gamma_{\nu \delta}^\rho \Gamma_{\mu \sigma}^\delta \ . \label{curvedRiemann}
\ee
Its  symmetries and Bianchi identities are
\be
R_{\rho \sigma \mu \nu} = g_{\rho \delta} R^{\delta}{}_{\sigma \mu \nu} = R_{([\rho \sigma][\mu \nu])}
\ , \ \ \ \
R^{\rho}{}_{[\sigma \mu \nu]} = 0 \ , \ \ \ \ \nabla_{[\mu} R_{\nu \lambda]}{}^\rho{}_\sigma = 0 \ . \label{Bi2}
\ee
Traces of the Riemann tensor give the Ricci tensor and scalar, respectively
\be
R_{\mu \nu} = R^{\rho}{}_{\mu \rho \nu} \ , \ \ \ \ \ \ R = g^{\mu \nu} R_{\mu \nu} \ .
\ee

The Chern-Simons three-form is defined as
\be
\Omega_{\mu \nu \rho} = \Gamma_{[\mu| \sigma}^\delta \partial_{|\nu} \Gamma_{\rho] \delta}^\sigma + \frac 2 3
\Gamma_{[\mu | \sigma}^\delta \Gamma_{|\nu| \lambda}^\sigma \Gamma_{|\rho] \delta}^\lambda \ ,
\ee
and it transforms under infinitesimal diffeomorphisms as
\be
\delta \Omega_{\mu \nu \rho} = L_\xi \Omega_{\mu \nu \rho} -
\partial_{[\mu} \left(\partial_{\nu|} \partial_{\sigma} \xi^{\delta} \Gamma_{|\rho] \delta}^\sigma\right) \ .
\ee

The Kalb-Ramond two-form transforms both with respect to gauge transformations, parameterized by a one-form $\xi_{\mu}$, and diffeomorphisms
\be
\delta B_{\mu \nu} = L_{\xi} B_{\mu \nu} + 2 \partial_{[\mu} \xi_{\nu]} + \partial_{[\mu|} \partial_\rho \xi^\sigma \Gamma_{|\nu]\sigma}^\rho \ .
\ee
The last term corresponds to a first order deformation know as the Green-Schwarz transformation \cite{Green:1984sg}. Due to this contribution, the standard
three-form curvature of the Kalb-Ramond two-form
\be
H_{\mu \nu \rho} = 3 \partial_{[\mu} B_{\nu \rho]} = \partial_{\mu} B_{\nu \rho} + \partial_{\nu} B_{\rho \mu} + \partial_{\rho} B_{\mu \nu} \ ,
\ee
fails to transform covariantly under diffeomorphisms. Instead, the proper covariant curvature tensor is given by
\be
\widehat H_{\mu \nu \rho} = H_{\mu \nu \rho} + 3 \Omega_{\mu \nu \rho} \ , \ \ \ \ \ \delta \widehat H_{\mu \nu \rho} = L_{\xi} \widehat H_{\mu \nu \rho} \ .
\ee
This tensor satisfies the following Bianchi identity
\be
\nabla_{[\mu} \widehat H_{\nu \rho \sigma]} = - \frac 3 4 R_{[\mu \nu}{}^{\delta \lambda} R_{\rho \sigma] \delta \lambda} \ .
\ee

\section{Covariant second order field redefinitions} \label{APP::FieldRedefs}

The lowest order action (\ref{OrdersAction}-\ref{L0}) is
\be
S^{(0)} = \int d^D x \sqrt{-g} e^{- 2 \phi} \left(R + 4 \nabla_\mu \nabla^\mu \phi - 4 \nabla_\mu \phi \nabla^\mu \phi -\frac 1 {12} H_{\mu \nu \rho} H^{\mu \nu \rho}\right) \ . \label{TwoDerivativeAction}
\ee
The equations of motion for the dilaton, two-form and metric are
\be
D\phi = 0 \ , \ \ \ DB_{\mu \nu} = 0 \ , \ \ \ Dg_{\mu \nu} = 0 \ ,
\ee
where
\bea
D\phi &=& R + 4\, \nabla_\mu \nabla^\mu \phi - 4\, \nabla_\mu \phi\, \nabla^\mu \phi -\frac 1 {12} H_{\mu \nu \rho} H^{\mu \nu \rho} \\
DB_{\mu \nu} &=& - 2\, \nabla_\rho \phi \, H^{\rho}{}_{\mu \nu} + \nabla^\rho H_{\rho \mu \nu} \\
Dg_{\mu \nu} &=& - \frac 1 2  g_{\mu \nu}\,  D\phi + R_{\mu \nu} + 2\, \nabla_\mu \nabla_\nu \phi - \frac 1 4 H_{\mu \rho \sigma} H_\nu{}^{\rho \sigma} \ .
\eea

We now explain that any six derivative term in the action that involves at least one of these quantities, can always be eliminated through a second order field redefinition. Indeed, consider a generic six derivative term depending on any of these quantities, then integrating by parts it can always be taken to the form
\be
S' = \int d^Dx \sqrt{-g} e^{-2 \phi} \left( Dg_{\mu \nu} \, T_1^{\mu \nu} + DB_{\mu \nu} \, T_2^{\mu \nu} + D\phi \, T_3\right) \ ,
\ee
for some $T_1^{\mu \nu}$, $T_2^{\mu \nu}$ and $T_3$ that depend on the Riemann tensor, three-form field strength, dilaton and their covariant derivatives. On the other hand, through a covariant second order field redefinition
\be
g_{\mu \nu} \to g_{\mu \nu} + \Delta_2 g_{\mu \nu} \ , \ \ \ B_{\mu \nu} \to B_{\mu \nu} + \Delta_2 B_{\mu \nu} \ , \ \ \ \phi \to \phi + \Delta_2 \phi \ ,
\ee
the two-derivative action (\ref{TwoDerivativeAction}) shifts as
\be
\Delta_2 S = \int d^D x \sqrt{-g} e^{-2 \phi} \left(Dg_{\mu \nu} \, \Delta_2 g^{\mu \nu} + DB_{\mu \nu} \, \Delta_2 B^{\mu \nu} + D\phi \, \Delta_2 \phi  \right) \ ,
\ee
up to boundary terms. Then, fixing the field redefinition to
\be
\Delta_2 g_{\mu \nu} = - T_{1 \mu \nu} \ , \ \ \ \Delta_2 B_{\mu \nu} = - T_{2 \mu \nu} \ , \ \ \ \Delta_2 \phi = - T_{3} \ ,
\ee
eliminates $S'$ from the total action.

\section{Covariant part of $F_2$} \label{APP::F2covariant}

We have delayed to this Appendix the discussion on why it is consistent to ignore the covariant second order contributions to $F_2$ in (\ref{f2covnoncov}).
Consider the most general  parameterization of a tensor with projections $F = \{F\}$
\be
F_{M N} = \left(\begin{matrix} g^{\mu \rho} F^+_{\rho \sigma} g^{\sigma \nu} & g^{\mu \rho} F^-_{\rho \nu} - g^{\mu \rho} F^+_{\rho \sigma} g^{\sigma \delta} B_{\delta \nu} \\
F^-_{\mu \rho} g^{\rho \nu} + B_{\mu \delta} g^{\delta \sigma} F^+_{\sigma \rho} g^{\rho \nu} &
F^+_{\mu \nu} - F^-_{\mu \rho} g^{\rho \sigma} B_{\sigma \nu} - F^-_{\nu \rho} g^{\rho \sigma} B_{\sigma \mu} - B_{\mu \rho} g^{\rho \sigma} F^+_{\sigma \delta} g^{\delta \epsilon} B_{\epsilon \nu}  \end{matrix}\right) \ . \label{ParamF2cov}
\ee
A symmetric tensor with projections $P F P^T$ ($\bar P F \bar P^T$) contains the same degrees of freedom as a symmetric matrix $\underline{F}_{\mu \nu}$ ($\overline{F}_{\mu \nu}$). Thus, a symmetric tensor with projection $F = \{F\} = P F P^T + \bar P F \bar P^T$ contains as degrees of freedom two symmetric matrices, which have been named in (\ref{ParamF2cov}) $F^\pm_{\mu \nu} = \overline{F}_{\mu \nu} \, \pm \, \underline{F}_{\mu \nu}$. Due to its projections $F_2^{\rm cov}$ must admit a parameterization of the form (\ref{ParamF2cov}), with $F^{\pm}_{\mu \nu}$ covariant under diffeomorphisms and gauge transformations.  We can then study how the action gets shifted when a non-vanishing $F_2^{\rm cov}$ is turned on. The result turns out to be simply
\be
S_{F_2^{\rm cov}} = \int d^D x\, \sqrt{-g}\, \nabla_{\mu} \left(- \frac 1 6\, e^{- 2 \phi}\, \nabla^\mu F^{+ \, \nu}_\nu\right) \ .
\ee
This is a covariant total derivative, and can then be dropped out from the action.

Let us now give a complementary argument on why it is possible to ignore the covariant part of $F_2$. We have checked with Cadabra \cite{Peeters:2007wn}, that the most general form of a covariant range two tensor with the projections of $F_2 = \left\{ F_2 \right\}$ takes the form
\be
F_{2MN}^{\rm cov} = a {\cal R}_{M P} {\cal R}^P{}_N + {\cal H}_{M N} \left(b {\cal R}^2 + c {\cal R}_{MN} {\cal R}^{M N} + d \nabla_{M}\nabla_{N}{\cal R}^{M N}\right)
\ee
where ${\cal R}$ and ${\cal R}_{M N} = {\cal R}_{\underline{M} \overline{N}} + {\cal R}_{\overline{M} \underline{N}}$ are the generalized Ricci scalar and tensor of DFT \cite{Ricci}. The covariant derivatives in the last term above are determined/covariant due to the projections of the Ricci tensor. The observation is that the Ricci scalar (tensor) encodes the lowest order equations of motion of the dilaton (metric and two-form), and then any covariant piece of $F_2$ can be ultimately eliminated from the action through field redefinitions.
\newpage
\section{Full action in terms of the double metric} \label{APP::DoubleMetricAction}

In this Appendix we show the full HSZ action written in terms of the double metric ${\cal M}_{M N}$ and dilaton $d$. Schematically it takes the form
\be
S = \int d^{2D}x  e^{- 2 d} \left(L^{(0)} + L^{(2)} + L^{(4)} + L^{(6)} \right) \ , \label{DoubleMetricAction}
\ee
where the supralabel $L^{(i)}$ indicates the number of derivatives that appear explicitly in each term. As can be verified in what follows, the action is at most cubic in the double metric. Explicitly one has
\begingroup\makeatletter\def\f@size{7}\check@mathfonts
\begin{dgroup*}
\begin{dmath*}
L^{(0)}= \frac{1}{2}\, {\cal M}^{M}\,_{M} - \frac{1}{6}\, {\cal M}^{M N} {\cal M}_{M}\,^{P} {\cal M}_{N P}
\end{dmath*}
\begin{dmath*}
L^{(2)}=  - \frac{3}{2}\, {\partial}^{M N}{{\cal M}_{M N}}\,  + 6\, {\cal M}^{M N} {\partial}_{M N}{d}\,  + 6\, {\partial}^{M}{{\cal M}_{M}\,^{N}}\,  {\partial}_{N}{d}\,  - 6\, {\cal M}^{M N} {\partial}_{M}{d}\,  {\partial}_{N}{d}\,  - \frac{1}{12}\, {\cal M}^{M N} {\cal M}^{P Q} {\partial}_{M N}{{\cal M}_{P Q}}\,  + \frac{1}{2}\, {\cal M}^{M N} {\cal M}^{P Q} {\partial}_{M P}{{\cal M}_{N Q}}\,  + {\cal M}^{M N} {\cal M}_{M}\,^{P} {\partial}_{N}\,^{Q}{{\cal M}_{P Q}}\,  + \frac{1}{2}\, {\cal M}^{M N} {\partial}^{P}{{\cal M}_{M P}}\,  {\partial}^{Q}{{\cal M}_{N Q}}\,  + \frac{1}{2}\, {\cal M}^{M N} {\partial}^{P}{{\cal M}_{M}\,^{Q}}\,  {\partial}_{Q}{{\cal M}_{N P}}\,  - \frac{1}{12}\, {\cal M}^{M N} {\partial}^{P}{{\cal M}_{M N}}\,  {\partial}^{Q}{{\cal M}_{P Q}}\,  + \frac{1}{2}\, {\cal M}^{M N} {\partial}_{M}{{\cal M}^{P Q}}\,  {\partial}_{P}{{\cal M}_{N Q}}\,  + \frac{1}{24}\, {\cal M}^{M N} {\partial}_{M}{{\cal M}^{P Q}}\,  {\partial}_{N}{{\cal M}_{P Q}}\,  + {\cal M}^{M N} {\partial}_{M}{{\cal M}_{N}\,^{P}}\,  {\partial}^{Q}{{\cal M}_{P Q}}\,  - 2\, {\cal M}^{M N} {\cal M}_{M}\,^{P} {\cal M}_{N}\,^{Q} {\partial}_{P Q}{d}\,  - 2\, {\cal M}^{M N} {\cal M}^{P Q} {\partial}_{M}{{\cal M}_{N P}}\,  {\partial}_{Q}{d}\,  + \frac{1}{6}\, {\cal M}^{M N} {\cal M}^{P Q} {\partial}_{M}{{\cal M}_{P Q}}\,  {\partial}_{N}{d}\,  - 2\, {\cal M}^{M N} {\cal M}_{M}\,^{P} {\partial}_{N}{{\cal M}_{P}\,^{Q}}\,  {\partial}_{Q}{d}\,  - 2\, {\cal M}^{M N} {\cal M}_{M}\,^{P} {\partial}^{Q}{{\cal M}_{N Q}}\,  {\partial}_{P}{d}\,  + 2\, {\cal M}^{M N} {\cal M}_{M}\,^{P} {\cal M}_{N}\,^{Q} {\partial}_{P}{d}\,  {\partial}_{Q}{d}\,
\end{dmath*}
\begin{dmath*}
L^{(4)}= {\cal M}_{M N} {\cal M}_{P Q} {\partial}_{I J K L}{{\cal M}_{R S}}\,  \left(\right. - \frac{1}{4}\, {\eta}^{M I} {\eta}^{N J} {\eta}^{P K} {\eta}^{Q R} {\eta}^{L S} - \frac{1}{2}\, {\eta}^{M P} {\eta}^{N I} {\eta}^{Q J} {\eta}^{K R} {\eta}^{L S}\left.\right) + {\cal M}_{M N} {\partial}_{P}{{\cal M}_{Q I}}\,  {\partial}_{J K L}{{\cal M}_{R S}}\,  \left(\right. - \frac{1}{4}\, {\eta}^{M J} {\eta}^{N R} {\eta}^{P Q} {\eta}^{I K} {\eta}^{L S} - \frac{1}{4}\, {\eta}^{M J} {\eta}^{N K} {\eta}^{P Q} {\eta}^{I R} {\eta}^{L S} - \frac{1}{2}\, {\eta}^{M J} {\eta}^{N R} {\eta}^{P S} {\eta}^{Q K} {\eta}^{I L} + \frac{1}{6}\, {\eta}^{M J} {\eta}^{N K} {\eta}^{P R} {\eta}^{Q L} {\eta}^{I S} - \frac{1}{2}\, {\eta}^{M Q} {\eta}^{N J} {\eta}^{P I} {\eta}^{K R} {\eta}^{L S} - {\eta}^{M Q} {\eta}^{N J} {\eta}^{P R} {\eta}^{I K} {\eta}^{L S} - \frac{1}{2}\, {\eta}^{M P} {\eta}^{N J} {\eta}^{Q K} {\eta}^{I R} {\eta}^{L S} - \frac{1}{2}\, {\eta}^{M P} {\eta}^{N Q} {\eta}^{I J} {\eta}^{K R} {\eta}^{L S}\left.\right) + {\cal M}_{M N} {\partial}_{P Q}{{\cal M}_{I J}}\,  {\partial}_{K L}{{\cal M}_{R S}}\,  \left(\right. - \frac{1}{2}\, {\eta}^{M P} {\eta}^{N R} {\eta}^{Q I} {\eta}^{J K} {\eta}^{L S} - \frac{1}{2}\, {\eta}^{M P} {\eta}^{N K} {\eta}^{Q I} {\eta}^{J R} {\eta}^{L S} - \frac{1}{4}\, {\eta}^{M P} {\eta}^{N R} {\eta}^{Q S} {\eta}^{I K} {\eta}^{J L} - \frac{1}{12}\, {\eta}^{M P} {\eta}^{N K} {\eta}^{Q R} {\eta}^{I L} {\eta}^{J S} - {\eta}^{M P} {\eta}^{N I} {\eta}^{Q R} {\eta}^{J K} {\eta}^{L S} + \frac{1}{4}\, {\eta}^{M P} {\eta}^{N Q} {\eta}^{I K} {\eta}^{J R} {\eta}^{L S}\left.\right) + {\partial}_{M}{{\cal M}_{N P}}\,  {\partial}_{Q}{{\cal M}_{I J}}\,  {\partial}_{K L}{{\cal M}_{R S}}\,  \left(\right. - \frac{1}{4}\, {\eta}^{M N} {\eta}^{P K} {\eta}^{Q I} {\eta}^{J R} {\eta}^{L S} - \frac{1}{2}\, {\eta}^{M N} {\eta}^{P R} {\eta}^{Q S} {\eta}^{I K} {\eta}^{J L} + \frac{1}{6}\, {\eta}^{M N} {\eta}^{P K} {\eta}^{Q R} {\eta}^{I L} {\eta}^{J S} - {\eta}^{M N} {\eta}^{P I} {\eta}^{Q R} {\eta}^{J K} {\eta}^{L S} + \frac{1}{4}\, {\eta}^{M N} {\eta}^{P Q} {\eta}^{I K} {\eta}^{J R} {\eta}^{L S} - \frac{1}{2}\, {\eta}^{M R} {\eta}^{N I} {\eta}^{P K} {\eta}^{Q S} {\eta}^{J L} - \frac{1}{2}\, {\eta}^{M R} {\eta}^{N Q} {\eta}^{P K} {\eta}^{I L} {\eta}^{J S} + \frac{1}{4}\, {\eta}^{M R} {\eta}^{N Q} {\eta}^{P S} {\eta}^{I K} {\eta}^{J L} - {\eta}^{M I} {\eta}^{N Q} {\eta}^{P K} {\eta}^{J R} {\eta}^{L S}\left.\right) + \frac{5}{2}\, {\cal M}_{M N} {\cal M}_{P Q} {\cal M}_{I J} {\partial}_{K L R S}{d}\,  {\eta}^{M P} {\eta}^{N K} {\eta}^{Q L} {\eta}^{I R} {\eta}^{J S} + {\cal M}_{M N} {\cal M}_{P Q} {\partial}_{I}{{\cal M}_{J K}}\,  {\partial}_{L R S}{d}\,  \left(\right.\frac{5}{2}\, {\eta}^{M I} {\eta}^{N J} {\eta}^{P L} {\eta}^{Q R} {\eta}^{K S} + 4\, {\eta}^{M I} {\eta}^{N L} {\eta}^{P J} {\eta}^{Q R} {\eta}^{K S} + 3\, {\eta}^{M P} {\eta}^{N I} {\eta}^{Q L} {\eta}^{J R} {\eta}^{K S} + \frac{5}{2}\, {\eta}^{M P} {\eta}^{N L} {\eta}^{Q R} {\eta}^{I J} {\eta}^{K S} + \frac{5}{2}\, {\eta}^{M J} {\eta}^{N L} {\eta}^{P R} {\eta}^{Q S} {\eta}^{I K}\left.\right) + {\cal M}_{M N} {\cal M}_{P Q} {\partial}_{I J}{{\cal M}_{K L}}\,  {\partial}_{R S}{d}\,  \left(\right.{\eta}^{M P} {\eta}^{N I} {\eta}^{Q J} {\eta}^{K R} {\eta}^{L S} + 3\, {\eta}^{M I} {\eta}^{N K} {\eta}^{P J} {\eta}^{Q R} {\eta}^{L S} + 5\, {\eta}^{M P} {\eta}^{N I} {\eta}^{Q R} {\eta}^{J K} {\eta}^{L S} + 3\, {\eta}^{M I} {\eta}^{N R} {\eta}^{P K} {\eta}^{Q S} {\eta}^{J L}\left.\right) + {\cal M}_{M N} {\cal M}_{P Q} {\partial}_{I J K}{{\cal M}_{L R}}\,  {\partial}_{S}{d}\,  \left(\right.\frac{1}{2}\, {\eta}^{M I} {\eta}^{N J} {\eta}^{P K} {\eta}^{Q L} {\eta}^{S R} + 2\, {\eta}^{M P} {\eta}^{N I} {\eta}^{Q J} {\eta}^{S L} {\eta}^{K R} + \frac{1}{2}\, {\eta}^{M S} {\eta}^{N I} {\eta}^{P J} {\eta}^{Q L} {\eta}^{K R} + {\eta}^{M P} {\eta}^{N S} {\eta}^{Q I} {\eta}^{J L} {\eta}^{K R} + \frac{1}{2}\, {\eta}^{M S} {\eta}^{N L} {\eta}^{P I} {\eta}^{Q J} {\eta}^{K R}\left.\right) + {\cal M}_{M N} {\partial}_{P}{{\cal M}_{Q I}}\,  {\partial}_{J}{{\cal M}_{K L}}\,  {\partial}_{R S}{d}\,  \left(\right.\frac{5}{2}\, {\eta}^{M Q} {\eta}^{N R} {\eta}^{P I} {\eta}^{J K} {\eta}^{L S} + 4\, {\eta}^{M Q} {\eta}^{N R} {\eta}^{P K} {\eta}^{I J} {\eta}^{L S} + {\eta}^{M P} {\eta}^{N R} {\eta}^{Q J} {\eta}^{I K} {\eta}^{L S} + 2\, {\eta}^{M P} {\eta}^{N K} {\eta}^{Q J} {\eta}^{I R} {\eta}^{L S} + {\eta}^{M P} {\eta}^{N J} {\eta}^{Q K} {\eta}^{I R} {\eta}^{L S} + 3\, {\eta}^{M P} {\eta}^{N R} {\eta}^{Q K} {\eta}^{I S} {\eta}^{J L} + {\eta}^{M P} {\eta}^{N K} {\eta}^{Q R} {\eta}^{I S} {\eta}^{J L} + {\eta}^{M P} {\eta}^{N Q} {\eta}^{I J} {\eta}^{K R} {\eta}^{L S} + \frac{5}{2}\, {\eta}^{M P} {\eta}^{N Q} {\eta}^{I R} {\eta}^{J K} {\eta}^{L S}\left.\right) + {\cal M}_{M N} {\partial}_{P}{{\cal M}_{Q I}}\,  {\partial}_{J K}{{\cal M}_{L R}}\,  {\partial}_{S}{d}\,  \left(\right.\frac{1}{2}\, {\eta}^{M S} {\eta}^{N L} {\eta}^{P Q} {\eta}^{I J} {\eta}^{K R} + \frac{1}{2}\, {\eta}^{M J} {\eta}^{N L} {\eta}^{P Q} {\eta}^{I K} {\eta}^{S R} + \frac{1}{2}\, {\eta}^{M J} {\eta}^{N K} {\eta}^{P Q} {\eta}^{I L} {\eta}^{S R} + \frac{1}{2}\, {\eta}^{M S} {\eta}^{N J} {\eta}^{P Q} {\eta}^{I L} {\eta}^{K R} + {\eta}^{M S} {\eta}^{N L} {\eta}^{P R} {\eta}^{Q J} {\eta}^{I K} - \frac{1}{2}\, {\eta}^{M J} {\eta}^{N K} {\eta}^{P L} {\eta}^{Q S} {\eta}^{I R} + 2\, {\eta}^{M J} {\eta}^{N L} {\eta}^{P R} {\eta}^{Q S} {\eta}^{I K} - \frac{1}{3}\, {\eta}^{M S} {\eta}^{N J} {\eta}^{P L} {\eta}^{Q K} {\eta}^{I R} + 2\, {\eta}^{M Q} {\eta}^{N J} {\eta}^{P I} {\eta}^{S L} {\eta}^{K R} + 2\, {\eta}^{M Q} {\eta}^{N S} {\eta}^{P L} {\eta}^{I J} {\eta}^{K R} + 2\, {\eta}^{M Q} {\eta}^{N J} {\eta}^{P L} {\eta}^{I K} {\eta}^{S R} + {\eta}^{M Q} {\eta}^{N J} {\eta}^{P L} {\eta}^{I S} {\eta}^{K R} - \frac{1}{2}\, {\eta}^{M P} {\eta}^{N S} {\eta}^{Q J} {\eta}^{I L} {\eta}^{K R} + {\eta}^{M P} {\eta}^{N L} {\eta}^{Q S} {\eta}^{I J} {\eta}^{K R} + {\eta}^{M P} {\eta}^{N J} {\eta}^{Q K} {\eta}^{I L} {\eta}^{S R} + 2\, {\eta}^{M P} {\eta}^{N J} {\eta}^{Q S} {\eta}^{I L} {\eta}^{K R} + 2\, {\eta}^{M P} {\eta}^{N Q} {\eta}^{I J} {\eta}^{S L} {\eta}^{K R}\left.\right) + {\partial}_{M}{{\cal M}_{N P}}\,  {\partial}_{Q}{{\cal M}_{I J}}\,  {\partial}_{K}{{\cal M}_{L R}}\,  {\partial}_{S}{d}\,  \left(\right.2\, {\eta}^{M N} {\eta}^{P I} {\eta}^{Q L} {\eta}^{J K} {\eta}^{R S} - \frac{1}{2}\, {\eta}^{M N} {\eta}^{P Q} {\eta}^{I K} {\eta}^{J L} {\eta}^{R S} + \frac{1}{2}\, {\eta}^{M N} {\eta}^{P Q} {\eta}^{I L} {\eta}^{J S} {\eta}^{K R} + 2\, {\eta}^{M I} {\eta}^{N Q} {\eta}^{P K} {\eta}^{J L} {\eta}^{R S}\left.\right) + {\cal M}_{M N} {\cal M}_{P Q} {\cal M}_{I J} {\partial}_{K}{d}\,  {\partial}_{L R S}{d}\,  \left(\right. - 5\, {\eta}^{M P} {\eta}^{N K} {\eta}^{Q L} {\eta}^{I R} {\eta}^{J S} - 5\, {\eta}^{M P} {\eta}^{N L} {\eta}^{Q R} {\eta}^{I K} {\eta}^{J S}\left.\right) - 8\, {\cal M}_{M N} {\cal M}_{P Q} {\cal M}_{I J} {\partial}_{K L}{d}\,  {\partial}_{R S}{d}\,  {\eta}^{M P} {\eta}^{N K} {\eta}^{Q R} {\eta}^{I L} {\eta}^{J S} + {\cal M}_{M N} {\cal M}_{P Q} {\partial}_{I}{{\cal M}_{J K}}\,  {\partial}_{L}{d}\,  {\partial}_{R S}{d}\,  \left(\right. - 5\, {\eta}^{M I} {\eta}^{N J} {\eta}^{P L} {\eta}^{Q R} {\eta}^{K S} - 2\, {\eta}^{M P} {\eta}^{N I} {\eta}^{Q L} {\eta}^{J R} {\eta}^{K S} - 6\, {\eta}^{M I} {\eta}^{N R} {\eta}^{P J} {\eta}^{Q L} {\eta}^{K S} - 10\, {\eta}^{M P} {\eta}^{N I} {\eta}^{Q R} {\eta}^{J L} {\eta}^{K S} - 6\, {\eta}^{M I} {\eta}^{N R} {\eta}^{P J} {\eta}^{Q S} {\eta}^{K L} - 5\, {\eta}^{M P} {\eta}^{N L} {\eta}^{Q R} {\eta}^{I J} {\eta}^{K S} - 5\, {\eta}^{M J} {\eta}^{N R} {\eta}^{P L} {\eta}^{Q S} {\eta}^{I K}\left.\right) + {\cal M}_{M N} {\cal M}_{P Q} {\partial}_{I J}{{\cal M}_{K L}}\,  {\partial}_{R}{d}\,  {\partial}_{S}{d}\,  \left(\right. - 2\, {\eta}^{M P} {\eta}^{N I} {\eta}^{Q J} {\eta}^{R K} {\eta}^{S L} - {\eta}^{M R} {\eta}^{N I} {\eta}^{P J} {\eta}^{Q K} {\eta}^{S L} - {\eta}^{M R} {\eta}^{N I} {\eta}^{P S} {\eta}^{Q K} {\eta}^{J L} - 4\, {\eta}^{M P} {\eta}^{N R} {\eta}^{Q I} {\eta}^{S K} {\eta}^{J L} - {\eta}^{M R} {\eta}^{N K} {\eta}^{P I} {\eta}^{Q J} {\eta}^{S L}\left.\right) + {\cal M}_{M N} {\partial}_{P}{{\cal M}_{Q I}}\,  {\partial}_{J}{{\cal M}_{K L}}\,  {\partial}_{R}{d}\,  {\partial}_{S}{d}\,  \left(\right. - 4\, {\eta}^{M Q} {\eta}^{N R} {\eta}^{P K} {\eta}^{I J} {\eta}^{L S} - {\eta}^{M Q} {\eta}^{N R} {\eta}^{P K} {\eta}^{I S} {\eta}^{J L} + {\eta}^{M P} {\eta}^{N R} {\eta}^{Q J} {\eta}^{I K} {\eta}^{L S} - 2\, {\eta}^{M P} {\eta}^{N K} {\eta}^{Q J} {\eta}^{I R} {\eta}^{L S} - 2\, {\eta}^{M P} {\eta}^{N J} {\eta}^{Q K} {\eta}^{I R} {\eta}^{L S} - {\eta}^{M P} {\eta}^{N R} {\eta}^{Q K} {\eta}^{I S} {\eta}^{J L} - 2\, {\eta}^{M P} {\eta}^{N K} {\eta}^{Q R} {\eta}^{I S} {\eta}^{J L} - 2\, {\eta}^{M P} {\eta}^{N Q} {\eta}^{I J} {\eta}^{K R} {\eta}^{L S}\left.\right) + 10\, {\cal M}_{M N} {\cal M}_{P Q} {\cal M}_{I J} {\partial}_{K}{d}\,  {\partial}_{L}{d}\,  {\partial}_{R S}{d}\,  {\eta}^{M P} {\eta}^{N K} {\eta}^{Q R} {\eta}^{I L} {\eta}^{J S} + {\cal M}_{M N} {\cal M}_{P Q} {\partial}_{I}{{\cal M}_{J K}}\,  {\partial}_{L}{d}\,  {\partial}_{R}{d}\,  {\partial}_{S}{d}\,  \left(\right.2\, {\eta}^{M I} {\eta}^{N L} {\eta}^{P J} {\eta}^{Q R} {\eta}^{K S} + 4\, {\eta}^{M P} {\eta}^{N I} {\eta}^{Q L} {\eta}^{J R} {\eta}^{K S}\left.\right)
\end{dmath*}
\end{dgroup*}
\endgroup

\begingroup\makeatletter\def\f@size{7}\check@mathfonts
\begin{dgroup*}
\begin{dmath*}
L^{(6)}= \frac{1}{8}\, {\cal M}_{M N} {\cal M}_{P Q} {\partial}_{I J K L R S}{{\cal M}_{G H}}\,  {\eta}^{M I} {\eta}^{N J} {\eta}^{P K} {\eta}^{Q L} {\eta}^{R G} {\eta}^{S H} + {\cal M}_{M N} {\partial}_{P}{{\cal M}_{Q I}}\,  {\partial}_{J K L R S}{{\cal M}_{G H}}\,  \left(\right.\frac{1}{4}\, {\eta}^{M J} {\eta}^{N K} {\eta}^{P Q} {\eta}^{I L} {\eta}^{R G} {\eta}^{S H} + \frac{1}{8}\, {\eta}^{M J} {\eta}^{N K} {\eta}^{P G} {\eta}^{Q L} {\eta}^{I R} {\eta}^{S H} + \frac{1}{4}\, {\eta}^{M P} {\eta}^{N J} {\eta}^{Q K} {\eta}^{I L} {\eta}^{R G} {\eta}^{S H}\left.\right) + {\cal M}_{M N} {\partial}_{P Q}{{\cal M}_{I J}}\,  {\partial}_{K L R S}{{\cal M}_{G H}}\,  \left(\right.\frac{1}{4}\, {\eta}^{M K} {\eta}^{N L} {\eta}^{P I} {\eta}^{Q G} {\eta}^{J R} {\eta}^{S H} - \frac{1}{24}\, {\eta}^{M K} {\eta}^{N L} {\eta}^{P G} {\eta}^{Q H} {\eta}^{I R} {\eta}^{J S} + \frac{3}{4}\, {\eta}^{M P} {\eta}^{N K} {\eta}^{Q I} {\eta}^{J L} {\eta}^{R G} {\eta}^{S H} + \frac{1}{4}\, {\eta}^{M P} {\eta}^{N K} {\eta}^{Q G} {\eta}^{I L} {\eta}^{J R} {\eta}^{S H} - \frac{1}{8}\, {\eta}^{M P} {\eta}^{N Q} {\eta}^{I K} {\eta}^{J L} {\eta}^{R G} {\eta}^{S H}\left.\right) + {\cal M}_{M N} {\partial}_{P Q I}{{\cal M}_{J K}}\,  {\partial}_{L R S}{{\cal M}_{G H}}\,  \left(\right.\frac{1}{8}\, {\eta}^{M P} {\eta}^{N L} {\eta}^{Q J} {\eta}^{I K} {\eta}^{R G} {\eta}^{S H} + \frac{1}{2}\, {\eta}^{M P} {\eta}^{N L} {\eta}^{Q J} {\eta}^{I G} {\eta}^{K R} {\eta}^{S H} + \frac{1}{48}\, {\eta}^{M P} {\eta}^{N L} {\eta}^{Q G} {\eta}^{I H} {\eta}^{J R} {\eta}^{K S} + \frac{1}{8}\, {\eta}^{M P} {\eta}^{N Q} {\eta}^{I J} {\eta}^{K L} {\eta}^{R G} {\eta}^{S H} - \frac{1}{8}\, {\eta}^{M P} {\eta}^{N Q} {\eta}^{I G} {\eta}^{J L} {\eta}^{K R} {\eta}^{S H}\left.\right) + {\partial}_{M}{{\cal M}_{N P}}\,  {\partial}_{Q}{{\cal M}_{I J}}\,  {\partial}_{K L R S}{{\cal M}_{G H}}\,  \left(\right.\frac{1}{8}\, {\eta}^{M N} {\eta}^{P K} {\eta}^{Q I} {\eta}^{J L} {\eta}^{R G} {\eta}^{S H} + \frac{1}{8}\, {\eta}^{M N} {\eta}^{P K} {\eta}^{Q G} {\eta}^{I L} {\eta}^{J R} {\eta}^{S H} - \frac{1}{8}\, {\eta}^{M N} {\eta}^{P Q} {\eta}^{I K} {\eta}^{J L} {\eta}^{R G} {\eta}^{S H} + \frac{1}{8}\, {\eta}^{M G} {\eta}^{N K} {\eta}^{P L} {\eta}^{Q H} {\eta}^{I R} {\eta}^{J S} + \frac{1}{2}\, {\eta}^{M I} {\eta}^{N Q} {\eta}^{P K} {\eta}^{J L} {\eta}^{R G} {\eta}^{S H}\left.\right) + {\partial}_{M}{{\cal M}_{N P}}\,  {\partial}_{Q I}{{\cal M}_{J K}}\,  {\partial}_{L R S}{{\cal M}_{G H}}\,  \left(\right.\frac{1}{4}\, {\eta}^{M N} {\eta}^{P L} {\eta}^{Q J} {\eta}^{I G} {\eta}^{K R} {\eta}^{S H} - \frac{1}{24}\, {\eta}^{M N} {\eta}^{P L} {\eta}^{Q G} {\eta}^{I H} {\eta}^{J R} {\eta}^{K S} + \frac{1}{8}\, {\eta}^{M N} {\eta}^{P Q} {\eta}^{I J} {\eta}^{K L} {\eta}^{R G} {\eta}^{S H} - \frac{1}{8}\, {\eta}^{M N} {\eta}^{P Q} {\eta}^{I G} {\eta}^{J L} {\eta}^{K R} {\eta}^{S H} + \frac{1}{2}\, {\eta}^{M G} {\eta}^{N L} {\eta}^{P R} {\eta}^{Q J} {\eta}^{I H} {\eta}^{K S} - \frac{1}{4}\, {\eta}^{M J} {\eta}^{N L} {\eta}^{P R} {\eta}^{Q K} {\eta}^{I G} {\eta}^{S H} - \frac{1}{8}\, {\eta}^{M J} {\eta}^{N L} {\eta}^{P R} {\eta}^{Q G} {\eta}^{I H} {\eta}^{K S} - \frac{1}{4}\, {\eta}^{M G} {\eta}^{N Q} {\eta}^{P I} {\eta}^{J L} {\eta}^{K R} {\eta}^{S H} + \frac{1}{4}\, {\eta}^{M J} {\eta}^{N Q} {\eta}^{P I} {\eta}^{K L} {\eta}^{R G} {\eta}^{S H} + \frac{1}{2}\, {\eta}^{M G} {\eta}^{N Q} {\eta}^{P L} {\eta}^{I J} {\eta}^{K R} {\eta}^{S H} + \frac{1}{2}\, {\eta}^{M J} {\eta}^{N Q} {\eta}^{P L} {\eta}^{I K} {\eta}^{R G} {\eta}^{S H} + {\eta}^{M J} {\eta}^{N Q} {\eta}^{P L} {\eta}^{I G} {\eta}^{K R} {\eta}^{S H} + \frac{1}{4}\, {\eta}^{M G} {\eta}^{N Q} {\eta}^{P L} {\eta}^{I H} {\eta}^{J R} {\eta}^{K S}\left.\right) + {\partial}_{M N}{{\cal M}_{P Q}}\,  {\partial}_{I J}{{\cal M}_{K L}}\,  {\partial}_{R S}{{\cal M}_{G H}}\,  \left(\right. - \frac{1}{24}\, {\eta}^{M G} {\eta}^{N H} {\eta}^{P I} {\eta}^{Q J} {\eta}^{K R} {\eta}^{L S} + \frac{1}{6}\, {\eta}^{M K} {\eta}^{N G} {\eta}^{P I} {\eta}^{Q R} {\eta}^{J H} {\eta}^{L S} + \frac{1}{2}\, {\eta}^{M P} {\eta}^{N K} {\eta}^{Q I} {\eta}^{J G} {\eta}^{L R} {\eta}^{S H}\left.\right) - \frac{1}{2}\, {\cal M}_{M N} {\cal M}_{P Q} {\cal M}_{I J} {\partial}_{K L R S G H}{d}\,  {\eta}^{M K} {\eta}^{N L} {\eta}^{P R} {\eta}^{Q S} {\eta}^{I G} {\eta}^{J H} + {\cal M}_{M N} {\cal M}_{P Q} {\partial}_{I}{{\cal M}_{J K}}\,  {\partial}_{L R S G H}{d}\,  \left(\right. - 3\, {\eta}^{M I} {\eta}^{N L} {\eta}^{P R} {\eta}^{Q S} {\eta}^{J G} {\eta}^{K H} - \frac{3}{2}\, {\eta}^{M L} {\eta}^{N R} {\eta}^{P S} {\eta}^{Q G} {\eta}^{I J} {\eta}^{K H}\left.\right) + {\cal M}_{M N} {\cal M}_{P Q} {\partial}_{I J}{{\cal M}_{K L}}\,  {\partial}_{R S G H}{d}\,  \left(\right. - \frac{1}{4}\, {\eta}^{M I} {\eta}^{N J} {\eta}^{P R} {\eta}^{Q S} {\eta}^{K G} {\eta}^{L H} - 2\, {\eta}^{M I} {\eta}^{N R} {\eta}^{P J} {\eta}^{Q S} {\eta}^{K G} {\eta}^{L H} - \frac{11}{2}\, {\eta}^{M I} {\eta}^{N R} {\eta}^{P S} {\eta}^{Q G} {\eta}^{J K} {\eta}^{L H}\left.\right) + {\cal M}_{M N} {\cal M}_{P Q} {\partial}_{I J K}{{\cal M}_{L R}}\,  {\partial}_{S G H}{d}\,  \left(\right. - {\eta}^{M I} {\eta}^{N J} {\eta}^{P K} {\eta}^{Q S} {\eta}^{L G} {\eta}^{R H} - \frac{3}{2}\, {\eta}^{M I} {\eta}^{N J} {\eta}^{P S} {\eta}^{Q G} {\eta}^{K L} {\eta}^{R H} - 4\, {\eta}^{M I} {\eta}^{N S} {\eta}^{P J} {\eta}^{Q G} {\eta}^{K L} {\eta}^{R H} - \frac{5}{4}\, {\eta}^{M I} {\eta}^{N S} {\eta}^{P G} {\eta}^{Q H} {\eta}^{J L} {\eta}^{K R}\left.\right) + {\cal M}_{M N} {\cal M}_{P Q} {\partial}_{I J K L}{{\cal M}_{R S}}\,  {\partial}_{G H}{d}\,  \left(\right. - \frac{1}{4}\, {\eta}^{M I} {\eta}^{N J} {\eta}^{P K} {\eta}^{Q L} {\eta}^{G R} {\eta}^{H S} - \frac{5}{2}\, {\eta}^{M G} {\eta}^{N I} {\eta}^{P J} {\eta}^{Q K} {\eta}^{H R} {\eta}^{L S} - \frac{3}{2}\, {\eta}^{M G} {\eta}^{N I} {\eta}^{P H} {\eta}^{Q J} {\eta}^{K R} {\eta}^{L S}\left.\right) + {\cal M}_{M N} {\cal M}_{P Q} {\partial}_{I J K L R}{{\cal M}_{S G}}\,  {\partial}_{H}{d}\,  \left(\right. - \frac{1}{2}\, {\eta}^{M I} {\eta}^{N J} {\eta}^{P K} {\eta}^{Q L} {\eta}^{H S} {\eta}^{R G} - \frac{1}{2}\, {\eta}^{M H} {\eta}^{N I} {\eta}^{P J} {\eta}^{Q K} {\eta}^{L S} {\eta}^{R G}\left.\right) + {\cal M}_{M N} {\partial}_{P}{{\cal M}_{Q I}}\,  {\partial}_{J}{{\cal M}_{K L}}\,  {\partial}_{R S G H}{d}\,  \left(\right. - \frac{1}{4}\, {\eta}^{M R} {\eta}^{N S} {\eta}^{P Q} {\eta}^{I J} {\eta}^{K G} {\eta}^{L H} - \frac{3}{2}\, {\eta}^{M R} {\eta}^{N S} {\eta}^{P Q} {\eta}^{I G} {\eta}^{J K} {\eta}^{L H} - \frac{5}{2}\, {\eta}^{M R} {\eta}^{N S} {\eta}^{P K} {\eta}^{Q J} {\eta}^{I G} {\eta}^{L H} - 2\, {\eta}^{M P} {\eta}^{N R} {\eta}^{Q J} {\eta}^{I S} {\eta}^{K G} {\eta}^{L H} - {\eta}^{M P} {\eta}^{N J} {\eta}^{Q R} {\eta}^{I S} {\eta}^{K G} {\eta}^{L H} - 3\, {\eta}^{M P} {\eta}^{N R} {\eta}^{Q S} {\eta}^{I G} {\eta}^{J K} {\eta}^{L H}\left.\right) + {\cal M}_{M N} {\partial}_{P}{{\cal M}_{Q I}}\,  {\partial}_{J K}{{\cal M}_{L R}}\,  {\partial}_{S G H}{d}\,  \left(\right. - \frac{3}{2}\, {\eta}^{M S} {\eta}^{N G} {\eta}^{P Q} {\eta}^{I J} {\eta}^{K L} {\eta}^{R H} - {\eta}^{M J} {\eta}^{N S} {\eta}^{P Q} {\eta}^{I K} {\eta}^{L G} {\eta}^{R H} - \frac{1}{4}\, {\eta}^{M J} {\eta}^{N K} {\eta}^{P Q} {\eta}^{I S} {\eta}^{L G} {\eta}^{R H} - \frac{11}{2}\, {\eta}^{M J} {\eta}^{N S} {\eta}^{P Q} {\eta}^{I G} {\eta}^{K L} {\eta}^{R H} - \frac{5}{2}\, {\eta}^{M S} {\eta}^{N G} {\eta}^{P L} {\eta}^{Q J} {\eta}^{I H} {\eta}^{K R} - \frac{5}{4}\, {\eta}^{M S} {\eta}^{N G} {\eta}^{P L} {\eta}^{Q J} {\eta}^{I K} {\eta}^{R H} - 4\, {\eta}^{M J} {\eta}^{N S} {\eta}^{P L} {\eta}^{Q K} {\eta}^{I G} {\eta}^{R H} - \frac{3}{2}\, {\eta}^{M J} {\eta}^{N S} {\eta}^{P L} {\eta}^{Q G} {\eta}^{I H} {\eta}^{K R} + \frac{1}{2}\, {\eta}^{M J} {\eta}^{N K} {\eta}^{P L} {\eta}^{Q S} {\eta}^{I G} {\eta}^{R H} - 4\, {\eta}^{M P} {\eta}^{N S} {\eta}^{Q J} {\eta}^{I G} {\eta}^{K L} {\eta}^{R H} - 2\, {\eta}^{M P} {\eta}^{N J} {\eta}^{Q K} {\eta}^{I S} {\eta}^{L G} {\eta}^{R H} - \frac{7}{2}\, {\eta}^{M P} {\eta}^{N J} {\eta}^{Q S} {\eta}^{I G} {\eta}^{K L} {\eta}^{R H}\left.\right) + {\cal M}_{M N} {\partial}_{P}{{\cal M}_{Q I}}\,  {\partial}_{J K L}{{\cal M}_{R S}}\,  {\partial}_{G H}{d}\,  \left(\right. - \frac{1}{2}\, {\eta}^{M J} {\eta}^{N K} {\eta}^{P Q} {\eta}^{I L} {\eta}^{G R} {\eta}^{H S} - \frac{3}{2}\, {\eta}^{M J} {\eta}^{N K} {\eta}^{P Q} {\eta}^{I G} {\eta}^{H R} {\eta}^{L S} - \frac{5}{2}\, {\eta}^{M G} {\eta}^{N J} {\eta}^{P Q} {\eta}^{I K} {\eta}^{H R} {\eta}^{L S} - \frac{5}{4}\, {\eta}^{M G} {\eta}^{N J} {\eta}^{P Q} {\eta}^{I H} {\eta}^{K R} {\eta}^{L S} - \frac{1}{4}\, {\eta}^{M J} {\eta}^{N K} {\eta}^{P R} {\eta}^{Q G} {\eta}^{I H} {\eta}^{L S} - \frac{3}{2}\, {\eta}^{M G} {\eta}^{N J} {\eta}^{P R} {\eta}^{Q K} {\eta}^{I L} {\eta}^{H S} - 3\, {\eta}^{M G} {\eta}^{N J} {\eta}^{P R} {\eta}^{Q H} {\eta}^{I K} {\eta}^{L S} - \frac{1}{2}\, {\eta}^{M P} {\eta}^{N G} {\eta}^{Q J} {\eta}^{I K} {\eta}^{H R} {\eta}^{L S} - \frac{3}{2}\, {\eta}^{M P} {\eta}^{N G} {\eta}^{Q H} {\eta}^{I J} {\eta}^{K R} {\eta}^{L S} - \frac{1}{2}\, {\eta}^{M P} {\eta}^{N J} {\eta}^{Q K} {\eta}^{I L} {\eta}^{G R} {\eta}^{H S} - 4\, {\eta}^{M P} {\eta}^{N J} {\eta}^{Q G} {\eta}^{I K} {\eta}^{H R} {\eta}^{L S} - \frac{1}{2}\, {\eta}^{M P} {\eta}^{N J} {\eta}^{Q G} {\eta}^{I H} {\eta}^{K R} {\eta}^{L S}\left.\right) + {\cal M}_{M N} {\partial}_{P}{{\cal M}_{Q I}}\,  {\partial}_{J K L R}{{\cal M}_{S G}}\,  {\partial}_{H}{d}\,  \left(\right. - {\eta}^{M J} {\eta}^{N K} {\eta}^{P Q} {\eta}^{I L} {\eta}^{H S} {\eta}^{R G} - \frac{1}{2}\, {\eta}^{M H} {\eta}^{N J} {\eta}^{P Q} {\eta}^{I K} {\eta}^{L S} {\eta}^{R G} - \frac{1}{4}\, {\eta}^{M J} {\eta}^{N K} {\eta}^{P S} {\eta}^{Q L} {\eta}^{I R} {\eta}^{H G} - \frac{1}{2}\, {\eta}^{M J} {\eta}^{N K} {\eta}^{P S} {\eta}^{Q H} {\eta}^{I L} {\eta}^{R G} - \frac{1}{4}\, {\eta}^{M H} {\eta}^{N J} {\eta}^{P S} {\eta}^{Q K} {\eta}^{I L} {\eta}^{R G} + \frac{1}{4}\, {\eta}^{M P} {\eta}^{N H} {\eta}^{Q J} {\eta}^{I K} {\eta}^{L S} {\eta}^{R G} - {\eta}^{M P} {\eta}^{N J} {\eta}^{Q K} {\eta}^{I L} {\eta}^{H S} {\eta}^{R G} - \frac{3}{2}\, {\eta}^{M P} {\eta}^{N J} {\eta}^{Q H} {\eta}^{I K} {\eta}^{L S} {\eta}^{R G}\left.\right) + {\cal M}_{M N} {\partial}_{P Q}{{\cal M}_{I J}}\,  {\partial}_{K L}{{\cal M}_{R S}}\,  {\partial}_{G H}{d}\,  \left(\right. - 3\, {\eta}^{M P} {\eta}^{N G} {\eta}^{Q I} {\eta}^{J K} {\eta}^{L R} {\eta}^{S H} - \frac{3}{2}\, {\eta}^{M P} {\eta}^{N K} {\eta}^{Q I} {\eta}^{J L} {\eta}^{R G} {\eta}^{S H} - 3\, {\eta}^{M P} {\eta}^{N K} {\eta}^{Q I} {\eta}^{J G} {\eta}^{L R} {\eta}^{S H} - 3\, {\eta}^{M P} {\eta}^{N G} {\eta}^{Q R} {\eta}^{I K} {\eta}^{J H} {\eta}^{L S} - \frac{1}{2}\, {\eta}^{M P} {\eta}^{N G} {\eta}^{Q R} {\eta}^{I K} {\eta}^{J L} {\eta}^{S H} - {\eta}^{M P} {\eta}^{N K} {\eta}^{Q R} {\eta}^{I L} {\eta}^{J G} {\eta}^{S H} + \frac{1}{4}\, {\eta}^{M P} {\eta}^{N Q} {\eta}^{I K} {\eta}^{J L} {\eta}^{R G} {\eta}^{S H} + \frac{1}{2}\, {\eta}^{M P} {\eta}^{N Q} {\eta}^{I K} {\eta}^{J G} {\eta}^{L R} {\eta}^{S H}\left.\right) + {\cal M}_{M N} {\partial}_{P Q}{{\cal M}_{I J}}\,  {\partial}_{K L R}{{\cal M}_{S G}}\,  {\partial}_{H}{d}\,  \left(\right. - \frac{1}{2}\, {\eta}^{M H} {\eta}^{N K} {\eta}^{P I} {\eta}^{Q S} {\eta}^{J L} {\eta}^{R G} + \frac{1}{12}\, {\eta}^{M H} {\eta}^{N K} {\eta}^{P S} {\eta}^{Q G} {\eta}^{I L} {\eta}^{J R} - \frac{1}{2}\, {\eta}^{M K} {\eta}^{N L} {\eta}^{H I} {\eta}^{P J} {\eta}^{Q S} {\eta}^{R G} + \frac{1}{4}\, {\eta}^{M K} {\eta}^{N L} {\eta}^{H I} {\eta}^{P S} {\eta}^{Q G} {\eta}^{J R} - \frac{1}{2}\, {\eta}^{M K} {\eta}^{N L} {\eta}^{H S} {\eta}^{P I} {\eta}^{Q G} {\eta}^{J R} - {\eta}^{M P} {\eta}^{N K} {\eta}^{H I} {\eta}^{Q J} {\eta}^{L S} {\eta}^{R G} - 2\, {\eta}^{M P} {\eta}^{N K} {\eta}^{H I} {\eta}^{Q S} {\eta}^{J L} {\eta}^{R G} - 3\, {\eta}^{M P} {\eta}^{N K} {\eta}^{H S} {\eta}^{Q I} {\eta}^{J L} {\eta}^{R G} - \frac{1}{2}\, {\eta}^{M P} {\eta}^{N K} {\eta}^{H S} {\eta}^{Q G} {\eta}^{I L} {\eta}^{J R} - \frac{1}{4}\, {\eta}^{M P} {\eta}^{N Q} {\eta}^{H I} {\eta}^{J K} {\eta}^{L S} {\eta}^{R G} + \frac{1}{2}\, {\eta}^{M P} {\eta}^{N Q} {\eta}^{H S} {\eta}^{I K} {\eta}^{J L} {\eta}^{R G} - \frac{1}{4}\, {\eta}^{M H} {\eta}^{N P} {\eta}^{Q I} {\eta}^{J K} {\eta}^{L S} {\eta}^{R G} + \frac{1}{4}\, {\eta}^{M H} {\eta}^{N P} {\eta}^{Q S} {\eta}^{I K} {\eta}^{J L} {\eta}^{R G}\left.\right) + {\partial}_{M}{{\cal M}_{N P}}\,  {\partial}_{Q}{{\cal M}_{I J}}\,  {\partial}_{K}{{\cal M}_{L R}}\,  {\partial}_{S G H}{d}\,  \left(\right. - \frac{1}{2}\, {\eta}^{M N} {\eta}^{P S} {\eta}^{Q I} {\eta}^{J G} {\eta}^{K L} {\eta}^{R H} - \frac{5}{2}\, {\eta}^{M N} {\eta}^{P S} {\eta}^{Q L} {\eta}^{I K} {\eta}^{J G} {\eta}^{R H} + \frac{1}{2}\, {\eta}^{M N} {\eta}^{P Q} {\eta}^{I K} {\eta}^{J S} {\eta}^{L G} {\eta}^{R H} - \frac{1}{4}\, {\eta}^{M N} {\eta}^{P Q} {\eta}^{I S} {\eta}^{J G} {\eta}^{K L} {\eta}^{R H} - 3\, {\eta}^{M I} {\eta}^{N Q} {\eta}^{P K} {\eta}^{J S} {\eta}^{L G} {\eta}^{R H}\left.\right) + {\partial}_{M}{{\cal M}_{N P}}\,  {\partial}_{Q}{{\cal M}_{I J}}\,  {\partial}_{K L}{{\cal M}_{R S}}\,  {\partial}_{G H}{d}\,  \left(\right. - \frac{1}{4}\, {\eta}^{M N} {\eta}^{P K} {\eta}^{Q I} {\eta}^{J L} {\eta}^{R G} {\eta}^{S H} - \frac{3}{2}\, {\eta}^{M N} {\eta}^{P K} {\eta}^{Q I} {\eta}^{J G} {\eta}^{L R} {\eta}^{S H} - \frac{5}{2}\, {\eta}^{M N} {\eta}^{P G} {\eta}^{Q R} {\eta}^{I K} {\eta}^{J H} {\eta}^{L S} - \frac{5}{4}\, {\eta}^{M N} {\eta}^{P G} {\eta}^{Q R} {\eta}^{I K} {\eta}^{J L} {\eta}^{S H} - \frac{1}{4}\, {\eta}^{M N} {\eta}^{P K} {\eta}^{Q R} {\eta}^{I G} {\eta}^{J H} {\eta}^{L S} + \frac{1}{4}\, {\eta}^{M N} {\eta}^{P Q} {\eta}^{I K} {\eta}^{J L} {\eta}^{R G} {\eta}^{S H}  + \frac{1}{2}\, {\eta}^{M N} {\eta}^{P Q} {\eta}^{I K} {\eta}^{J G} {\eta}^{L R} {\eta}^{S H} - \frac{1}{2}\, {\eta}^{M R} {\eta}^{N K} {\eta}^{P L} {\eta}^{Q S} {\eta}^{I G} {\eta}^{J H} - 2\, {\eta}^{M R} {\eta}^{N K} {\eta}^{P G} {\eta}^{Q S} {\eta}^{I L} {\eta}^{J H} - 2\, {\eta}^{M R} {\eta}^{N Q} {\eta}^{P K} {\eta}^{I L} {\eta}^{J G} {\eta}^{S H} - {\eta}^{M R} {\eta}^{N Q} {\eta}^{P K} {\eta}^{I G} {\eta}^{J H} {\eta}^{L S}
\end{dmath*}
\begin{dmath*}
{\vphantom{L^{(6)}=} + {\eta}^{M R} {\eta}^{N Q} {\eta}^{P G} {\eta}^{I K} {\eta}^{J L} {\eta}^{S H} - {\eta}^{M I} {\eta}^{N Q} {\eta}^{P K} {\eta}^{J L} {\eta}^{R G} {\eta}^{S H} - 5\, {\eta}^{M I} {\eta}^{N Q} {\eta}^{P K} {\eta}^{J G} {\eta}^{L R} {\eta}^{S H}\left.\right) + {\partial}_{M}{{\cal M}_{N P}}\,  {\partial}_{Q}{{\cal M}_{I J}}\,  {\partial}_{K L R}{{\cal M}_{S G}}\,  {\partial}_{H}{d}\,  \left(\right. \ \ \ \ \ \ \ } - \frac{1}{2}\, {\eta}^{M N} {\eta}^{P K} {\eta}^{Q I} {\eta}^{J L} {\eta}^{H S} {\eta}^{R G} - \frac{1}{4}\, {\eta}^{M N} {\eta}^{P K} {\eta}^{Q S} {\eta}^{I L} {\eta}^{J R} {\eta}^{H G} - \frac{1}{2}\, {\eta}^{M N} {\eta}^{P K} {\eta}^{Q S} {\eta}^{I H} {\eta}^{J L} {\eta}^{R G} + \frac{1}{2}\, {\eta}^{M N} {\eta}^{P Q} {\eta}^{I K} {\eta}^{J L} {\eta}^{H S} {\eta}^{R G} - \frac{1}{4}\, {\eta}^{M N} {\eta}^{P Q} {\eta}^{I H} {\eta}^{J K} {\eta}^{L S} {\eta}^{R G} - {\eta}^{M S} {\eta}^{N H} {\eta}^{P K} {\eta}^{Q G} {\eta}^{I L} {\eta}^{J R} - {\eta}^{M S} {\eta}^{N Q} {\eta}^{P K} {\eta}^{I H} {\eta}^{J L} {\eta}^{R G} + \frac{1}{2}\, {\eta}^{M S} {\eta}^{N Q} {\eta}^{P H} {\eta}^{I K} {\eta}^{J L} {\eta}^{R G} - 2\, {\eta}^{M I} {\eta}^{N Q} {\eta}^{P K} {\eta}^{J L} {\eta}^{H S} {\eta}^{R G} - {\eta}^{M I} {\eta}^{N Q} {\eta}^{P H} {\eta}^{J K} {\eta}^{L S} {\eta}^{R G}\left.\right) + {\partial}_{M}{{\cal M}_{N P}}\,  {\partial}_{Q I}{{\cal M}_{J K}}\,  {\partial}_{L R}{{\cal M}_{S G}}\,  {\partial}_{H}{d}\,  \left(\right. - \frac{1}{2}\, {\eta}^{M N} {\eta}^{P Q} {\eta}^{H J} {\eta}^{I S} {\eta}^{K L} {\eta}^{R G} - \frac{1}{2}\, {\eta}^{M N} {\eta}^{P Q} {\eta}^{H S} {\eta}^{I J} {\eta}^{K L} {\eta}^{R G} + \frac{1}{4}\, {\eta}^{M N} {\eta}^{P Q} {\eta}^{H S} {\eta}^{I G} {\eta}^{J L} {\eta}^{K R} - 2\, {\eta}^{M J} {\eta}^{N H} {\eta}^{P Q} {\eta}^{I S} {\eta}^{K L} {\eta}^{R G} + \frac{1}{2}\, {\eta}^{M S} {\eta}^{N Q} {\eta}^{P I} {\eta}^{H J} {\eta}^{K L} {\eta}^{R G} - {\eta}^{M J} {\eta}^{N Q} {\eta}^{P I} {\eta}^{H S} {\eta}^{K L} {\eta}^{R G} + \frac{1}{2}\, {\eta}^{M S} {\eta}^{N Q} {\eta}^{P I} {\eta}^{H G} {\eta}^{J L} {\eta}^{K R} - {\eta}^{M J} {\eta}^{N Q} {\eta}^{P L} {\eta}^{H K} {\eta}^{I S} {\eta}^{R G} - 2\, {\eta}^{M J} {\eta}^{N Q} {\eta}^{P L} {\eta}^{H S} {\eta}^{I G} {\eta}^{K R} - 2\, {\eta}^{M J} {\eta}^{N Q} {\eta}^{P L} {\eta}^{H S} {\eta}^{I K} {\eta}^{R G}\left.\right) + 3\, {\cal M}_{M N} {\cal M}_{P Q} {\cal M}_{I J} {\partial}_{K}{d}\,  {\partial}_{L R S G H}{d}\,  {\eta}^{M K} {\eta}^{N L} {\eta}^{P R} {\eta}^{Q S} {\eta}^{I G} {\eta}^{J H} + 11\, {\cal M}_{M N} {\cal M}_{P Q} {\cal M}_{I J} {\partial}_{K L}{d}\,  {\partial}_{R S G H}{d}\,  {\eta}^{M K} {\eta}^{N R} {\eta}^{P L} {\eta}^{Q S} {\eta}^{I G} {\eta}^{J H} + {\cal M}_{M N} {\cal M}_{P Q} {\cal M}_{I J} {\partial}_{K L R}{d}\,  {\partial}_{S G H}{d}\,  \left(\right.4\, {\eta}^{M K} {\eta}^{N L} {\eta}^{P R} {\eta}^{Q S} {\eta}^{I G} {\eta}^{J H} + 4\, {\eta}^{M K} {\eta}^{N S} {\eta}^{P L} {\eta}^{Q G} {\eta}^{I R} {\eta}^{J H}\left.\right) + {\cal M}_{M N} {\cal M}_{P Q} {\partial}_{I}{{\cal M}_{J K}}\,  {\partial}_{L}{d}\,  {\partial}_{R S G H}{d}\,  \left(\right.\frac{1}{2}\, {\eta}^{M I} {\eta}^{N L} {\eta}^{P R} {\eta}^{Q S} {\eta}^{J G} {\eta}^{K H} + 6\, {\eta}^{M I} {\eta}^{N R} {\eta}^{P L} {\eta}^{Q S} {\eta}^{J G} {\eta}^{K H} + 11\, {\eta}^{M I} {\eta}^{N R} {\eta}^{P S} {\eta}^{Q G} {\eta}^{J L} {\eta}^{K H} + 6\, {\eta}^{M L} {\eta}^{N R} {\eta}^{P S} {\eta}^{Q G} {\eta}^{I J} {\eta}^{K H}\left.\right) + {\cal M}_{M N} {\cal M}_{P Q} {\partial}_{I}{{\cal M}_{J K}}\,  {\partial}_{L R}{d}\,  {\partial}_{S G H}{d}\,  \left(\right.10\, {\eta}^{M I} {\eta}^{N L} {\eta}^{P R} {\eta}^{Q S} {\eta}^{J G} {\eta}^{K H} + 11\, {\eta}^{M I} {\eta}^{N L} {\eta}^{P S} {\eta}^{Q G} {\eta}^{J R} {\eta}^{K H} + 24\, {\eta}^{M I} {\eta}^{N S} {\eta}^{P L} {\eta}^{Q G} {\eta}^{J R} {\eta}^{K H} + \frac{5}{2}\, {\eta}^{M I} {\eta}^{N S} {\eta}^{P G} {\eta}^{Q H} {\eta}^{J L} {\eta}^{K R} + 11\, {\eta}^{M L} {\eta}^{N S} {\eta}^{P R} {\eta}^{Q G} {\eta}^{I J} {\eta}^{K H} + 8\, {\eta}^{M L} {\eta}^{N S} {\eta}^{P G} {\eta}^{Q H} {\eta}^{I J} {\eta}^{K R}\left.\right) + {\cal M}_{M N} {\cal M}_{P Q} {\partial}_{I J}{{\cal M}_{K L}}\,  {\partial}_{R}{d}\,  {\partial}_{S G H}{d}\,  \left(\right.3\, {\eta}^{M I} {\eta}^{N J} {\eta}^{P S} {\eta}^{Q G} {\eta}^{R K} {\eta}^{L H} + 8\, {\eta}^{M I} {\eta}^{N S} {\eta}^{P J} {\eta}^{Q G} {\eta}^{R K} {\eta}^{L H} + 5\, {\eta}^{M I} {\eta}^{N S} {\eta}^{P G} {\eta}^{Q H} {\eta}^{R K} {\eta}^{J L} + 2\, {\eta}^{M R} {\eta}^{N I} {\eta}^{P J} {\eta}^{Q S} {\eta}^{K G} {\eta}^{L H} + 3\, {\eta}^{M R} {\eta}^{N I} {\eta}^{P S} {\eta}^{Q G} {\eta}^{J K} {\eta}^{L H} + \frac{1}{2}\, {\eta}^{M R} {\eta}^{N S} {\eta}^{P I} {\eta}^{Q J} {\eta}^{K G} {\eta}^{L H} + 11\, {\eta}^{M R} {\eta}^{N S} {\eta}^{P I} {\eta}^{Q G} {\eta}^{J K} {\eta}^{L H}\left.\right) + {\cal M}_{M N} {\cal M}_{P Q} {\partial}_{I J}{{\cal M}_{K L}}\,  {\partial}_{R S}{d}\,  {\partial}_{G H}{d}\,  \left(\right.2\, {\eta}^{M I} {\eta}^{N J} {\eta}^{P R} {\eta}^{Q G} {\eta}^{K S} {\eta}^{L H} + 3\, {\eta}^{M I} {\eta}^{N R} {\eta}^{P J} {\eta}^{Q S} {\eta}^{K G} {\eta}^{L H} + 8\, {\eta}^{M I} {\eta}^{N R} {\eta}^{P J} {\eta}^{Q G} {\eta}^{K S} {\eta}^{L H} + 18\, {\eta}^{M I} {\eta}^{N R} {\eta}^{P S} {\eta}^{Q G} {\eta}^{J K} {\eta}^{L H}\left.\right) + {\cal M}_{M N} {\cal M}_{P Q} {\partial}_{I J K}{{\cal M}_{L R}}\,  {\partial}_{S}{d}\,  {\partial}_{G H}{d}\,  \left(\right.{\eta}^{M S} {\eta}^{N I} {\eta}^{P J} {\eta}^{Q K} {\eta}^{G L} {\eta}^{H R} + 5\, {\eta}^{M G} {\eta}^{N I} {\eta}^{P J} {\eta}^{Q K} {\eta}^{S L} {\eta}^{H R} + 5\, {\eta}^{M S} {\eta}^{N I} {\eta}^{P G} {\eta}^{Q J} {\eta}^{H L} {\eta}^{K R} + 6\, {\eta}^{M G} {\eta}^{N I} {\eta}^{P H} {\eta}^{Q J} {\eta}^{S L} {\eta}^{K R} + 3\, {\eta}^{M S} {\eta}^{N G} {\eta}^{P I} {\eta}^{Q J} {\eta}^{H L} {\eta}^{K R} + \frac{5}{2}\, {\eta}^{M S} {\eta}^{N G} {\eta}^{P H} {\eta}^{Q I} {\eta}^{J L} {\eta}^{K R}\left.\right) + {\cal M}_{M N} {\cal M}_{P Q} {\partial}_{I J K L}{{\cal M}_{R S}}\,  {\partial}_{G}{d}\,  {\partial}_{H}{d}\,  \left(\right.\frac{1}{2}\, {\eta}^{M I} {\eta}^{N J} {\eta}^{P K} {\eta}^{Q L} {\eta}^{G R} {\eta}^{H S} + 2\, {\eta}^{M G} {\eta}^{N I} {\eta}^{P J} {\eta}^{Q K} {\eta}^{H R} {\eta}^{L S} + \frac{1}{2}\, {\eta}^{M G} {\eta}^{N I} {\eta}^{P H} {\eta}^{Q J} {\eta}^{K R} {\eta}^{L S}\left.\right) + {\cal M}_{M N} {\partial}_{P}{{\cal M}_{Q I}}\,  {\partial}_{J}{{\cal M}_{K L}}\,  {\partial}_{R}{d}\,  {\partial}_{S G H}{d}\,  \left(\right.\frac{1}{2}\, {\eta}^{M R} {\eta}^{N S} {\eta}^{P Q} {\eta}^{I J} {\eta}^{K G} {\eta}^{L H} + 3\, {\eta}^{M S} {\eta}^{N G} {\eta}^{P Q} {\eta}^{I J} {\eta}^{K R} {\eta}^{L H} + 3\, {\eta}^{M R} {\eta}^{N S} {\eta}^{P Q} {\eta}^{I G} {\eta}^{J K} {\eta}^{L H} + 5\, {\eta}^{M R} {\eta}^{N S} {\eta}^{P K} {\eta}^{Q J} {\eta}^{I G} {\eta}^{L H} + 5\, {\eta}^{M S} {\eta}^{N G} {\eta}^{P K} {\eta}^{Q J} {\eta}^{I R} {\eta}^{L H} + 3\, {\eta}^{M P} {\eta}^{N S} {\eta}^{Q J} {\eta}^{I R} {\eta}^{K G} {\eta}^{L H} - {\eta}^{M P} {\eta}^{N R} {\eta}^{Q J} {\eta}^{I S} {\eta}^{K G} {\eta}^{L H} + 8\, {\eta}^{M P} {\eta}^{N S} {\eta}^{Q J} {\eta}^{I G} {\eta}^{K R} {\eta}^{L H} + 7\, {\eta}^{M P} {\eta}^{N J} {\eta}^{Q R} {\eta}^{I S} {\eta}^{K G} {\eta}^{L H} + 11\, {\eta}^{M P} {\eta}^{N S} {\eta}^{Q R} {\eta}^{I G} {\eta}^{J K} {\eta}^{L H} + \frac{1}{2}\, {\eta}^{M P} {\eta}^{N R} {\eta}^{Q S} {\eta}^{I G} {\eta}^{J K} {\eta}^{L H}\left.\right) + {\cal M}_{M N} {\partial}_{P}{{\cal M}_{Q I}}\,  {\partial}_{J}{{\cal M}_{K L}}\,  {\partial}_{R S}{d}\,  {\partial}_{G H}{d}\,  \left(\right.2\, {\eta}^{M R} {\eta}^{N G} {\eta}^{P Q} {\eta}^{I J} {\eta}^{K S} {\eta}^{L H} + 4\, {\eta}^{M R} {\eta}^{N G} {\eta}^{P Q} {\eta}^{I S} {\eta}^{J K} {\eta}^{L H} + 8\, {\eta}^{M R} {\eta}^{N G} {\eta}^{P K} {\eta}^{Q J} {\eta}^{I S} {\eta}^{L H} + 3\, {\eta}^{M P} {\eta}^{N R} {\eta}^{Q J} {\eta}^{I S} {\eta}^{K G} {\eta}^{L H} + 8\, {\eta}^{M P} {\eta}^{N R} {\eta}^{Q J} {\eta}^{I G} {\eta}^{K S} {\eta}^{L H} + \frac{1}{2}\, {\eta}^{M P} {\eta}^{N J} {\eta}^{Q R} {\eta}^{I S} {\eta}^{K G} {\eta}^{L H} + 4\, {\eta}^{M P} {\eta}^{N J} {\eta}^{Q R} {\eta}^{I G} {\eta}^{K S} {\eta}^{L H} + 11\, {\eta}^{M P} {\eta}^{N R} {\eta}^{Q S} {\eta}^{I G} {\eta}^{J K} {\eta}^{L H} + \frac{5}{2}\, {\eta}^{M P} {\eta}^{N R} {\eta}^{Q G} {\eta}^{I H} {\eta}^{J K} {\eta}^{L S}\left.\right) + {\cal M}_{M N} {\partial}_{P}{{\cal M}_{Q I}}\,  {\partial}_{J K}{{\cal M}_{L R}}\,  {\partial}_{S}{d}\,  {\partial}_{G H}{d}\,  \left(\right.3\, {\eta}^{M S} {\eta}^{N G} {\eta}^{P Q} {\eta}^{I J} {\eta}^{K L} {\eta}^{R H} + 5\, {\eta}^{M J} {\eta}^{N G} {\eta}^{P Q} {\eta}^{I K} {\eta}^{S L} {\eta}^{R H} + 5\, {\eta}^{M J} {\eta}^{N G} {\eta}^{P Q} {\eta}^{I H} {\eta}^{S L} {\eta}^{K R} + 3\, {\eta}^{M J} {\eta}^{N K} {\eta}^{P Q} {\eta}^{I G} {\eta}^{S L} {\eta}^{R H} + {\eta}^{M S} {\eta}^{N J} {\eta}^{P Q} {\eta}^{I K} {\eta}^{L G} {\eta}^{R H} + 3\, {\eta}^{M S} {\eta}^{N J} {\eta}^{P Q} {\eta}^{I G} {\eta}^{K L} {\eta}^{R H} + 5\, {\eta}^{M S} {\eta}^{N G} {\eta}^{P L} {\eta}^{Q J} {\eta}^{I H} {\eta}^{K R} + \frac{5}{2}\, {\eta}^{M S} {\eta}^{N G} {\eta}^{P L} {\eta}^{Q J} {\eta}^{I K} {\eta}^{R H} + 6\, {\eta}^{M J} {\eta}^{N G} {\eta}^{P L} {\eta}^{Q K} {\eta}^{I H} {\eta}^{S R} + 6\, {\eta}^{M J} {\eta}^{N G} {\eta}^{P L} {\eta}^{Q S} {\eta}^{I H} {\eta}^{K R} - {\eta}^{M J} {\eta}^{N K} {\eta}^{P L} {\eta}^{Q S} {\eta}^{I G} {\eta}^{R H} + \frac{1}{2}\, {\eta}^{M J} {\eta}^{N K} {\eta}^{P L} {\eta}^{Q G} {\eta}^{I H} {\eta}^{S R} + 6\, {\eta}^{M J} {\eta}^{N G} {\eta}^{P L} {\eta}^{Q S} {\eta}^{I K} {\eta}^{R H} + \frac{1}{2}\, {\eta}^{M S} {\eta}^{N J} {\eta}^{P L} {\eta}^{Q G} {\eta}^{I H} {\eta}^{K R} - \frac{1}{2}\, {\eta}^{M P} {\eta}^{N S} {\eta}^{Q J} {\eta}^{I K} {\eta}^{L G} {\eta}^{R H} + {\eta}^{M P} {\eta}^{N G} {\eta}^{Q J} {\eta}^{I K} {\eta}^{S L} {\eta}^{R H} + 6\, {\eta}^{M P} {\eta}^{N G} {\eta}^{Q J} {\eta}^{I H} {\eta}^{S L} {\eta}^{K R} - {\eta}^{M P} {\eta}^{N S} {\eta}^{Q J} {\eta}^{I G} {\eta}^{K L} {\eta}^{R H} + 6\, {\eta}^{M P} {\eta}^{N G} {\eta}^{Q S} {\eta}^{I J} {\eta}^{K L} {\eta}^{R H}  + 8\, {\eta}^{M P} {\eta}^{N J} {\eta}^{Q K} {\eta}^{I G} {\eta}^{S L} {\eta}^{R H} + 3\, {\eta}^{M P} {\eta}^{N J} {\eta}^{Q S} {\eta}^{I K} {\eta}^{L G} {\eta}^{R H} + 2\, {\eta}^{M P} {\eta}^{N J} {\eta}^{Q G} {\eta}^{I H} {\eta}^{S L} {\eta}^{K R} + 12\, {\eta}^{M P} {\eta}^{N J} {\eta}^{Q S} {\eta}^{I G} {\eta}^{K L} {\eta}^{R H}\left.\right) + {\cal M}_{M N} {\partial}_{P}{{\cal M}_{Q I}}\,  {\partial}_{J K L}{{\cal M}_{R S}}\,  {\partial}_{G}{d}\,  {\partial}_{H}{d}\,  \left(\right.{\eta}^{M J} {\eta}^{N K} {\eta}^{P Q} {\eta}^{I L} {\eta}^{G R} {\eta}^{H S} + 2\, {\eta}^{M G} {\eta}^{N J} {\eta}^{P Q} {\eta}^{I K} {\eta}^{H R} {\eta}^{L S} + \frac{1}{2}\, {\eta}^{M J} {\eta}^{N K} {\eta}^{P R} {\eta}^{Q G} {\eta}^{I H} {\eta}^{L S} + {\eta}^{M J} {\eta}^{N K} {\eta}^{P R} {\eta}^{Q G} {\eta}^{I L} {\eta}^{H S} + \frac{1}{2}\, {\eta}^{M G} {\eta}^{N J} {\eta}^{P R} {\eta}^{Q K} {\eta}^{I L} {\eta}^{H S} + {\eta}^{M G} {\eta}^{N J} {\eta}^{P R} {\eta}^{Q H} {\eta}^{I K} {\eta}^{L S} - {\eta}^{M P} {\eta}^{N G} {\eta}^{Q J} {\eta}^{I K} {\eta}^{H R} {\eta}^{L S} + \frac{1}{2}\, {\eta}^{M P} {\eta}^{N G} {\eta}^{Q H} {\eta}^{I J} {\eta}^{K R} {\eta}^{L S} + {\eta}^{M P} {\eta}^{N J} {\eta}^{Q K} {\eta}^{I L} {\eta}^{G R} {\eta}^{H S} + 6\, {\eta}^{M P} {\eta}^{N J} {\eta}^{Q G} {\eta}^{I K} {\eta}^{H R} {\eta}^{L S} + {\eta}^{M P} {\eta}^{N J} {\eta}^{Q G} {\eta}^{I H} {\eta}^{K R} {\eta}^{L S}\left.\right) + {\cal M}_{M N} {\partial}_{P Q}{{\cal M}_{I J}}\,  {\partial}_{K L}{{\cal M}_{R S}}\,  {\partial}_{G}{d}\,  {\partial}_{H}{d}\,  \left(\right.{\eta}^{M P} {\eta}^{N Q} {\eta}^{G I} {\eta}^{H R} {\eta}^{J K} {\eta}^{L S} - \frac{1}{2}\, {\eta}^{M P} {\eta}^{N Q} {\eta}^{G R} {\eta}^{H S} {\eta}^{I K} {\eta}^{J L} + 3\, {\eta}^{M P} {\eta}^{N K} {\eta}^{G I} {\eta}^{H J} {\eta}^{Q R} {\eta}^{L S} + 2\, {\eta}^{M P} {\eta}^{N K} {\eta}^{G I} {\eta}^{H R} {\eta}^{Q S} {\eta}^{J L} + 2\, {\eta}^{M P} {\eta}^{N K} {\eta}^{G I} {\eta}^{H R} {\eta}^{Q J} {\eta}^{L S} + {\eta}^{M G} {\eta}^{N P} {\eta}^{H I} {\eta}^{Q R} {\eta}^{J K} {\eta}^{L S} + {\eta}^{M G} {\eta}^{N P} {\eta}^{H R} {\eta}^{Q I} {\eta}^{J K} {\eta}^{L S} - \frac{1}{2}\, {\eta}^{M G} {\eta}^{N P} {\eta}^{H R} {\eta}^{Q S} {\eta}^{I K} {\eta}^{J L}\left.\right) + {\partial}_{M}{{\cal M}_{N P}}\,  {\partial}_{Q}{{\cal M}_{I J}}\,  {\partial}_{K}{{\cal M}_{L R}}\,  {\partial}_{S}{d}\,  {\partial}_{G H}{d}\,  \left(\right.5\, {\eta}^{M N} {\eta}^{P G} {\eta}^{Q L} {\eta}^{I K} {\eta}^{J S} {\eta}^{R H} + \frac{1}{2}\, {\eta}^{M N} {\eta}^{P Q} {\eta}^{I K} {\eta}^{J S} {\eta}^{L G} {\eta}^{R H} - {\eta}^{M N} {\eta}^{P Q} {\eta}^{I K} {\eta}^{J G} {\eta}^{L S} {\eta}^{R H} + 3\, {\eta}^{M N} {\eta}^{P Q} {\eta}^{I S} {\eta}^{J G} {\eta}^{K L} {\eta}^{R H} + 2\, {\eta}^{M I} {\eta}^{N Q} {\eta}^{P K} {\eta}^{J S} {\eta}^{L G} {\eta}^{R H} + 10\, {\eta}^{M I} {\eta}^{N Q} {\eta}^{P K} {\eta}^{J G} {\eta}^{L S} {\eta}^{R H}\left.\right)
 \end{dmath*}
\begin{dmath*}
{+ {\partial}_{M}{{\cal M}_{N P}}\,  {\partial}_{Q}{{\cal M}_{I J}}\,  {\partial}_{K L}{{\cal M}_{R S}}\,  {\partial}_{G}{d}\,  {\partial}_{H}{d}\,  \left(\right.\frac{1}{2}\, {\eta}^{M N} {\eta}^{P K} {\eta}^{Q I} {\eta}^{J L} {\eta}^{G R} {\eta}^{H S} + \frac{1}{2}\, {\eta}^{M N} {\eta}^{P K} {\eta}^{Q R} {\eta}^{I G} {\eta}^{J H} {\eta}^{L S} + {\eta}^{M N} {\eta}^{P K} {\eta}^{Q R} {\eta}^{I G} {\eta}^{J L} {\eta}^{H S} \ \ \ \ \ \ \ \ \ \ \  } - \frac{1}{2}\, {\eta}^{M N} {\eta}^{P Q} {\eta}^{I K} {\eta}^{J L} {\eta}^{G R} {\eta}^{H S} + {\eta}^{M N} {\eta}^{P Q} {\eta}^{I G} {\eta}^{J K} {\eta}^{H R} {\eta}^{L S} + 2\, {\eta}^{M R} {\eta}^{N G} {\eta}^{P K} {\eta}^{Q S} {\eta}^{I H} {\eta}^{J L} + {\eta}^{M R} {\eta}^{N G} {\eta}^{P H} {\eta}^{Q S} {\eta}^{I K} {\eta}^{J L} + 2\, {\eta}^{M R} {\eta}^{N Q} {\eta}^{P K} {\eta}^{I G} {\eta}^{J H} {\eta}^{L S} + 2\, {\eta}^{M R} {\eta}^{N Q} {\eta}^{P K} {\eta}^{I G} {\eta}^{J L} {\eta}^{H S} - {\eta}^{M R} {\eta}^{N Q} {\eta}^{P G} {\eta}^{I K} {\eta}^{J L} {\eta}^{H S} + 2\, {\eta}^{M I} {\eta}^{N Q} {\eta}^{P K} {\eta}^{J L} {\eta}^{G R} {\eta}^{H S} + 4\, {\eta}^{M I} {\eta}^{N Q} {\eta}^{P G} {\eta}^{J K} {\eta}^{H R} {\eta}^{L S}\left.\right)
 - 6\, {\cal M}_{M N} {\cal M}_{P Q} {\cal M}_{I J} {\partial}_{K}{d}\,  {\partial}_{L}{d}\,  {\partial}_{R S G H}{d}\,  {\eta}^{M K} {\eta}^{N R} {\eta}^{P L} {\eta}^{Q S} {\eta}^{I G} {\eta}^{J H} + {\cal M}_{M N} {\cal M}_{P Q} {\cal M}_{I J} {\partial}_{K}{d}\,  {\partial}_{L R}{d}\,  {\partial}_{S G H}{d}\,  \left(\right. - 16\, {\eta}^{M K} {\eta}^{N L} {\eta}^{P R} {\eta}^{Q S} {\eta}^{I G} {\eta}^{J H} - 22\, {\eta}^{M K} {\eta}^{N S} {\eta}^{P L} {\eta}^{Q G} {\eta}^{I R} {\eta}^{J H}\left.\right)  - 12\, {\cal M}_{M N} {\cal M}_{P Q} {\cal M}_{I J} {\partial}_{K L}{d}\,  {\partial}_{R S}{d}\,  {\partial}_{G H}{d}\,  {\eta}^{M K} {\eta}^{N R} {\eta}^{P L} {\eta}^{Q G} {\eta}^{I S} {\eta}^{J H} + {\cal M}_{M N} {\cal M}_{P Q} {\partial}_{I}{{\cal M}_{J K}}\,  {\partial}_{L}{d}\,  {\partial}_{R}{d}\,  {\partial}_{S G H}{d}\,  \left(\right. - {\eta}^{M I} {\eta}^{N L} {\eta}^{P R} {\eta}^{Q S} {\eta}^{J G} {\eta}^{K H} - 6\, {\eta}^{M I} {\eta}^{N L} {\eta}^{P S} {\eta}^{Q G} {\eta}^{J R} {\eta}^{K H} - 22\, {\eta}^{M I} {\eta}^{N S} {\eta}^{P L} {\eta}^{Q G} {\eta}^{J R} {\eta}^{K H} - 5\, {\eta}^{M I} {\eta}^{N S} {\eta}^{P G} {\eta}^{Q H} {\eta}^{J L} {\eta}^{K R} - 6\, {\eta}^{M L} {\eta}^{N S} {\eta}^{P R} {\eta}^{Q G} {\eta}^{I J} {\eta}^{K H}\left.\right) + {\cal M}_{M N} {\cal M}_{P Q} {\partial}_{I}{{\cal M}_{J K}}\,  {\partial}_{L}{d}\,  {\partial}_{R S}{d}\,  {\partial}_{G H}{d}\,  \left(\right. - 4\, {\eta}^{M I} {\eta}^{N L} {\eta}^{P R} {\eta}^{Q G} {\eta}^{J S} {\eta}^{K H} - 5\, {\eta}^{M I} {\eta}^{N R} {\eta}^{P L} {\eta}^{Q S} {\eta}^{J G} {\eta}^{K H} - 22\, {\eta}^{M I} {\eta}^{N R} {\eta}^{P L} {\eta}^{Q G} {\eta}^{J S} {\eta}^{K H} - 36\, {\eta}^{M I} {\eta}^{N R} {\eta}^{P S} {\eta}^{Q G} {\eta}^{J L} {\eta}^{K H} - 16\, {\eta}^{M L} {\eta}^{N R} {\eta}^{P S} {\eta}^{Q G} {\eta}^{I J} {\eta}^{K H}\left.\right) + {\cal M}_{M N} {\cal M}_{P Q} {\partial}_{I J}{{\cal M}_{K L}}\,  {\partial}_{R}{d}\,  {\partial}_{S}{d}\,  {\partial}_{G H}{d}\,  \left(\right. - 6\, {\eta}^{M I} {\eta}^{N G} {\eta}^{P J} {\eta}^{Q H} {\eta}^{R K} {\eta}^{S L} - 10\, {\eta}^{M R} {\eta}^{N I} {\eta}^{P J} {\eta}^{Q G} {\eta}^{S K} {\eta}^{L H} - {\eta}^{M R} {\eta}^{N I} {\eta}^{P S} {\eta}^{Q J} {\eta}^{K G} {\eta}^{L H} - 6\, {\eta}^{M R} {\eta}^{N I} {\eta}^{P S} {\eta}^{Q G} {\eta}^{J K} {\eta}^{L H} - 6\, {\eta}^{M R} {\eta}^{N G} {\eta}^{P I} {\eta}^{Q J} {\eta}^{S K} {\eta}^{L H} - 10\, {\eta}^{M R} {\eta}^{N G} {\eta}^{P I} {\eta}^{Q H} {\eta}^{S K} {\eta}^{J L}\left.\right) + {\cal M}_{M N} {\cal M}_{P Q} {\partial}_{I J K}{{\cal M}_{L R}}\,  {\partial}_{S}{d}\,  {\partial}_{G}{d}\,  {\partial}_{H}{d}\,  \left(\right. - 2\, {\eta}^{M S} {\eta}^{N I} {\eta}^{P J} {\eta}^{Q K} {\eta}^{G L} {\eta}^{H R} - 2\, {\eta}^{M S} {\eta}^{N I} {\eta}^{P G} {\eta}^{Q J} {\eta}^{H L} {\eta}^{K R}\left.\right)
   \end{dmath*}
\begin{dmath*}
{+ {\cal M}_{M N} {\partial}_{P}{{\cal M}_{Q I}}\,  {\partial}_{J}{{\cal M}_{K L}}\,  {\partial}_{R}{d}\,  {\partial}_{S}{d}\,  {\partial}_{G H}{d}\,  \left(\right. - 6\, {\eta}^{M R} {\eta}^{N G} {\eta}^{P Q} {\eta}^{I J} {\eta}^{K S} {\eta}^{L H} - 10\, {\eta}^{M R} {\eta}^{N G} {\eta}^{P K} {\eta}^{Q J} {\eta}^{I S} {\eta}^{L H} - 12\, {\eta}^{M P} {\eta}^{N G} {\eta}^{Q J} {\eta}^{I R} {\eta}^{K S} {\eta}^{L H} \ \ \ \ \ \ \ } - {\eta}^{M P} {\eta}^{N R} {\eta}^{Q J} {\eta}^{I S} {\eta}^{K G} {\eta}^{L H} + 2\, {\eta}^{M P} {\eta}^{N R} {\eta}^{Q J} {\eta}^{I G} {\eta}^{K S} {\eta}^{L H} - 6\, {\eta}^{M P} {\eta}^{N G} {\eta}^{Q J} {\eta}^{I H} {\eta}^{K R} {\eta}^{L S} - 2\, {\eta}^{M P} {\eta}^{N J} {\eta}^{Q R} {\eta}^{I S} {\eta}^{K G} {\eta}^{L H} - 12\, {\eta}^{M P} {\eta}^{N J} {\eta}^{Q R} {\eta}^{I G} {\eta}^{K S} {\eta}^{L H} - 5\, {\eta}^{M P} {\eta}^{N G} {\eta}^{Q R} {\eta}^{I S} {\eta}^{J K} {\eta}^{L H} - 6\, {\eta}^{M P} {\eta}^{N R} {\eta}^{Q S} {\eta}^{I G} {\eta}^{J K} {\eta}^{L H}\left.\right) + {\cal M}_{M N} {\partial}_{P}{{\cal M}_{Q I}}\,  {\partial}_{J K}{{\cal M}_{L R}}\,  {\partial}_{S}{d}\,  {\partial}_{G}{d}\,  {\partial}_{H}{d}\,  \left(\right. - 2\, {\eta}^{M S} {\eta}^{N J} {\eta}^{P Q} {\eta}^{I K} {\eta}^{G L} {\eta}^{H R} - {\eta}^{M J} {\eta}^{N K} {\eta}^{P L} {\eta}^{Q S} {\eta}^{I G} {\eta}^{H R} - {\eta}^{M S} {\eta}^{N J} {\eta}^{P L} {\eta}^{Q G} {\eta}^{I H} {\eta}^{K R} - 2\, {\eta}^{M S} {\eta}^{N J} {\eta}^{P L} {\eta}^{Q G} {\eta}^{I K} {\eta}^{H R} + {\eta}^{M P} {\eta}^{N S} {\eta}^{Q J} {\eta}^{I K} {\eta}^{G L} {\eta}^{H R} - 2\, {\eta}^{M P} {\eta}^{N S} {\eta}^{Q G} {\eta}^{I J} {\eta}^{H L} {\eta}^{K R} - 6\, {\eta}^{M P} {\eta}^{N J} {\eta}^{Q S} {\eta}^{I K} {\eta}^{G L} {\eta}^{H R} - 4\, {\eta}^{M P} {\eta}^{N J} {\eta}^{Q S} {\eta}^{I G} {\eta}^{H L} {\eta}^{K R}\left.\right) + {\partial}_{M}{{\cal M}_{N P}}\,  {\partial}_{Q}{{\cal M}_{I J}}\,  {\partial}_{K}{{\cal M}_{L R}}\,  {\partial}_{S}{d}\,  {\partial}_{G}{d}\,  {\partial}_{H}{d}\,  \left(\right. - {\eta}^{M N} {\eta}^{P Q} {\eta}^{I K} {\eta}^{J S} {\eta}^{L G} {\eta}^{R H} - 4\, {\eta}^{M I} {\eta}^{N Q} {\eta}^{P K} {\eta}^{J S} {\eta}^{L G} {\eta}^{R H}\left.\right) + 4\, {\cal M}_{M N} {\cal M}_{P Q} {\cal M}_{I J} {\partial}_{K}{d}\,  {\partial}_{L}{d}\,  {\partial}_{R}{d}\,  {\partial}_{S G H}{d}\,  {\eta}^{M K} {\eta}^{N S} {\eta}^{P L} {\eta}^{Q G} {\eta}^{I R} {\eta}^{J H} + 16\, {\cal M}_{M N} {\cal M}_{P Q} {\cal M}_{I J} {\partial}_{K}{d}\,  {\partial}_{L}{d}\,  {\partial}_{R S}{d}\,  {\partial}_{G H}{d}\,  {\eta}^{M K} {\eta}^{N R} {\eta}^{P L} {\eta}^{Q G} {\eta}^{I S} {\eta}^{J H} + {\cal M}_{M N} {\cal M}_{P Q} {\partial}_{I}{{\cal M}_{J K}}\,  {\partial}_{L}{d}\,  {\partial}_{R}{d}\,  {\partial}_{S}{d}\,  {\partial}_{G H}{d}\,  \left(\right.12\, {\eta}^{M I} {\eta}^{N L} {\eta}^{P R} {\eta}^{Q G} {\eta}^{J S} {\eta}^{K H} + 10\, {\eta}^{M I} {\eta}^{N G} {\eta}^{P L} {\eta}^{Q H} {\eta}^{J R} {\eta}^{K S}\left.\right) + 2\, {\cal M}_{M N} {\cal M}_{P Q} {\partial}_{I J}{{\cal M}_{K L}}\,  {\partial}_{R}{d}\,  {\partial}_{S}{d}\,  {\partial}_{G}{d}\,  {\partial}_{H}{d}\,  {\eta}^{M R} {\eta}^{N I} {\eta}^{P S} {\eta}^{Q J} {\eta}^{G K} {\eta}^{H L} + {\cal M}_{M N} {\partial}_{P}{{\cal M}_{Q I}}\,  {\partial}_{J}{{\cal M}_{K L}}\,  {\partial}_{R}{d}\,  {\partial}_{S}{d}\,  {\partial}_{G}{d}\,  {\partial}_{H}{d}\,  \left(\right.2\, {\eta}^{M P} {\eta}^{N R} {\eta}^{Q J} {\eta}^{I S} {\eta}^{K G} {\eta}^{L H} + 2\, {\eta}^{M P} {\eta}^{N J} {\eta}^{Q R} {\eta}^{I S} {\eta}^{K G} {\eta}^{L H}\left.\right)
\end{dmath*}
\end{dgroup*}
\endgroup

\newpage
\section{Some relevant intermediate results} \label{APP::Results}
Here we show the explicit form of certain results, which due to their length may generate confusion if displayed in the main text.

The second order transformation of the duality covariant metric $\delta^{(2)}\widehat g_{\mu \nu}$ in (\ref{transfG}) has a shorter expression when contracted with some symmetric tensor $S^{\mu \nu}$
\be
\delta^{(2)} \widehat g = S^{\mu \nu} \delta^{(2)}\widehat g_{\mu \nu} \ . \label{d2tildeGa}
\ee
This should not be confused with the second order transformation of the determinant of the metric. The above contraction is only a device to shorten the output, and allows to determine $\delta^{(2)}\widehat g_{\mu \nu}$ from $\delta^{(2)} \widehat g$ simply by factorizing $S^{\mu \nu}$ out, and symmetrizing the remnant. The result is given by
\begingroup\makeatletter\def\f@size{7}\check@mathfonts
\begin{dgroup*}

\begin{dmath*}
{ + \frac{1}{16}\, {S}^{\delta \mu} {\widehat g}^{\alpha \nu} {\widehat g}^{\beta \rho} {\widehat g}^{\tau \sigma} {\widehat g}^{\gamma \kappa} {\widehat g}^{\epsilon \pi} + \frac{1}{4}\, {S}^{\delta \mu} {\widehat g}^{\alpha \nu} {\widehat g}^{\beta \sigma} {\widehat g}^{\tau \gamma} {\widehat g}^{\rho \kappa} {\widehat g}^{\epsilon \pi} - \frac{1}{4}\, {S}^{\delta \mu} {\widehat g}^{\alpha \nu} {\widehat g}^{\beta \sigma} {\widehat g}^{\tau \kappa} {\widehat g}^{\rho \gamma} {\widehat g}^{\epsilon \pi} + \frac{1}{4}\, {S}^{\delta \mu} {\widehat g}^{\alpha \sigma} {\widehat g}^{\beta \gamma} {\widehat g}^{\tau \kappa} {\widehat g}^{\nu \epsilon} {\widehat g}^{\rho \pi} - \frac{1}{4}\, {S}^{\delta \mu} {\widehat g}^{\alpha \sigma} {\widehat g}^{\beta \nu} {\widehat g}^{\tau \gamma} {\widehat g}^{\rho \kappa} {\widehat g}^{\epsilon \pi} \ \ \ \ \ \ \ \ \ \  } + \frac{1}{4}\, {S}^{\delta \mu} {\widehat g}^{\alpha \sigma} {\widehat g}^{\beta \nu} {\widehat g}^{\tau \kappa} {\widehat g}^{\rho \gamma} {\widehat g}^{\epsilon \pi}  - \frac{1}{16}\, {S}^{\delta \nu} {\widehat g}^{\alpha \mu} {\widehat g}^{\beta \rho} {\widehat g}^{\tau \sigma} {\widehat g}^{\gamma \kappa} {\widehat g}^{\epsilon \pi} + \frac{1}{16}\, {S}^{\delta \nu} {\widehat g}^{\alpha \rho} {\widehat g}^{\beta \mu} {\widehat g}^{\tau \sigma} {\widehat g}^{\gamma \kappa} {\widehat g}^{\epsilon \pi} - \frac{1}{16}\, {S}^{\delta \nu} {\widehat g}^{\alpha \sigma} {\widehat g}^{\beta \mu} {\widehat g}^{\tau \rho} {\widehat g}^{\gamma \kappa} {\widehat g}^{\epsilon \pi}\left.\right) + {\partial}_{\mu}{{\widehat g}_{\nu \rho}}\,  {\partial}_{\sigma}{{\widehat g}_{\gamma \epsilon}}\,  {\widehat B}_{\delta \lambda} {\partial}_{\alpha}{{\widehat B}_{\beta \tau}}\,  {\partial}_{\kappa \pi}{{\xi}^{\delta}}\,  \left(\right. - \frac{1}{16}\, {S}^{\alpha \kappa} {\widehat g}^{\lambda \nu} {\widehat g}^{\beta \sigma} {\widehat g}^{\tau \gamma} {\widehat g}^{\mu \epsilon} {\widehat g}^{\rho \pi} + \frac{1}{16}\, {S}^{\alpha \mu} {\widehat g}^{\lambda \gamma} {\widehat g}^{\beta \sigma} {\widehat g}^{\tau \epsilon} {\widehat g}^{\nu \kappa} {\widehat g}^{\rho \pi} - \frac{1}{16}\, {S}^{\alpha \mu} {\widehat g}^{\lambda \nu} {\widehat g}^{\beta \sigma} {\widehat g}^{\tau \gamma} {\widehat g}^{\rho \kappa} {\widehat g}^{\epsilon \pi} - \frac{1}{16}\, {S}^{\beta \kappa} {\widehat g}^{\lambda \nu} {\widehat g}^{\alpha \gamma} {\widehat g}^{\tau \epsilon} {\widehat g}^{\mu \sigma} {\widehat g}^{\rho \pi} - \frac{1}{16}\, {S}^{\beta \kappa} {\widehat g}^{\lambda \nu} {\widehat g}^{\alpha \gamma} {\widehat g}^{\tau \sigma} {\widehat g}^{\mu \epsilon} {\widehat g}^{\rho \pi} + \frac{1}{16}\, {S}^{\beta \kappa} {\widehat g}^{\lambda \nu} {\widehat g}^{\alpha \sigma} {\widehat g}^{\tau \gamma} {\widehat g}^{\mu \epsilon} {\widehat g}^{\rho \pi} + \frac{1}{16}\, {S}^{\beta \mu} {\widehat g}^{\lambda \gamma} {\widehat g}^{\alpha \epsilon} {\widehat g}^{\tau \sigma} {\widehat g}^{\nu \kappa} {\widehat g}^{\rho \pi} - \frac{1}{16}\, {S}^{\beta \mu} {\widehat g}^{\lambda \gamma} {\widehat g}^{\alpha \sigma} {\widehat g}^{\tau \epsilon} {\widehat g}^{\nu \kappa} {\widehat g}^{\rho \pi} - \frac{1}{16}\, {S}^{\beta \mu} {\widehat g}^{\lambda \nu} {\widehat g}^{\alpha \gamma} {\widehat g}^{\tau \epsilon} {\widehat g}^{\rho \kappa} {\widehat g}^{\sigma \pi} - \frac{1}{16}\, {S}^{\beta \mu} {\widehat g}^{\lambda \nu} {\widehat g}^{\alpha \gamma} {\widehat g}^{\tau \sigma} {\widehat g}^{\rho \kappa} {\widehat g}^{\epsilon \pi} + \frac{1}{16}\, {S}^{\beta \mu} {\widehat g}^{\lambda \nu} {\widehat g}^{\alpha \sigma} {\widehat g}^{\tau \gamma} {\widehat g}^{\rho \kappa} {\widehat g}^{\epsilon \pi} + \frac{1}{16}\, {S}^{\beta \mu} {\widehat g}^{\lambda \sigma} {\widehat g}^{\alpha \gamma} {\widehat g}^{\tau \epsilon} {\widehat g}^{\nu \kappa} {\widehat g}^{\rho \pi} + \frac{1}{16}\, {S}^{\mu \gamma} {\widehat g}^{\lambda \alpha} {\widehat g}^{\beta \sigma} {\widehat g}^{\tau \epsilon} {\widehat g}^{\nu \kappa} {\widehat g}^{\rho \pi} + \frac{1}{16}\, {S}^{\mu \gamma} {\widehat g}^{\lambda \beta} {\widehat g}^{\alpha \epsilon} {\widehat g}^{\tau \sigma} {\widehat g}^{\nu \kappa} {\widehat g}^{\rho \pi} - \frac{1}{16}\, {S}^{\mu \gamma} {\widehat g}^{\lambda \beta} {\widehat g}^{\alpha \sigma} {\widehat g}^{\tau \epsilon} {\widehat g}^{\nu \kappa} {\widehat g}^{\rho \pi} + \frac{1}{16}\, {S}^{\mu \gamma} {\widehat g}^{\lambda \nu} {\widehat g}^{\alpha \epsilon} {\widehat g}^{\beta \sigma} {\widehat g}^{\tau \kappa} {\widehat g}^{\rho \pi}
\end{dmath*}
\begin{dmath*}
{ - \frac{1}{16}\, {S}^{\mu \gamma} {\widehat g}^{\lambda \nu} {\widehat g}^{\alpha \kappa} {\widehat g}^{\beta \sigma} {\widehat g}^{\tau \epsilon} {\widehat g}^{\rho \pi} - \frac{1}{16}\, {S}^{\mu \gamma} {\widehat g}^{\lambda \nu} {\widehat g}^{\alpha \sigma} {\widehat g}^{\beta \epsilon} {\widehat g}^{\tau \kappa} {\widehat g}^{\rho \pi} + \frac{1}{8}\, {S}^{\mu \kappa} {\widehat g}^{\lambda \alpha} {\widehat g}^{\beta \nu} {\widehat g}^{\tau \gamma} {\widehat g}^{\rho \sigma} {\widehat g}^{\epsilon \pi}  - \frac{1}{8}\, {S}^{\mu \kappa} {\widehat g}^{\lambda \alpha} {\widehat g}^{\beta \nu} {\widehat g}^{\tau \sigma} {\widehat g}^{\rho \gamma} {\widehat g}^{\epsilon \pi}
+ \frac{1}{8}\, {S}^{\mu \kappa} {\widehat g}^{\lambda \alpha} {\widehat g}^{\beta \sigma} {\widehat g}^{\tau \gamma} {\widehat g}^{\nu \epsilon} {\widehat g}^{\rho \pi} } + \frac{1}{8}\, {S}^{\mu \kappa} {\widehat g}^{\lambda \beta} {\widehat g}^{\alpha \gamma} {\widehat g}^{\tau \epsilon} {\widehat g}^{\nu \sigma} {\widehat g}^{\rho \pi} + \frac{1}{8}\, {S}^{\mu \kappa} {\widehat g}^{\lambda \beta} {\widehat g}^{\alpha \gamma} {\widehat g}^{\tau \nu} {\widehat g}^{\rho \epsilon} {\widehat g}^{\sigma \pi} + \frac{1}{8}\, {S}^{\mu \kappa} {\widehat g}^{\lambda \beta} {\widehat g}^{\alpha \gamma} {\widehat g}^{\tau \nu} {\widehat g}^{\rho \sigma} {\widehat g}^{\epsilon \pi} + \frac{1}{8}\, {S}^{\mu \kappa} {\widehat g}^{\lambda \beta} {\widehat g}^{\alpha \gamma} {\widehat g}^{\tau \sigma} {\widehat g}^{\nu \epsilon} {\widehat g}^{\rho \pi} + \frac{1}{8}\, {S}^{\mu \kappa} {\widehat g}^{\lambda \beta} {\widehat g}^{\alpha \nu} {\widehat g}^{\tau \gamma} {\widehat g}^{\rho \epsilon} {\widehat g}^{\sigma \pi} - \frac{1}{8}\, {S}^{\mu \kappa} {\widehat g}^{\lambda \beta} {\widehat g}^{\alpha \nu} {\widehat g}^{\tau \gamma} {\widehat g}^{\rho \sigma} {\widehat g}^{\epsilon \pi} + \frac{1}{8}\, {S}^{\mu \kappa} {\widehat g}^{\lambda \beta} {\widehat g}^{\alpha \nu} {\widehat g}^{\tau \sigma} {\widehat g}^{\rho \gamma} {\widehat g}^{\epsilon \pi} - \frac{1}{8}\, {S}^{\mu \kappa} {\widehat g}^{\lambda \beta} {\widehat g}^{\alpha \sigma} {\widehat g}^{\tau \gamma} {\widehat g}^{\nu \epsilon} {\widehat g}^{\rho \pi} - \frac{1}{8}\, {S}^{\mu \kappa} {\widehat g}^{\lambda \beta} {\widehat g}^{\alpha \sigma} {\widehat g}^{\tau \nu} {\widehat g}^{\rho \gamma} {\widehat g}^{\epsilon \pi} - \frac{1}{8}\, {S}^{\mu \kappa} {\widehat g}^{\lambda \gamma} {\widehat g}^{\alpha \epsilon} {\widehat g}^{\beta \nu} {\widehat g}^{\tau \pi} {\widehat g}^{\rho \sigma} + \frac{1}{8}\, {S}^{\mu \kappa} {\widehat g}^{\lambda \gamma} {\widehat g}^{\alpha \epsilon} {\widehat g}^{\beta \nu} {\widehat g}^{\tau \sigma} {\widehat g}^{\rho \pi} + \frac{1}{8}\, {S}^{\mu \kappa} {\widehat g}^{\lambda \gamma} {\widehat g}^{\alpha \nu} {\widehat g}^{\beta \epsilon} {\widehat g}^{\tau \pi} {\widehat g}^{\rho \sigma} + \frac{1}{16}\, {S}^{\mu \kappa} {\widehat g}^{\lambda \gamma} {\widehat g}^{\alpha \nu} {\widehat g}^{\beta \rho} {\widehat g}^{\tau \sigma} {\widehat g}^{\epsilon \pi} + \frac{1}{8}\, {S}^{\mu \kappa} {\widehat g}^{\lambda \gamma} {\widehat g}^{\alpha \nu} {\widehat g}^{\beta \sigma} {\widehat g}^{\tau \epsilon} {\widehat g}^{\rho \pi} - \frac{1}{8}\, {S}^{\mu \kappa} {\widehat g}^{\lambda \gamma} {\widehat g}^{\alpha \nu} {\widehat g}^{\beta \sigma} {\widehat g}^{\tau \pi} {\widehat g}^{\rho \epsilon} + \frac{1}{8}\, {S}^{\mu \kappa} {\widehat g}^{\lambda \gamma} {\widehat g}^{\alpha \pi} {\widehat g}^{\beta \nu} {\widehat g}^{\tau \epsilon} {\widehat g}^{\rho \sigma} - \frac{1}{8}\, {S}^{\mu \kappa} {\widehat g}^{\lambda \gamma} {\widehat g}^{\alpha \pi} {\widehat g}^{\beta \nu} {\widehat g}^{\tau \sigma} {\widehat g}^{\rho \epsilon} - \frac{1}{8}\, {S}^{\mu \kappa} {\widehat g}^{\lambda \gamma} {\widehat g}^{\alpha \sigma} {\widehat g}^{\beta \nu} {\widehat g}^{\tau \epsilon} {\widehat g}^{\rho \pi}%
 + \frac{1}{8}\, {S}^{\mu \kappa} {\widehat g}^{\lambda \gamma} {\widehat g}^{\alpha \sigma} {\widehat g}^{\beta \nu} {\widehat g}^{\tau \pi} {\widehat g}^{\rho \epsilon} - \frac{1}{8}\, {S}^{\mu \kappa} {\widehat g}^{\lambda \nu} {\widehat g}^{\alpha \gamma} {\widehat g}^{\beta \epsilon} {\widehat g}^{\tau \pi} {\widehat g}^{\rho \sigma} + \frac{1}{8}\, {S}^{\mu \kappa} {\widehat g}^{\lambda \nu} {\widehat g}^{\alpha \gamma} {\widehat g}^{\beta \rho} {\widehat g}^{\tau \epsilon} {\widehat g}^{\sigma \pi} + \frac{1}{8}\, {S}^{\mu \kappa} {\widehat g}^{\lambda \nu} {\widehat g}^{\alpha \gamma} {\widehat g}^{\beta \rho} {\widehat g}^{\tau \sigma} {\widehat g}^{\epsilon \pi} - \frac{1}{8}\, {S}^{\mu \kappa} {\widehat g}^{\lambda \nu} {\widehat g}^{\alpha \gamma} {\widehat g}^{\beta \sigma} {\widehat g}^{\tau \pi} {\widehat g}^{\rho \epsilon} + \frac{1}{8}\, {S}^{\mu \kappa} {\widehat g}^{\lambda \nu} {\widehat g}^{\alpha \pi} {\widehat g}^{\beta \sigma} {\widehat g}^{\tau \gamma} {\widehat g}^{\rho \epsilon} + \frac{1}{8}\, {S}^{\mu \kappa} {\widehat g}^{\lambda \nu} {\widehat g}^{\alpha \rho} {\widehat g}^{\beta \sigma} {\widehat g}^{\tau \gamma} {\widehat g}^{\epsilon \pi} + \frac{1}{8}\, {S}^{\mu \kappa} {\widehat g}^{\lambda \nu} {\widehat g}^{\alpha \sigma} {\widehat g}^{\beta \gamma} {\widehat g}^{\tau \pi} {\widehat g}^{\rho \epsilon} - \frac{1}{8}\, {S}^{\mu \kappa} {\widehat g}^{\lambda \nu} {\widehat g}^{\alpha \sigma} {\widehat g}^{\beta \rho} {\widehat g}^{\tau \gamma} {\widehat g}^{\epsilon \pi} + \frac{1}{8}\, {S}^{\mu \kappa} {\widehat g}^{\lambda \sigma} {\widehat g}^{\alpha \gamma} {\widehat g}^{\beta \nu} {\widehat g}^{\tau \epsilon} {\widehat g}^{\rho \pi} - \frac{1}{8}\, {S}^{\mu \kappa} {\widehat g}^{\lambda \sigma} {\widehat g}^{\alpha \gamma} {\widehat g}^{\beta \nu} {\widehat g}^{\tau \pi} {\widehat g}^{\rho \epsilon} - \frac{1}{8}\, {S}^{\mu \kappa} {\widehat g}^{\lambda \sigma} {\widehat g}^{\alpha \nu} {\widehat g}^{\beta \gamma} {\widehat g}^{\tau \pi} {\widehat g}^{\rho \epsilon} + \frac{1}{16}\, {S}^{\mu \sigma} {\widehat g}^{\lambda \beta} {\widehat g}^{\alpha \nu} {\widehat g}^{\tau \rho} {\widehat g}^{\gamma \kappa} {\widehat g}^{\epsilon \pi} + \frac{1}{16}\, {S}^{\mu \sigma} {\widehat g}^{\lambda \nu} {\widehat g}^{\alpha \gamma} {\widehat g}^{\beta \epsilon} {\widehat g}^{\tau \kappa} {\widehat g}^{\rho \pi} - \frac{1}{16}\, {S}^{\nu \kappa} {\widehat g}^{\lambda \gamma} {\widehat g}^{\alpha \mu} {\widehat g}^{\beta \rho} {\widehat g}^{\tau \sigma} {\widehat g}^{\epsilon \pi} + \frac{1}{16}\, {S}^{\nu \kappa} {\widehat g}^{\lambda \gamma} {\widehat g}^{\alpha \rho} {\widehat g}^{\beta \mu} {\widehat g}^{\tau \sigma} {\widehat g}^{\epsilon \pi} - \frac{1}{16}\, {S}^{\nu \kappa} {\widehat g}^{\lambda \gamma} {\widehat g}^{\alpha \sigma} {\widehat g}^{\beta \mu} {\widehat g}^{\tau \rho} {\widehat g}^{\epsilon \pi}\left.\right) + {\widehat g}_{\mu \nu} {\partial}_{\rho}{{\widehat g}_{\sigma \gamma}}\,  {\partial}_{\epsilon}{{\widehat B}_{\delta \lambda}}\,  {\partial}_{\alpha}{{\widehat B}_{\beta \tau}}\,  {\partial}_{\kappa \pi}{{\xi}^{\nu}}\,  \left(\right.\frac{1}{16}\, {S}^{\delta \mu} {\widehat g}^{\epsilon \alpha} {\widehat g}^{\lambda \beta} {\widehat g}^{\tau \rho} {\widehat g}^{\sigma \kappa} {\widehat g}^{\gamma \pi} - \frac{1}{16}\, {S}^{\delta \mu} {\widehat g}^{\epsilon \beta} {\widehat g}^{\lambda \alpha} {\widehat g}^{\tau \rho} {\widehat g}^{\sigma \kappa} {\widehat g}^{\gamma \pi} + \frac{1}{16}\, {S}^{\delta \mu} {\widehat g}^{\epsilon \beta} {\widehat g}^{\lambda \tau} {\widehat g}^{\alpha \rho} {\widehat g}^{\sigma \kappa} {\widehat g}^{\gamma \pi} - \frac{1}{16}\, {S}^{\epsilon \mu} {\widehat g}^{\delta \alpha} {\widehat g}^{\lambda \beta} {\widehat g}^{\tau \rho} {\widehat g}^{\sigma \kappa} {\widehat g}^{\gamma \pi} - \frac{1}{32}\, {S}^{\epsilon \mu} {\widehat g}^{\delta \beta} {\widehat g}^{\lambda \tau} {\widehat g}^{\alpha \rho} {\widehat g}^{\sigma \kappa} {\widehat g}^{\gamma \pi}\left.\right) + {\widehat g}_{\mu \nu} {\partial}_{\rho}{{\widehat g}_{\sigma \gamma}}\,  {\partial}_{\epsilon}{{\widehat g}_{\delta \lambda}}\,  {\partial}_{\alpha}{{\widehat g}_{\beta \tau}}\,  {\partial}_{\kappa \pi}{{\xi}^{\nu}}\,  \left(\right. - \frac{1}{32}\, {S}^{\rho \mu} {\widehat g}^{\sigma \delta} {\widehat g}^{\gamma \lambda} {\widehat g}^{\epsilon \alpha} {\widehat g}^{\beta \kappa} {\widehat g}^{\tau \pi} + \frac{1}{16}\, {S}^{\rho \mu} {\widehat g}^{\sigma \epsilon} {\widehat g}^{\gamma \delta} {\widehat g}^{\lambda \alpha} {\widehat g}^{\beta \kappa} {\widehat g}^{\tau \pi} + \frac{1}{16}\, {S}^{\sigma \mu} {\widehat g}^{\rho \delta} {\widehat g}^{\gamma \epsilon} {\widehat g}^{\lambda \alpha} {\widehat g}^{\beta \kappa} {\widehat g}^{\tau \pi} + \frac{1}{16}\, {S}^{\sigma \mu} {\widehat g}^{\rho \delta} {\widehat g}^{\gamma \lambda} {\widehat g}^{\epsilon \alpha} {\widehat g}^{\beta \kappa} {\widehat g}^{\tau \pi} - \frac{1}{16}\, {S}^{\sigma \mu} {\widehat g}^{\rho \epsilon} {\widehat g}^{\gamma \delta} {\widehat g}^{\lambda \alpha} {\widehat g}^{\beta \kappa} {\widehat g}^{\tau \pi}\left.\right)
\end{dmath*}
\end{dgroup*}
\endgroup

The second order transformation of the duality covariant two-form $\delta^{(2)}\widehat B_{\mu \nu}$ in (\ref{transfB}) also has a shorter expression when contracted in this case with some antisymmetric tensor $A^{\mu \nu}$
\be
\delta^{(2)} \widehat B = A^{\mu \nu} \delta^{(2)}\widehat B_{\mu \nu} \ .\label{d2tildeBa}
\ee
The result is
\begingroup\makeatletter\def\f@size{7}\check@mathfonts
\begin{dgroup*}
\begin{dmath*}
\delta^{(2)} \widehat B =  - \frac{1}{4}\, {\partial}_{\mu}{{g}_{\nu \rho}}\,  {\partial}_{\sigma \gamma}{{B}_{\epsilon \delta}}\,  {\partial}_{\lambda \alpha}{{\xi}^{\gamma}}\,  {A}^{\sigma \lambda} {g}^{\mu \epsilon} {g}^{\nu \delta} {g}^{\rho \alpha} + {\partial}_{\mu}{{g}_{\nu \rho}}\,  {\partial}_{\sigma \gamma}{{B}_{\epsilon \delta}}\,  {\partial}_{\lambda \alpha}{{\xi}^{\sigma}}\,  \left(\right. - \frac{1}{4}\, {A}^{\epsilon \lambda} {g}^{\mu \alpha} {g}^{\nu \gamma} {g}^{\rho \delta} - \frac{1}{4}\, {A}^{\epsilon \lambda} {g}^{\mu \delta} {g}^{\nu \gamma} {g}^{\rho \alpha} + \frac{1}{4}\, {A}^{\epsilon \lambda} {g}^{\mu \gamma} {g}^{\nu \delta} {g}^{\rho \alpha} + \frac{1}{4}\, {A}^{\mu \lambda} {g}^{\nu \gamma} {g}^{\rho \epsilon} {g}^{\delta \alpha} - \frac{1}{4}\, {A}^{\nu \lambda} {g}^{\mu \epsilon} {g}^{\rho \delta} {g}^{\gamma \alpha} + \frac{1}{4}\, {A}^{\nu \lambda} {g}^{\mu \epsilon} {g}^{\rho \gamma} {g}^{\delta \alpha} - \frac{1}{4}\, {A}^{\nu \lambda} {g}^{\mu \gamma} {g}^{\rho \epsilon} {g}^{\delta \alpha}\left.\right) + \frac{1}{4}\, {\partial}_{\mu \nu}{{g}_{\rho \sigma}}\,  {\partial}_{\gamma}{{B}_{\epsilon \delta}}\,  {\partial}_{\lambda \alpha}{{\xi}^{\nu}}\,  {A}^{\lambda \mu} {g}^{\gamma \rho} {g}^{\epsilon \alpha} {g}^{\delta \sigma} + {\partial}_{\mu \nu}{{g}_{\rho \sigma}}\,  {\partial}_{\gamma}{{B}_{\epsilon \delta}}\,  {\partial}_{\lambda \alpha}{{\xi}^{\mu}}\,  \left(\right.\frac{1}{4}\, {A}^{\epsilon \lambda} {g}^{\gamma \nu} {g}^{\delta \rho} {g}^{\alpha \sigma} - \frac{1}{4}\, {A}^{\epsilon \lambda} {g}^{\gamma \rho} {g}^{\delta \nu} {g}^{\alpha \sigma} - \frac{1}{4}\, {A}^{\epsilon \lambda} {g}^{\gamma \rho} {g}^{\delta \sigma} {g}^{\alpha \nu} - \frac{1}{4}\, {A}^{\gamma \lambda} {g}^{\epsilon \nu} {g}^{\delta \rho} {g}^{\alpha \sigma} + \frac{1}{4}\, {A}^{\lambda \rho} {g}^{\gamma \alpha} {g}^{\epsilon \nu} {g}^{\delta \sigma} - \frac{1}{4}\, {A}^{\lambda \rho} {g}^{\gamma \nu} {g}^{\epsilon \alpha} {g}^{\delta \sigma} + \frac{1}{4}\, {A}^{\lambda \rho} {g}^{\gamma \sigma} {g}^{\epsilon \alpha} {g}^{\delta \nu}\left.\right) + {\partial}_{\mu}{{B}_{\nu \rho}}\,  {\partial}_{\sigma}{{B}_{\gamma \epsilon}}\,  {\partial}_{\delta}{{B}_{\lambda \alpha}}\,  {\partial}_{\beta \tau}{{\xi}^{\mu}}\,  \left(\right.\frac{1}{8}\, {A}^{\nu \gamma} {g}^{\rho \beta} {g}^{\sigma \delta} {g}^{\epsilon \lambda} {g}^{\alpha \tau} + \frac{1}{8}\, {A}^{\nu \gamma} {g}^{\rho \beta} {g}^{\sigma \lambda} {g}^{\epsilon \alpha} {g}^{\delta \tau} - \frac{1}{8}\, {A}^{\nu \gamma} {g}^{\rho \beta} {g}^{\sigma \lambda} {g}^{\epsilon \delta} {g}^{\alpha \tau}\left.\right) + {\partial}_{\mu}{{B}_{\nu \rho}}\,  {\partial}_{\sigma}{{B}_{\gamma \epsilon}}\,  {\partial}_{\delta}{{B}_{\lambda \alpha}}\,  {\partial}_{\beta \tau}{{\xi}^{\nu}}\,  \left(\right. - \frac{1}{16}\, {A}^{\mu \gamma} {g}^{\rho \beta} {g}^{\sigma \delta} {g}^{\epsilon \lambda} {g}^{\alpha \tau} - \frac{1}{16}\, {A}^{\mu \gamma} {g}^{\rho \beta} {g}^{\sigma \lambda} {g}^{\epsilon \alpha} {g}^{\delta \tau} + \frac{1}{16}\, {A}^{\mu \gamma} {g}^{\rho \beta} {g}^{\sigma \lambda} {g}^{\epsilon \delta} {g}^{\alpha \tau} + \frac{1}{16}\, {A}^{\mu \sigma} {g}^{\rho \beta} {g}^{\gamma \delta} {g}^{\epsilon \lambda} {g}^{\alpha \tau} + \frac{1}{32}\, {A}^{\mu \sigma} {g}^{\rho \beta} {g}^{\gamma \lambda} {g}^{\epsilon \alpha} {g}^{\delta \tau}\left.\right) + {\partial}_{\mu}{{B}_{\nu \rho}}\,  {\partial}_{\sigma}{{B}_{\gamma \epsilon}}\,  {\partial}_{\delta}{{B}_{\lambda \alpha}}\,  {\partial}_{\beta \tau}{{\xi}^{\gamma}}\,  \left(\right. - \frac{1}{16}\, {A}^{\mu \beta} {g}^{\nu \delta} {g}^{\rho \lambda} {g}^{\sigma \alpha} {g}^{\epsilon \tau} - \frac{1}{32}\, {A}^{\mu \beta} {g}^{\nu \lambda} {g}^{\rho \alpha} {g}^{\sigma \delta} {g}^{\epsilon \tau} + \frac{1}{16}\, {A}^{\nu \beta} {g}^{\mu \delta} {g}^{\rho \lambda} {g}^{\sigma \alpha} {g}^{\epsilon \tau} + \frac{1}{16}\, {A}^{\nu \beta} {g}^{\mu \lambda} {g}^{\rho \alpha} {g}^{\sigma \delta} {g}^{\epsilon \tau} - \frac{1}{16}\, {A}^{\nu \beta} {g}^{\mu \lambda} {g}^{\rho \delta} {g}^{\sigma \alpha} {g}^{\epsilon \tau}\left.\right) + {\partial}_{\mu}{{B}_{\nu \rho}}\,  {\partial}_{\sigma}{{B}_{\gamma \epsilon}}\,  {\partial}_{\delta}{{B}_{\lambda \alpha}}\,  {\partial}_{\beta \tau}{{\xi}^{\sigma}}\,  \left(\right.\frac{1}{8}\, {A}^{\mu \beta} {g}^{\nu \delta} {g}^{\rho \lambda} {g}^{\gamma \alpha} {g}^{\epsilon \tau} + \frac{1}{16}\, {A}^{\mu \beta} {g}^{\nu \lambda} {g}^{\rho \alpha} {g}^{\gamma \delta} {g}^{\epsilon \tau} + \frac{1}{8}\, {A}^{\mu \gamma} {g}^{\nu \delta} {g}^{\rho \lambda} {g}^{\epsilon \beta} {g}^{\alpha \tau} + \frac{1}{16}\, {A}^{\mu \gamma} {g}^{\nu \lambda} {g}^{\rho \alpha} {g}^{\epsilon \beta} {g}^{\delta \tau} - \frac{1}{8}\, {A}^{\nu \beta} {g}^{\mu \delta} {g}^{\rho \lambda} {g}^{\gamma \alpha} {g}^{\epsilon \tau} - \frac{1}{8}\, {A}^{\nu \beta} {g}^{\mu \lambda} {g}^{\rho \alpha} {g}^{\gamma \delta} {g}^{\epsilon \tau} + \frac{1}{8}\, {A}^{\nu \beta} {g}^{\mu \lambda} {g}^{\rho \delta} {g}^{\gamma \alpha} {g}^{\epsilon \tau}\left.\right) + {\partial}_{\mu}{{g}_{\nu \rho}}\,  {\partial}_{\sigma}{{B}_{\gamma \epsilon}}\,  {\partial}_{\delta}{{B}_{\lambda \alpha}}\,  {\partial}_{\beta \tau}{{\xi}_{\kappa}}\,  \left(\right. - \frac{1}{16}\, {A}^{\gamma \beta} {g}^{\sigma \delta} {g}^{\epsilon \lambda} {g}^{\alpha \mu} {g}^{\nu \tau} {g}^{\rho \kappa} - \frac{1}{16}\, {A}^{\gamma \beta} {g}^{\sigma \lambda} {g}^{\epsilon \alpha} {g}^{\delta \mu} {g}^{\nu \tau} {g}^{\rho \kappa} + \frac{1}{16}\, {A}^{\gamma \beta} {g}^{\sigma \lambda} {g}^{\epsilon \delta} {g}^{\alpha \mu} {g}^{\nu \tau} {g}^{\rho \kappa} + \frac{1}{16}\, {A}^{\gamma \kappa} {g}^{\sigma \delta} {g}^{\epsilon \lambda} {g}^{\alpha \mu} {g}^{\nu \beta} {g}^{\rho \tau} + \frac{1}{16}\, {A}^{\gamma \kappa} {g}^{\sigma \lambda} {g}^{\epsilon \alpha} {g}^{\delta \mu} {g}^{\nu \beta} {g}^{\rho \tau} - \frac{1}{16}\, {A}^{\gamma \kappa} {g}^{\sigma \lambda} {g}^{\epsilon \delta} {g}^{\alpha \mu} {g}^{\nu \beta} {g}^{\rho \tau} - \frac{1}{16}\, {A}^{\gamma \mu} {g}^{\sigma \delta} {g}^{\epsilon \lambda} {g}^{\alpha \beta} {g}^{\nu \tau} {g}^{\rho \kappa} + \frac{1}{16}\, {A}^{\gamma \mu} {g}^{\sigma \delta} {g}^{\epsilon \lambda} {g}^{\alpha \kappa} {g}^{\nu \beta} {g}^{\rho \tau} - \frac{1}{16}\, {A}^{\gamma \mu} {g}^{\sigma \lambda} {g}^{\epsilon \alpha} {g}^{\delta \beta} {g}^{\nu \tau} {g}^{\rho \kappa} + \frac{1}{16}\, {A}^{\gamma \mu} {g}^{\sigma \lambda} {g}^{\epsilon \alpha} {g}^{\delta \kappa} {g}^{\nu \beta} {g}^{\rho \tau} + \frac{1}{16}\, {A}^{\gamma \mu} {g}^{\sigma \lambda} {g}^{\epsilon \delta} {g}^{\alpha \beta} {g}^{\nu \tau} {g}^{\rho \kappa} - \frac{1}{16}\, {A}^{\gamma \mu} {g}^{\sigma \lambda} {g}^{\epsilon \delta} {g}^{\alpha \kappa} {g}^{\nu \beta} {g}^{\rho \tau} + \frac{1}{16}\, {A}^{\sigma \beta} {g}^{\gamma \delta} {g}^{\epsilon \lambda} {g}^{\alpha \mu} {g}^{\nu \tau} {g}^{\rho \kappa} + \frac{1}{32}\, {A}^{\sigma \beta} {g}^{\gamma \lambda} {g}^{\epsilon \alpha} {g}^{\delta \mu} {g}^{\nu \tau} {g}^{\rho \kappa} - \frac{1}{16}\, {A}^{\sigma \kappa} {g}^{\gamma \delta} {g}^{\epsilon \lambda} {g}^{\alpha \mu} {g}^{\nu \beta} {g}^{\rho \tau} - \frac{1}{32}\, {A}^{\sigma \kappa} {g}^{\gamma \lambda} {g}^{\epsilon \alpha} {g}^{\delta \mu} {g}^{\nu \beta} {g}^{\rho \tau} + \frac{1}{16}\, {A}^{\sigma \mu} {g}^{\gamma \delta} {g}^{\epsilon \lambda} {g}^{\alpha \beta} {g}^{\nu \tau} {g}^{\rho \kappa} - \frac{1}{16}\, {A}^{\sigma \mu} {g}^{\gamma \delta} {g}^{\epsilon \lambda} {g}^{\alpha \kappa} {g}^{\nu \beta} {g}^{\rho \tau} + \frac{1}{32}\, {A}^{\sigma \mu} {g}^{\gamma \lambda} {g}^{\epsilon \alpha} {g}^{\delta \beta} {g}^{\nu \tau} {g}^{\rho \kappa}  - \frac{1}{32}\, {A}^{\sigma \mu} {g}^{\gamma \lambda} {g}^{\epsilon \alpha} {g}^{\delta \kappa} {g}^{\nu \beta} {g}^{\rho \tau}\left.\right) + {\partial}_{\mu}{{g}_{\nu \rho}}\,  {\partial}_{\sigma}{{g}_{\gamma \epsilon}}\,  {\partial}_{\delta}{{B}_{\lambda \alpha}}\,  {\partial}_{\beta \tau}{{\xi}^{\gamma}}\,  \left(\right. - \frac{1}{16}\, {A}^{\delta \mu} {g}^{\lambda \sigma} {g}^{\alpha \epsilon} {g}^{\nu \beta} {g}^{\rho \tau} - \frac{1}{16}\, {A}^{\lambda \mu} {g}^{\delta \epsilon} {g}^{\alpha \sigma} {g}^{\nu \beta} {g}^{\rho \tau} + \frac{1}{16}\, {A}^{\lambda \mu} {g}^{\delta \sigma} {g}^{\alpha \epsilon} {g}^{\nu \beta} {g}^{\rho \tau}\left.\right) + {\partial}_{\mu}{{g}_{\nu \rho}}\,  {\partial}_{\sigma}{{g}_{\gamma \epsilon}}\,  {\partial}_{\delta}{{B}_{\lambda \alpha}}\,  {\partial}_{\beta \tau}{{\xi}^{\delta}}\,  \left(\right. - \frac{1}{16}\, {A}^{\lambda \mu} {g}^{\alpha \beta} {g}^{\nu \gamma} {g}^{\rho \epsilon} {g}^{\sigma \tau} + \frac{1}{8}\, {A}^{\lambda \mu} {g}^{\alpha \beta} {g}^{\nu \sigma} {g}^{\rho \gamma} {g}^{\epsilon \tau} + \frac{1}{8}\, {A}^{\lambda \nu} {g}^{\alpha \beta} {g}^{\mu \gamma} {g}^{\rho \epsilon} {g}^{\sigma \tau} + \frac{1}{8}\, {A}^{\lambda \nu} {g}^{\alpha \beta} {g}^{\mu \gamma} {g}^{\rho \sigma} {g}^{\epsilon \tau} - \frac{1}{8}\, {A}^{\lambda \nu} {g}^{\alpha \beta} {g}^{\mu \sigma} {g}^{\rho \gamma} {g}^{\epsilon \tau} - \frac{1}{8}\, {A}^{\mu \beta} {g}^{\lambda \gamma} {g}^{\alpha \tau} {g}^{\nu \sigma} {g}^{\rho \epsilon} + \frac{1}{16}\, {A}^{\mu \beta} {g}^{\lambda \sigma} {g}^{\alpha \tau} {g}^{\nu \gamma} {g}^{\rho \epsilon} + \frac{1}{16}\, {A}^{\mu \gamma} {g}^{\lambda \sigma} {g}^{\alpha \epsilon} {g}^{\nu \beta} {g}^{\rho \tau} - \frac{1}{8}\, {A}^{\nu \beta} {g}^{\lambda \gamma} {g}^{\alpha \tau} {g}^{\mu \epsilon} {g}^{\rho \sigma} + \frac{1}{8}\, {A}^{\nu \beta} {g}^{\lambda \gamma} {g}^{\alpha \tau} {g}^{\mu \sigma} {g}^{\rho \epsilon} - \frac{1}{8}\, {A}^{\nu \beta} {g}^{\lambda \sigma} {g}^{\alpha \tau} {g}^{\mu \gamma} {g}^{\rho \epsilon}\left.\right) + {\partial}_{\mu}{{g}_{\nu \rho}}\,  {\partial}_{\sigma}{{g}_{\gamma \epsilon}}\,  {\partial}_{\delta}{{B}_{\lambda \alpha}}\,  {\partial}_{\beta \tau}{{\xi}^{\lambda}}\,  \left(\right.\frac{1}{32}\, {A}^{\delta \mu} {g}^{\alpha \beta} {g}^{\nu \gamma} {g}^{\rho \epsilon} {g}^{\sigma \tau} - \frac{1}{16}\, {A}^{\delta \mu} {g}^{\alpha \beta} {g}^{\nu \sigma} {g}^{\rho \gamma} {g}^{\epsilon \tau} - \frac{1}{16}\, {A}^{\delta \nu} {g}^{\alpha \beta} {g}^{\mu \gamma} {g}^{\rho \epsilon} {g}^{\sigma \tau} - \frac{1}{16}\, {A}^{\delta \nu} {g}^{\alpha \beta} {g}^{\mu \gamma} {g}^{\rho \sigma} {g}^{\epsilon \tau} + \frac{1}{16}\, {A}^{\delta \nu} {g}^{\alpha \beta} {g}^{\mu \sigma} {g}^{\rho \gamma} {g}^{\epsilon \tau} + \frac{1}{16}\, {A}^{\mu \beta} {g}^{\delta \gamma} {g}^{\alpha \tau} {g}^{\nu \sigma} {g}^{\rho \epsilon} - \frac{1}{32}\, {A}^{\mu \beta} {g}^{\delta \sigma} {g}^{\alpha \tau} {g}^{\nu \gamma} {g}^{\rho \epsilon} + \frac{1}{16}\, {A}^{\mu \gamma} {g}^{\delta \epsilon} {g}^{\alpha \sigma} {g}^{\nu \beta} {g}^{\rho \tau} - \frac{1}{16}\, {A}^{\mu \gamma} {g}^{\delta \sigma} {g}^{\alpha \epsilon} {g}^{\nu \beta} {g}^{\rho \tau} - \frac{1}{16}\, {A}^{\mu \sigma} {g}^{\delta \nu} {g}^{\alpha \rho} {g}^{\gamma \beta} {g}^{\epsilon \tau} + \frac{1}{16}\, {A}^{\nu \beta} {g}^{\delta \gamma} {g}^{\alpha \tau} {g}^{\mu \epsilon} {g}^{\rho \sigma} - \frac{1}{16}\, {A}^{\nu \beta} {g}^{\delta \gamma} {g}^{\alpha \tau} {g}^{\mu \sigma} {g}^{\rho \epsilon} + \frac{1}{16}\, {A}^{\nu \beta} {g}^{\delta \sigma} {g}^{\alpha \tau} {g}^{\mu \gamma} {g}^{\rho \epsilon}\left.\right) + {\partial}_{\mu}{{g}_{\nu \rho}}\,  {\partial}_{\sigma}{{g}_{\gamma \epsilon}}\,  {\partial}_{\delta}{{B}_{\lambda \alpha}}\,  {\partial}_{\beta \tau}{{\xi}^{\sigma}}\,  \left(\right. - \frac{1}{16}\, {A}^{\lambda \mu} {g}^{\delta \gamma} {g}^{\alpha \epsilon} {g}^{\nu \beta} {g}^{\rho \tau} - \frac{1}{4}\, {A}^{\mu \beta} {g}^{\delta \gamma} {g}^{\lambda \nu} {g}^{\alpha \tau} {g}^{\rho \epsilon} - \frac{1}{4}\, {A}^{\mu \beta} {g}^{\delta \nu} {g}^{\lambda \gamma} {g}^{\alpha \tau} {g}^{\rho \epsilon} - \frac{1}{8}\, {A}^{\mu \beta} {g}^{\delta \nu} {g}^{\lambda \rho} {g}^{\alpha \gamma} {g}^{\epsilon \tau} - \frac{1}{8}\, {A}^{\mu \gamma} {g}^{\delta \nu} {g}^{\lambda \rho} {g}^{\alpha \beta} {g}^{\epsilon \tau} + \frac{1}{8}\, {A}^{\nu \beta} {g}^{\delta \gamma} {g}^{\lambda \mu} {g}^{\alpha \rho} {g}^{\epsilon \tau} - \frac{1}{4}\, {A}^{\nu \beta} {g}^{\delta \gamma} {g}^{\lambda \mu} {g}^{\alpha \tau} {g}^{\rho \epsilon} + \frac{1}{4}\, {A}^{\nu \beta} {g}^{\delta \gamma} {g}^{\lambda \rho} {g}^{\alpha \tau} {g}^{\mu \epsilon} + \frac{1}{4}\, {A}^{\nu \beta} {g}^{\delta \mu} {g}^{\lambda \gamma} {g}^{\alpha \tau} {g}^{\rho \epsilon} + \frac{1}{8}\, {A}^{\nu \beta} {g}^{\delta \mu} {g}^{\lambda \rho} {g}^{\alpha \gamma} {g}^{\epsilon \tau} - \frac{1}{4}\, {A}^{\nu \beta} {g}^{\delta \rho} {g}^{\lambda \gamma} {g}^{\alpha \tau} {g}^{\mu \epsilon} - \frac{1}{8}\, {A}^{\nu \beta} {g}^{\delta \rho} {g}^{\lambda \mu} {g}^{\alpha \gamma} {g}^{\epsilon \tau} + \frac{1}{4}\, {A}^{\nu \beta} {g}^{\delta \tau} {g}^{\lambda \mu} {g}^{\alpha \gamma} {g}^{\rho \epsilon} - \frac{1}{4}\, {A}^{\nu \beta} {g}^{\delta \tau} {g}^{\lambda \rho} {g}^{\alpha \gamma} {g}^{\mu \epsilon}\left.\right) + {\partial}_{\mu}{{g}_{\nu \rho}}\,  {\partial}_{\sigma}{{g}_{\gamma \epsilon}}\,  {\partial}_{\delta}{{B}_{\lambda \alpha}}\,  {\partial}_{\beta \tau}{{\xi}^{\mu}}\,  \left(\right.\frac{1}{4}\, {A}^{\delta \beta} {g}^{\lambda \nu} {g}^{\alpha \gamma} {g}^{\rho \sigma} {g}^{\epsilon \tau} - \frac{1}{4}\, {A}^{\delta \beta} {g}^{\lambda \nu} {g}^{\alpha \sigma} {g}^{\rho \gamma} {g}^{\epsilon \tau} + \frac{1}{8}\, {A}^{\delta \beta} {g}^{\lambda \sigma} {g}^{\alpha \gamma} {g}^{\nu \epsilon} {g}^{\rho \tau} + \frac{1}{8}\, {A}^{\delta \nu} {g}^{\lambda \sigma} {g}^{\alpha \gamma} {g}^{\rho \beta} {g}^{\epsilon \tau} + \frac{1}{8}\, {A}^{\lambda \beta} {g}^{\delta \gamma} {g}^{\alpha \epsilon} {g}^{\nu \sigma} {g}^{\rho \tau} + \frac{1}{4}\, {A}^{\lambda \beta} {g}^{\delta \gamma} {g}^{\alpha \nu} {g}^{\rho \epsilon} {g}^{\sigma \tau} + \frac{1}{4}\, {A}^{\lambda \beta} {g}^{\delta \gamma} {g}^{\alpha \nu} {g}^{\rho \sigma} {g}^{\epsilon \tau}
  \end{dmath*}
\begin{dmath*}
{ + \frac{1}{8}\, {A}^{\lambda \beta} {g}^{\delta \gamma} {g}^{\alpha \sigma} {g}^{\nu \epsilon} {g}^{\rho \tau} + \frac{1}{4}\, {A}^{\lambda \beta} {g}^{\delta \nu} {g}^{\alpha \gamma} {g}^{\rho \epsilon} {g}^{\sigma \tau} - \frac{1}{4}\, {A}^{\lambda \beta} {g}^{\delta \nu} {g}^{\alpha \gamma} {g}^{\rho \sigma} {g}^{\epsilon \tau} + \frac{1}{4}\, {A}^{\lambda \beta} {g}^{\delta \nu} {g}^{\alpha \sigma} {g}^{\rho \gamma} {g}^{\epsilon \tau} - \frac{1}{8}\, {A}^{\lambda \beta} {g}^{\delta \sigma} {g}^{\alpha \gamma} {g}^{\nu \epsilon} {g}^{\rho \tau} - \frac{1}{4}\, {A}^{\lambda \beta} {g}^{\delta \sigma} {g}^{\alpha \nu} {g}^{\rho \gamma} {g}^{\epsilon \tau} \ \ \ \ \ \ } + \frac{1}{8}\, {A}^{\lambda \nu} {g}^{\delta \gamma} {g}^{\alpha \epsilon} {g}^{\rho \beta} {g}^{\sigma \tau} + \frac{1}{8}\, {A}^{\lambda \nu} {g}^{\delta \gamma} {g}^{\alpha \sigma} {g}^{\rho \beta} {g}^{\epsilon \tau} - \frac{1}{8}\, {A}^{\lambda \nu} {g}^{\delta \sigma} {g}^{\alpha \gamma} {g}^{\rho \beta} {g}^{\epsilon \tau} - \frac{1}{4}\, {A}^{\nu \beta} {g}^{\delta \gamma} {g}^{\lambda \epsilon} {g}^{\alpha \tau} {g}^{\rho \sigma} + \frac{1}{4}\, {A}^{\nu \beta} {g}^{\delta \gamma} {g}^{\lambda \rho} {g}^{\alpha \epsilon} {g}^{\sigma \tau} + \frac{1}{4}\, {A}^{\nu \beta} {g}^{\delta \gamma} {g}^{\lambda \rho} {g}^{\alpha \sigma} {g}^{\epsilon \tau} - \frac{1}{4}\, {A}^{\nu \beta} {g}^{\delta \gamma} {g}^{\lambda \sigma} {g}^{\alpha \tau} {g}^{\rho \epsilon} + \frac{1}{4}\, {A}^{\nu \beta} {g}^{\delta \rho} {g}^{\lambda \sigma} {g}^{\alpha \gamma} {g}^{\epsilon \tau} + \frac{1}{4}\, {A}^{\nu \beta} {g}^{\delta \sigma} {g}^{\lambda \gamma} {g}^{\alpha \tau} {g}^{\rho \epsilon} - \frac{1}{4}\, {A}^{\nu \beta} {g}^{\delta \sigma} {g}^{\lambda \rho} {g}^{\alpha \gamma} {g}^{\epsilon \tau} + \frac{1}{4}\, {A}^{\nu \beta} {g}^{\delta \tau} {g}^{\lambda \sigma} {g}^{\alpha \gamma} {g}^{\rho \epsilon} - \frac{1}{8}\, {A}^{\nu \gamma} {g}^{\delta \beta} {g}^{\lambda \sigma} {g}^{\alpha \epsilon} {g}^{\rho \tau} + \frac{1}{8}\, {A}^{\nu \gamma} {g}^{\delta \epsilon} {g}^{\lambda \sigma} {g}^{\alpha \beta} {g}^{\rho \tau} - \frac{1}{8}\, {A}^{\nu \gamma} {g}^{\delta \sigma} {g}^{\lambda \epsilon} {g}^{\alpha \beta} {g}^{\rho \tau}\left.\right) + {\partial}_{\mu}{{g}_{\nu \rho}}\,  {\partial}_{\sigma}{{g}_{\gamma \epsilon}}\,  {\partial}_{\delta}{{g}_{\lambda \alpha}}\,  {\partial}_{\beta \tau}{{\xi}_{\kappa}}\,  \left(\right.\frac{1}{32}\, {A}^{\mu \beta} {g}^{\nu \gamma} {g}^{\rho \epsilon} {g}^{\sigma \delta} {g}^{\lambda \tau} {g}^{\alpha \kappa} - \frac{1}{16}\, {A}^{\mu \beta} {g}^{\nu \sigma} {g}^{\rho \gamma} {g}^{\epsilon \delta} {g}^{\lambda \tau} {g}^{\alpha \kappa} - \frac{1}{16}\, {A}^{\mu \gamma} {g}^{\nu \beta} {g}^{\rho \kappa} {g}^{\sigma \delta} {g}^{\epsilon \lambda} {g}^{\alpha \tau} + \frac{1}{16}\, {A}^{\mu \gamma} {g}^{\nu \beta} {g}^{\rho \kappa} {g}^{\sigma \lambda} {g}^{\epsilon \alpha} {g}^{\delta \tau} + \frac{1}{16}\, {A}^{\mu \gamma} {g}^{\nu \beta} {g}^{\rho \kappa} {g}^{\sigma \lambda} {g}^{\epsilon \delta} {g}^{\alpha \tau} + \frac{1}{16}\, {A}^{\mu \gamma} {g}^{\nu \beta} {g}^{\rho \tau} {g}^{\sigma \delta} {g}^{\epsilon \lambda} {g}^{\alpha \kappa} - \frac{1}{16}\, {A}^{\mu \gamma} {g}^{\nu \beta} {g}^{\rho \tau} {g}^{\sigma \lambda} {g}^{\epsilon \alpha} {g}^{\delta \kappa} - \frac{1}{16}\, {A}^{\mu \gamma} {g}^{\nu \beta} {g}^{\rho \tau} {g}^{\sigma \lambda} {g}^{\epsilon \delta} {g}^{\alpha \kappa} - \frac{1}{32}\, {A}^{\mu \kappa} {g}^{\nu \gamma} {g}^{\rho \epsilon} {g}^{\sigma \delta} {g}^{\lambda \beta} {g}^{\alpha \tau} + \frac{1}{16}\, {A}^{\mu \kappa} {g}^{\nu \sigma} {g}^{\rho \gamma} {g}^{\epsilon \delta} {g}^{\lambda \beta} {g}^{\alpha \tau} - \frac{1}{16}\, {A}^{\mu \sigma} {g}^{\nu \delta} {g}^{\rho \lambda} {g}^{\gamma \beta} {g}^{\epsilon \kappa} {g}^{\alpha \tau} + \frac{1}{16}\, {A}^{\mu \sigma} {g}^{\nu \delta} {g}^{\rho \lambda} {g}^{\gamma \beta} {g}^{\epsilon \tau} {g}^{\alpha \kappa} + \frac{1}{32}\, {A}^{\mu \sigma} {g}^{\nu \lambda} {g}^{\rho \alpha} {g}^{\gamma \beta} {g}^{\epsilon \kappa} {g}^{\delta \tau} - \frac{1}{32}\, {A}^{\mu \sigma} {g}^{\nu \lambda} {g}^{\rho \alpha} {g}^{\gamma \beta} {g}^{\epsilon \tau} {g}^{\delta \kappa} - \frac{1}{16}\, {A}^{\nu \beta} {g}^{\mu \gamma} {g}^{\rho \epsilon} {g}^{\sigma \delta} {g}^{\lambda \tau} {g}^{\alpha \kappa} - \frac{1}{16}\, {A}^{\nu \beta} {g}^{\mu \gamma} {g}^{\rho \sigma} {g}^{\epsilon \delta} {g}^{\lambda \tau} {g}^{\alpha \kappa} + \frac{1}{16}\, {A}^{\nu \beta} {g}^{\mu \sigma} {g}^{\rho \gamma} {g}^{\epsilon \delta} {g}^{\lambda \tau} {g}^{\alpha \kappa} + \frac{1}{16}\, {A}^{\nu \kappa} {g}^{\mu \gamma} {g}^{\rho \epsilon} {g}^{\sigma \delta} {g}^{\lambda \beta} {g}^{\alpha \tau} + \frac{1}{16}\, {A}^{\nu \kappa} {g}^{\mu \gamma} {g}^{\rho \sigma} {g}^{\epsilon \delta} {g}^{\lambda \beta} {g}^{\alpha \tau}%
 - \frac{1}{16}\, {A}^{\nu \kappa} {g}^{\mu \sigma} {g}^{\rho \gamma} {g}^{\epsilon \delta} {g}^{\lambda \beta} {g}^{\alpha \tau}\left.\right) + {g}_{\mu \nu} {\partial}_{\rho}{{g}_{\sigma \gamma}}\,  {\partial}_{\epsilon \delta}{{B}_{\lambda \alpha}}\,  {\partial}_{\beta \tau}{{\xi}^{\nu}}\,  \left(\right. - \frac{1}{4}\, {A}^{\epsilon \mu} {g}^{\rho \beta} {g}^{\sigma \delta} {g}^{\gamma \lambda} {g}^{\alpha \tau} + \frac{1}{4}\, {A}^{\epsilon \mu} {g}^{\rho \delta} {g}^{\sigma \lambda} {g}^{\gamma \beta} {g}^{\alpha \tau} + \frac{1}{4}\, {A}^{\epsilon \mu} {g}^{\rho \lambda} {g}^{\sigma \alpha} {g}^{\gamma \beta} {g}^{\delta \tau} - \frac{1}{4}\, {A}^{\epsilon \mu} {g}^{\rho \lambda} {g}^{\sigma \delta} {g}^{\gamma \beta} {g}^{\alpha \tau}\left.\right) + {g}_{\mu \nu} {\partial}_{\rho \sigma}{{g}_{\gamma \epsilon}}\,  {\partial}_{\delta}{{B}_{\lambda \alpha}}\,  {\partial}_{\beta \tau}{{\xi}^{\nu}}\,  \left(\right.\frac{1}{4}\, {A}^{\rho \mu} {g}^{\delta \beta} {g}^{\lambda \sigma} {g}^{\alpha \gamma} {g}^{\tau \epsilon} + \frac{1}{4}\, {A}^{\rho \mu} {g}^{\delta \gamma} {g}^{\lambda \beta} {g}^{\alpha \epsilon} {g}^{\tau \sigma} + \frac{1}{4}\, {A}^{\rho \mu} {g}^{\delta \gamma} {g}^{\lambda \beta} {g}^{\alpha \sigma} {g}^{\tau \epsilon} - \frac{1}{4}\, {A}^{\rho \mu} {g}^{\delta \sigma} {g}^{\lambda \beta} {g}^{\alpha \gamma} {g}^{\tau \epsilon}\left.\right) + {\partial}_{\mu}{{g}_{\nu \rho}}\,  {B}_{\sigma \gamma} {\partial}_{\epsilon}{{B}_{\delta \lambda}}\,  {\partial}_{\alpha}{{B}_{\beta \tau}}\,  {\partial}_{\kappa \pi}{{\xi}^{\gamma}}\,  \left(\right.\frac{1}{16}\, {A}^{\sigma \delta} {g}^{\epsilon \alpha} {g}^{\lambda \beta} {g}^{\tau \mu} {g}^{\nu \kappa} {g}^{\rho \pi} - \frac{1}{16}\, {A}^{\sigma \delta} {g}^{\epsilon \beta} {g}^{\lambda \alpha} {g}^{\tau \mu} {g}^{\nu \kappa} {g}^{\rho \pi} + \frac{1}{16}\, {A}^{\sigma \delta} {g}^{\epsilon \beta} {g}^{\lambda \tau} {g}^{\alpha \mu} {g}^{\nu \kappa} {g}^{\rho \pi} - \frac{1}{16}\, {A}^{\sigma \epsilon} {g}^{\delta \alpha} {g}^{\lambda \beta} {g}^{\tau \mu} {g}^{\nu \kappa} {g}^{\rho \pi} - \frac{1}{32}\, {A}^{\sigma \epsilon} {g}^{\delta \beta} {g}^{\lambda \tau} {g}^{\alpha \mu} {g}^{\nu \kappa} {g}^{\rho \pi}\left.\right) + {\partial}_{\mu}{{g}_{\nu \rho}}\,  {B}_{\sigma \gamma} {\partial}_{\epsilon}{{B}_{\delta \lambda}}\,  {\partial}_{\alpha}{{B}_{\beta \tau}}\,  {\partial}_{\kappa \pi}{{\xi}^{\sigma}}\,  \left(\right. - \frac{1}{16}\, {A}^{\delta \kappa} {g}^{\gamma \nu} {g}^{\epsilon \alpha} {g}^{\lambda \beta} {g}^{\tau \mu} {g}^{\rho \pi} + \frac{1}{16}\, {A}^{\delta \kappa} {g}^{\gamma \nu} {g}^{\epsilon \beta} {g}^{\lambda \alpha} {g}^{\tau \mu} {g}^{\rho \pi} - \frac{1}{16}\, {A}^{\delta \kappa} {g}^{\gamma \nu} {g}^{\epsilon \beta} {g}^{\lambda \tau} {g}^{\alpha \mu} {g}^{\rho \pi} + \frac{1}{16}\, {A}^{\delta \mu} {g}^{\gamma \alpha} {g}^{\epsilon \beta} {g}^{\lambda \tau} {g}^{\nu \kappa} {g}^{\rho \pi} - \frac{1}{16}\, {A}^{\delta \mu} {g}^{\gamma \beta} {g}^{\epsilon \alpha} {g}^{\lambda \tau} {g}^{\nu \kappa} {g}^{\rho \pi} + \frac{1}{16}\, {A}^{\delta \mu} {g}^{\gamma \beta} {g}^{\epsilon \tau} {g}^{\lambda \alpha} {g}^{\nu \kappa} {g}^{\rho \pi} - \frac{1}{16}\, {A}^{\delta \mu} {g}^{\gamma \nu} {g}^{\epsilon \alpha} {g}^{\lambda \beta} {g}^{\tau \kappa} {g}^{\rho \pi} + \frac{1}{16}\, {A}^{\delta \mu} {g}^{\gamma \nu} {g}^{\epsilon \beta} {g}^{\lambda \alpha} {g}^{\tau \kappa} {g}^{\rho \pi} - \frac{1}{16}\, {A}^{\delta \mu} {g}^{\gamma \nu} {g}^{\epsilon \beta} {g}^{\lambda \tau} {g}^{\alpha \kappa} {g}^{\rho \pi} + \frac{1}{16}\, {A}^{\epsilon \kappa} {g}^{\gamma \nu} {g}^{\delta \alpha} {g}^{\lambda \beta} {g}^{\tau \mu} {g}^{\rho \pi} + \frac{1}{32}\, {A}^{\epsilon \kappa} {g}^{\gamma \nu} {g}^{\delta \beta} {g}^{\lambda \tau} {g}^{\alpha \mu} {g}^{\rho \pi} - \frac{1}{32}\, {A}^{\epsilon \mu} {g}^{\gamma \alpha} {g}^{\delta \beta} {g}^{\lambda \tau} {g}^{\nu \kappa} {g}^{\rho \pi} + \frac{1}{16}\, {A}^{\epsilon \mu} {g}^{\gamma \beta} {g}^{\delta \alpha} {g}^{\lambda \tau} {g}^{\nu \kappa} {g}^{\rho \pi} + \frac{1}{16}\, {A}^{\epsilon \mu} {g}^{\gamma \nu} {g}^{\delta \alpha} {g}^{\lambda \beta} {g}^{\tau \kappa} {g}^{\rho \pi} + \frac{1}{32}\, {A}^{\epsilon \mu} {g}^{\gamma \nu} {g}^{\delta \beta} {g}^{\lambda \tau} {g}^{\alpha \kappa} {g}^{\rho \pi}\left.\right)  + {\partial}_{\mu}{{g}_{\nu \rho}}\,  {\partial}_{\sigma}{{g}_{\gamma \epsilon}}\,  {\partial}_{\delta}{{g}_{\lambda \alpha}}\,  {B}_{\beta \tau} {\partial}_{\kappa \pi}{{\xi}^{\tau}}\,  \left(\right. - \frac{1}{32}\, {A}^{\beta \mu} {g}^{\nu \gamma} {g}^{\rho \epsilon} {g}^{\sigma \delta} {g}^{\lambda \kappa} {g}^{\alpha \pi} + \frac{1}{16}\, {A}^{\beta \mu} {g}^{\nu \sigma} {g}^{\rho \gamma} {g}^{\epsilon \delta} {g}^{\lambda \kappa} {g}^{\alpha \pi} + \frac{1}{16}\, {A}^{\beta \nu} {g}^{\mu \gamma} {g}^{\rho \epsilon} {g}^{\sigma \delta} {g}^{\lambda \kappa} {g}^{\alpha \pi} + \frac{1}{16}\, {A}^{\beta \nu} {g}^{\mu \gamma} {g}^{\rho \sigma} {g}^{\epsilon \delta} {g}^{\lambda \kappa} {g}^{\alpha \pi} - \frac{1}{16}\, {A}^{\beta \nu} {g}^{\mu \sigma} {g}^{\rho \gamma} {g}^{\epsilon \delta} {g}^{\lambda \kappa} {g}^{\alpha \pi}\left.\right) + {\partial}_{\mu}{{g}_{\nu \rho}}\,  {\partial}_{\sigma}{{g}_{\gamma \epsilon}}\,  {\partial}_{\delta}{{g}_{\lambda \alpha}}\,  {B}_{\beta \tau} {\partial}_{\kappa \pi}{{\xi}^{\beta}}\,  \left(\right. - \frac{1}{16}\, {A}^{\mu \gamma} {g}^{\tau \delta} {g}^{\nu \kappa} {g}^{\rho \pi} {g}^{\sigma \lambda} {g}^{\epsilon \alpha} - \frac{1}{16}\, {A}^{\mu \gamma} {g}^{\tau \lambda} {g}^{\nu \kappa} {g}^{\rho \pi} {g}^{\sigma \alpha} {g}^{\epsilon \delta} + \frac{1}{16}\, {A}^{\mu \gamma} {g}^{\tau \lambda} {g}^{\nu \kappa} {g}^{\rho \pi} {g}^{\sigma \delta} {g}^{\epsilon \alpha} - \frac{1}{16}\, {A}^{\mu \gamma} {g}^{\tau \nu} {g}^{\rho \kappa} {g}^{\sigma \delta} {g}^{\epsilon \lambda} {g}^{\alpha \pi} + \frac{1}{16}\, {A}^{\mu \gamma} {g}^{\tau \nu} {g}^{\rho \kappa} {g}^{\sigma \lambda} {g}^{\epsilon \alpha} {g}^{\delta \pi} + \frac{1}{16}\, {A}^{\mu \gamma} {g}^{\tau \nu} {g}^{\rho \kappa} {g}^{\sigma \lambda} {g}^{\epsilon \delta} {g}^{\alpha \pi} - \frac{1}{16}\, {A}^{\mu \kappa} {g}^{\tau \gamma} {g}^{\nu \delta} {g}^{\rho \lambda} {g}^{\sigma \alpha} {g}^{\epsilon \pi} + \frac{1}{32}\, {A}^{\mu \kappa} {g}^{\tau \gamma} {g}^{\nu \lambda} {g}^{\rho \alpha} {g}^{\sigma \delta} {g}^{\epsilon \pi} - \frac{1}{32}\, {A}^{\mu \sigma} {g}^{\tau \delta} {g}^{\nu \lambda} {g}^{\rho \alpha} {g}^{\gamma \kappa} {g}^{\epsilon \pi} + \frac{1}{16}\, {A}^{\mu \sigma} {g}^{\tau \lambda} {g}^{\nu \delta} {g}^{\rho \alpha} {g}^{\gamma \kappa} {g}^{\epsilon \pi} + \frac{1}{16}\, {A}^{\mu \sigma} {g}^{\tau \nu} {g}^{\rho \kappa} {g}^{\gamma \delta} {g}^{\epsilon \lambda} {g}^{\alpha \pi} - \frac{1}{32}\, {A}^{\mu \sigma} {g}^{\tau \nu} {g}^{\rho \kappa} {g}^{\gamma \lambda} {g}^{\epsilon \alpha} {g}^{\delta \pi} + \frac{1}{16}\, {A}^{\nu \kappa} {g}^{\tau \gamma} {g}^{\mu \delta} {g}^{\rho \lambda} {g}^{\sigma \alpha} {g}^{\epsilon \pi} - \frac{1}{16}\, {A}^{\nu \kappa} {g}^{\tau \gamma} {g}^{\mu \lambda} {g}^{\rho \alpha} {g}^{\sigma \delta} {g}^{\epsilon \pi} - \frac{1}{16}\, {A}^{\nu \kappa} {g}^{\tau \gamma} {g}^{\mu \lambda} {g}^{\rho \delta} {g}^{\sigma \alpha} {g}^{\epsilon \pi}\left.\right) + {g}_{\mu \nu} {\partial}_{\rho}{{g}_{\sigma \gamma}}\,  {\partial}_{\epsilon}{{g}_{\delta \lambda}}\,  {\partial}_{\alpha}{{B}_{\beta \tau}}\,  {\partial}_{\kappa \pi}{{\xi}^{\nu}}\,  \left(\right.\frac{1}{16}\, {A}^{\alpha \mu} {g}^{\beta \rho} {g}^{\tau \sigma} {g}^{\gamma \epsilon} {g}^{\delta \kappa} {g}^{\lambda \pi} - \frac{1}{16}\, {A}^{\beta \mu} {g}^{\alpha \rho} {g}^{\tau \sigma} {g}^{\gamma \epsilon} {g}^{\delta \kappa} {g}^{\lambda \pi} + \frac{1}{16}\, {A}^{\beta \mu} {g}^{\alpha \sigma} {g}^{\tau \gamma} {g}^{\rho \epsilon} {g}^{\delta \kappa} {g}^{\lambda \pi} + \frac{1}{16}\, {A}^{\beta \mu} {g}^{\alpha \sigma} {g}^{\tau \rho} {g}^{\gamma \epsilon} {g}^{\delta \kappa} {g}^{\lambda \pi} + \frac{1}{4}\, {A}^{\rho \mu} {g}^{\alpha \delta} {g}^{\beta \epsilon} {g}^{\tau \kappa} {g}^{\sigma \lambda} {g}^{\gamma \pi} + \frac{1}{4}\, {A}^{\rho \mu} {g}^{\alpha \delta} {g}^{\beta \lambda} {g}^{\tau \kappa} {g}^{\sigma \epsilon} {g}^{\gamma \pi} - \frac{1}{4}\, {A}^{\rho \mu} {g}^{\alpha \delta} {g}^{\beta \sigma} {g}^{\tau \epsilon} {g}^{\gamma \kappa} {g}^{\lambda \pi} + \frac{1}{4}\, {A}^{\rho \mu} {g}^{\alpha \delta} {g}^{\beta \sigma} {g}^{\tau \kappa} {g}^{\gamma \epsilon} {g}^{\lambda \pi} + \frac{1}{4}\, {A}^{\rho \mu} {g}^{\alpha \delta} {g}^{\beta \sigma} {g}^{\tau \kappa} {g}^{\gamma \lambda} {g}^{\epsilon \pi} - \frac{1}{4}\, {A}^{\rho \mu} {g}^{\alpha \delta} {g}^{\beta \sigma} {g}^{\tau \lambda} {g}^{\gamma \kappa} {g}^{\epsilon \pi} - \frac{1}{4}\, {A}^{\rho \mu} {g}^{\alpha \epsilon} {g}^{\beta \delta} {g}^{\tau \kappa} {g}^{\sigma \lambda} {g}^{\gamma \pi} + \frac{1}{4}\, {A}^{\rho \mu} {g}^{\alpha \epsilon} {g}^{\beta \sigma} {g}^{\tau \delta} {g}^{\gamma \kappa} {g}^{\lambda \pi} - \frac{1}{4}\, {A}^{\rho \mu} {g}^{\alpha \epsilon} {g}^{\beta \sigma} {g}^{\tau \kappa} {g}^{\gamma \delta} {g}^{\lambda \pi} - \frac{1}{4}\, {A}^{\rho \mu} {g}^{\alpha \kappa} {g}^{\beta \epsilon} {g}^{\tau \delta} {g}^{\sigma \lambda} {g}^{\gamma \pi}
\end{dmath*}
\begin{dmath*}
- \frac{1}{4}\, {A}^{\rho \mu} {g}^{\alpha \kappa} {g}^{\beta \sigma} {g}^{\tau \delta} {g}^{\gamma \epsilon} {g}^{\lambda \pi} + \frac{1}{4}\, {A}^{\rho \mu} {g}^{\alpha \kappa} {g}^{\beta \sigma} {g}^{\tau \epsilon} {g}^{\gamma \delta} {g}^{\lambda \pi} - \frac{1}{4}\, {A}^{\rho \mu} {g}^{\alpha \sigma} {g}^{\beta \delta} {g}^{\tau \kappa} {g}^{\gamma \epsilon} {g}^{\lambda \pi} + \frac{1}{4}\, {A}^{\rho \mu} {g}^{\alpha \sigma} {g}^{\beta \delta} {g}^{\tau \kappa} {g}^{\gamma \lambda} {g}^{\epsilon \pi} - \frac{1}{4}\, {A}^{\rho \mu} {g}^{\alpha \sigma} {g}^{\beta \epsilon} {g}^{\tau \delta} {g}^{\gamma \kappa} {g}^{\lambda \pi}%
 + \frac{1}{4}\, {A}^{\rho \mu} {g}^{\alpha \sigma} {g}^{\beta \epsilon} {g}^{\tau \kappa} {g}^{\gamma \delta} {g}^{\lambda \pi} - \frac{1}{16}\, {A}^{\rho \mu} {g}^{\alpha \sigma} {g}^{\beta \gamma} {g}^{\tau \epsilon} {g}^{\delta \kappa} {g}^{\lambda \pi} + \frac{1}{16}\, {A}^{\sigma \mu} {g}^{\alpha \epsilon} {g}^{\beta \rho} {g}^{\tau \gamma} {g}^{\delta \kappa} {g}^{\lambda \pi} - \frac{1}{16}\, {A}^{\sigma \mu} {g}^{\alpha \gamma} {g}^{\beta \rho} {g}^{\tau \epsilon} {g}^{\delta \kappa} {g}^{\lambda \pi} + \frac{1}{16}\, {A}^{\sigma \mu} {g}^{\alpha \rho} {g}^{\beta \gamma} {g}^{\tau \epsilon} {g}^{\delta \kappa} {g}^{\lambda \pi}\left.\right)
\end{dmath*}
\end{dgroup*}
\endgroup

The second order field redefinition for the duality covariant metric $\Delta_2 \widehat g_{\mu \nu}$ in (\ref{DeltaG}) also has a shorter expression when contracted with some symmetric tensor $S^{\mu \nu}$
\be
\Delta_2  g = S^{\mu \nu} \Delta_2  g_{\mu \nu} \ . \label{Delta2tildeGa}
\ee
The result is
\begingroup\makeatletter\def\f@size{7}\check@mathfonts
\begin{dgroup*}
\begin{dmath*}
\Delta_2  g = {\partial}_{\mu \nu}{{B}_{\rho \sigma}}\,  {\partial}_{\gamma \epsilon}{{B}_{\delta \lambda}}\,  \left(\right. - \frac{1}{4}\, {S}^{\mu \delta} {g}^{\nu \gamma} {g}^{\rho \epsilon} {g}^{\sigma \lambda} + \frac{1}{16}\, {S}^{\mu \gamma} {g}^{\nu \epsilon} {g}^{\rho \delta} {g}^{\sigma \lambda} + \frac{1}{8}\, {S}^{\rho \delta} {g}^{\mu \gamma} {g}^{\nu \epsilon} {g}^{\sigma \lambda} - \frac{1}{8}\, {S}^{\rho \delta} {g}^{\mu \gamma} {g}^{\nu \lambda} {g}^{\sigma \epsilon}\left.\right) + {\partial}_{\mu \nu}{{g}_{\rho \sigma}}\,  {\partial}_{\gamma \epsilon}{{g}_{\delta \lambda}}\,  \left(\right. - \frac{1}{4}\, {S}^{\mu \delta} {g}^{\nu \gamma} {g}^{\rho \epsilon} {g}^{\sigma \lambda} + \frac{1}{4}\, {S}^{\mu \delta} {g}^{\nu \lambda} {g}^{\rho \gamma} {g}^{\sigma \epsilon} + \frac{1}{16}\, {S}^{\mu \gamma} {g}^{\nu \epsilon} {g}^{\rho \delta} {g}^{\sigma \lambda} + \frac{1}{8}\, {S}^{\rho \delta} {g}^{\mu \gamma} {g}^{\nu \epsilon} {g}^{\sigma \lambda} - \frac{1}{8}\, {S}^{\rho \delta} {g}^{\mu \gamma} {g}^{\nu \lambda} {g}^{\sigma \epsilon}\left.\right) + {\partial}_{\mu}{{g}_{\nu \rho}}\,  {\partial}_{\sigma}{{B}_{\gamma \epsilon}}\,  {\partial}_{\delta \lambda}{{B}_{\alpha \beta}}\,  \left(\right. - \frac{1}{8}\, {S}^{\gamma \alpha} {g}^{\sigma \beta} {g}^{\epsilon \mu} {g}^{\nu \delta} {g}^{\rho \lambda} + \frac{1}{4}\, {S}^{\gamma \alpha} {g}^{\sigma \beta} {g}^{\epsilon \nu} {g}^{\mu \delta} {g}^{\rho \lambda} + \frac{1}{8}\, {S}^{\gamma \alpha} {g}^{\sigma \delta} {g}^{\epsilon \mu} {g}^{\nu \lambda} {g}^{\rho \beta} - \frac{1}{8}\, {S}^{\gamma \alpha} {g}^{\sigma \delta} {g}^{\epsilon \nu} {g}^{\mu \beta} {g}^{\rho \lambda} - \frac{1}{8}\, {S}^{\gamma \alpha} {g}^{\sigma \delta} {g}^{\epsilon \nu} {g}^{\mu \lambda} {g}^{\rho \beta} + \frac{1}{8}\, {S}^{\gamma \alpha} {g}^{\sigma \mu} {g}^{\epsilon \beta} {g}^{\nu \delta} {g}^{\rho \lambda} - \frac{1}{8}\, {S}^{\gamma \alpha} {g}^{\sigma \mu} {g}^{\epsilon \delta} {g}^{\nu \lambda} {g}^{\rho \beta} - \frac{1}{4}\, {S}^{\gamma \alpha} {g}^{\sigma \nu} {g}^{\epsilon \beta} {g}^{\mu \delta} {g}^{\rho \lambda} + \frac{1}{8}\, {S}^{\gamma \alpha} {g}^{\sigma \nu} {g}^{\epsilon \delta} {g}^{\mu \beta} {g}^{\rho \lambda} + \frac{1}{8}\, {S}^{\gamma \alpha} {g}^{\sigma \nu} {g}^{\epsilon \delta} {g}^{\mu \lambda} {g}^{\rho \beta} - \frac{1}{8}\, {S}^{\gamma \delta} {g}^{\sigma \alpha} {g}^{\epsilon \mu} {g}^{\nu \lambda} {g}^{\rho \beta} + \frac{1}{8}\, {S}^{\gamma \delta} {g}^{\sigma \alpha} {g}^{\epsilon \nu} {g}^{\mu \beta} {g}^{\rho \lambda} + \frac{1}{8}\, {S}^{\gamma \delta} {g}^{\sigma \alpha} {g}^{\epsilon \nu} {g}^{\mu \lambda} {g}^{\rho \beta} + \frac{1}{8}\, {S}^{\gamma \delta} {g}^{\sigma \mu} {g}^{\epsilon \alpha} {g}^{\nu \lambda} {g}^{\rho \beta} - \frac{1}{8}\, {S}^{\gamma \delta} {g}^{\sigma \nu} {g}^{\epsilon \alpha} {g}^{\mu \beta} {g}^{\rho \lambda} - \frac{1}{8}\, {S}^{\gamma \delta} {g}^{\sigma \nu} {g}^{\epsilon \alpha} {g}^{\mu \lambda} {g}^{\rho \beta} - \frac{1}{8}\, {S}^{\gamma \mu} {g}^{\sigma \alpha} {g}^{\epsilon \beta} {g}^{\nu \delta} {g}^{\rho \lambda} + \frac{1}{8}\, {S}^{\gamma \mu} {g}^{\sigma \alpha} {g}^{\epsilon \delta} {g}^{\nu \lambda} {g}^{\rho \beta} - \frac{1}{8}\, {S}^{\gamma \mu} {g}^{\sigma \delta} {g}^{\epsilon \alpha} {g}^{\nu \lambda} {g}^{\rho \beta}  + \frac{1}{8}\, {S}^{\gamma \nu} {g}^{\sigma \alpha} {g}^{\epsilon \delta} {g}^{\mu \beta} {g}^{\rho \lambda} - \frac{1}{8}\, {S}^{\gamma \nu} {g}^{\sigma \alpha} {g}^{\epsilon \delta} {g}^{\mu \lambda} {g}^{\rho \beta} - \frac{1}{8}\, {S}^{\gamma \nu} {g}^{\sigma \delta} {g}^{\epsilon \alpha} {g}^{\mu \beta} {g}^{\rho \lambda} + \frac{1}{8}\, {S}^{\gamma \nu} {g}^{\sigma \delta} {g}^{\epsilon \alpha} {g}^{\mu \lambda} {g}^{\rho \beta} + \frac{1}{8}\, {S}^{\mu \delta} {g}^{\sigma \alpha} {g}^{\gamma \nu} {g}^{\epsilon \beta} {g}^{\rho \lambda} - \frac{1}{8}\, {S}^{\mu \delta} {g}^{\sigma \alpha} {g}^{\gamma \nu} {g}^{\epsilon \lambda} {g}^{\rho \beta} + \frac{1}{8}\, {S}^{\mu \delta} {g}^{\sigma \lambda} {g}^{\gamma \nu} {g}^{\epsilon \alpha} {g}^{\rho \beta} - \frac{1}{16}\, {S}^{\mu \delta} {g}^{\sigma \nu} {g}^{\gamma \alpha} {g}^{\epsilon \beta} {g}^{\rho \lambda} - \frac{1}{8}\, {S}^{\mu \delta} {g}^{\sigma \nu} {g}^{\gamma \lambda} {g}^{\epsilon \alpha} {g}^{\rho \beta} - \frac{1}{4}\, {S}^{\nu \alpha} {g}^{\sigma \beta} {g}^{\gamma \mu} {g}^{\epsilon \delta} {g}^{\rho \lambda} + \frac{1}{4}\, {S}^{\nu \alpha} {g}^{\sigma \beta} {g}^{\gamma \rho} {g}^{\epsilon \delta} {g}^{\mu \lambda} + \frac{1}{4}\, {S}^{\nu \alpha} {g}^{\sigma \delta} {g}^{\gamma \mu} {g}^{\epsilon \beta} {g}^{\rho \lambda} - \frac{1}{4}\, {S}^{\nu \alpha} {g}^{\sigma \delta} {g}^{\gamma \rho} {g}^{\epsilon \beta} {g}^{\mu \lambda} - \frac{1}{4}\, {S}^{\nu \alpha} {g}^{\sigma \mu} {g}^{\gamma \delta} {g}^{\epsilon \beta} {g}^{\rho \lambda} + \frac{1}{4}\, {S}^{\nu \alpha} {g}^{\sigma \rho} {g}^{\gamma \delta} {g}^{\epsilon \beta} {g}^{\mu \lambda} - \frac{1}{4}\, {S}^{\nu \delta} {g}^{\sigma \alpha} {g}^{\gamma \mu} {g}^{\epsilon \beta} {g}^{\rho \lambda} + \frac{1}{4}\, {S}^{\nu \delta} {g}^{\sigma \alpha} {g}^{\gamma \rho} {g}^{\epsilon \beta} {g}^{\mu \lambda} + \frac{1}{8}\, {S}^{\nu \delta} {g}^{\sigma \mu} {g}^{\gamma \alpha} {g}^{\epsilon \beta} {g}^{\rho \lambda} - \frac{1}{8}\, {S}^{\nu \delta} {g}^{\sigma \rho} {g}^{\gamma \alpha} {g}^{\epsilon \beta} {g}^{\mu \lambda} - \frac{1}{8}\, {S}^{\sigma \alpha} {g}^{\gamma \mu} {g}^{\epsilon \beta} {g}^{\nu \delta} {g}^{\rho \lambda}
 \end{dmath*}

\begin{dmath*}
{ - \frac{1}{8}\, {S}^{\lambda \mu} {g}^{\delta \gamma} {g}^{\alpha \tau} {g}^{\beta \nu} {g}^{\kappa \epsilon} {g}^{\rho \sigma} + \frac{1}{8}\, {S}^{\lambda \mu} {g}^{\delta \gamma} {g}^{\alpha \tau} {g}^{\beta \nu} {g}^{\kappa \sigma} {g}^{\rho \epsilon} - \frac{1}{8}\, {S}^{\lambda \mu} {g}^{\delta \gamma} {g}^{\alpha \tau} {g}^{\beta \sigma} {g}^{\kappa \nu} {g}^{\rho \epsilon} - \frac{1}{16}\, {S}^{\lambda \mu} {g}^{\delta \sigma} {g}^{\alpha \gamma} {g}^{\beta \nu} {g}^{\tau \rho} {g}^{\kappa \epsilon} - \frac{1}{16}\, {S}^{\lambda \mu} {g}^{\delta \tau} {g}^{\alpha \beta} {g}^{\kappa \gamma} {g}^{\nu \sigma} {g}^{\rho \epsilon} \ \ \ \ \ \ } + \frac{1}{32}\, {S}^{\lambda \mu} {g}^{\delta \tau} {g}^{\alpha \beta} {g}^{\kappa \sigma} {g}^{\nu \gamma} {g}^{\rho \epsilon} - \frac{1}{8}\, {S}^{\lambda \mu} {g}^{\delta \tau} {g}^{\alpha \gamma} {g}^{\beta \epsilon} {g}^{\kappa \nu} {g}^{\rho \sigma} + \frac{1}{8}\, {S}^{\lambda \mu} {g}^{\delta \tau} {g}^{\alpha \gamma} {g}^{\beta \nu} {g}^{\kappa \epsilon} {g}^{\rho \sigma} - \frac{1}{8}\, {S}^{\lambda \mu} {g}^{\delta \tau} {g}^{\alpha \gamma} {g}^{\beta \nu} {g}^{\kappa \sigma} {g}^{\rho \epsilon} + \frac{1}{8}\, {S}^{\lambda \mu} {g}^{\delta \tau} {g}^{\alpha \gamma} {g}^{\beta \sigma} {g}^{\kappa \nu} {g}^{\rho \epsilon} + \frac{1}{16}\, {S}^{\lambda \mu} {g}^{\delta \tau} {g}^{\alpha \kappa} {g}^{\beta \gamma} {g}^{\nu \sigma} {g}^{\rho \epsilon} - \frac{1}{32}\, {S}^{\lambda \mu} {g}^{\delta \tau} {g}^{\alpha \kappa} {g}^{\beta \sigma} {g}^{\nu \gamma} {g}^{\rho \epsilon} - \frac{1}{4}\, {S}^{\lambda \nu} {g}^{\delta \beta} {g}^{\alpha \gamma} {g}^{\tau \mu} {g}^{\kappa \epsilon} {g}^{\rho \sigma} + \frac{1}{4}\, {S}^{\lambda \nu} {g}^{\delta \beta} {g}^{\alpha \gamma} {g}^{\tau \rho} {g}^{\kappa \epsilon} {g}^{\mu \sigma} + \frac{1}{8}\, {S}^{\lambda \nu} {g}^{\delta \beta} {g}^{\alpha \sigma} {g}^{\tau \mu} {g}^{\kappa \gamma} {g}^{\rho \epsilon} - \frac{1}{8}\, {S}^{\lambda \nu} {g}^{\delta \beta} {g}^{\alpha \sigma} {g}^{\tau \rho} {g}^{\kappa \gamma} {g}^{\mu \epsilon} + \frac{1}{16}\, {S}^{\lambda \nu} {g}^{\delta \beta} {g}^{\alpha \tau} {g}^{\kappa \gamma} {g}^{\mu \epsilon} {g}^{\rho \sigma} - \frac{1}{16}\, {S}^{\lambda \nu} {g}^{\delta \beta} {g}^{\alpha \tau} {g}^{\kappa \gamma} {g}^{\mu \sigma} {g}^{\rho \epsilon} + \frac{1}{16}\, {S}^{\lambda \nu} {g}^{\delta \beta} {g}^{\alpha \tau} {g}^{\kappa \sigma} {g}^{\mu \gamma} {g}^{\rho \epsilon} + \frac{1}{4}\, {S}^{\lambda \nu} {g}^{\delta \gamma} {g}^{\alpha \beta} {g}^{\tau \mu} {g}^{\kappa \epsilon} {g}^{\rho \sigma} - \frac{1}{4}\, {S}^{\lambda \nu} {g}^{\delta \gamma} {g}^{\alpha \beta} {g}^{\tau \rho} {g}^{\kappa \epsilon} {g}^{\mu \sigma} - \frac{1}{16}\, {S}^{\lambda \nu} {g}^{\delta \gamma} {g}^{\alpha \epsilon} {g}^{\beta \mu} {g}^{\tau \rho} {g}^{\kappa \sigma} + \frac{1}{16}\, {S}^{\lambda \nu} {g}^{\delta \gamma} {g}^{\alpha \epsilon} {g}^{\beta \rho} {g}^{\tau \mu} {g}^{\kappa \sigma} - \frac{1}{16}\, {S}^{\lambda \nu} {g}^{\delta \gamma} {g}^{\alpha \epsilon} {g}^{\beta \sigma} {g}^{\tau \mu} {g}^{\kappa \rho} - \frac{1}{16}\, {S}^{\lambda \nu} {g}^{\delta \gamma} {g}^{\alpha \sigma} {g}^{\beta \epsilon} {g}^{\tau \mu} {g}^{\kappa \rho} - \frac{1}{16}\, {S}^{\lambda \nu} {g}^{\delta \gamma} {g}^{\alpha \sigma} {g}^{\beta \mu} {g}^{\tau \rho} {g}^{\kappa \epsilon} + \frac{1}{16}\, {S}^{\lambda \nu} {g}^{\delta \gamma} {g}^{\alpha \sigma} {g}^{\beta \rho} {g}^{\tau \mu} {g}^{\kappa \epsilon} + \frac{1}{4}\, {S}^{\lambda \nu} {g}^{\delta \gamma} {g}^{\alpha \tau} {g}^{\beta \epsilon} {g}^{\kappa \mu} {g}^{\rho \sigma} - \frac{1}{4}\, {S}^{\lambda \nu} {g}^{\delta \gamma} {g}^{\alpha \tau} {g}^{\beta \epsilon} {g}^{\kappa \rho} {g}^{\mu \sigma} - \frac{1}{4}\, {S}^{\lambda \nu} {g}^{\delta \gamma} {g}^{\alpha \tau} {g}^{\beta \mu} {g}^{\kappa \epsilon} {g}^{\rho \sigma}%
 \end{dmath*}
\begin{dmath*}
{ + \frac{1}{4}\, {S}^{\lambda \nu} {g}^{\delta \gamma} {g}^{\alpha \tau} {g}^{\beta \rho} {g}^{\kappa \epsilon} {g}^{\mu \sigma} - \frac{1}{8}\, {S}^{\lambda \nu} {g}^{\delta \sigma} {g}^{\alpha \beta} {g}^{\tau \mu} {g}^{\kappa \gamma} {g}^{\rho \epsilon} + \frac{1}{8}\, {S}^{\lambda \nu} {g}^{\delta \sigma} {g}^{\alpha \beta} {g}^{\tau \rho} {g}^{\kappa \gamma} {g}^{\mu \epsilon} + \frac{1}{16}\, {S}^{\lambda \nu} {g}^{\delta \sigma} {g}^{\alpha \gamma} {g}^{\beta \epsilon} {g}^{\tau \mu} {g}^{\kappa \rho} + \frac{1}{16}\, {S}^{\lambda \nu} {g}^{\delta \sigma} {g}^{\alpha \gamma} {g}^{\beta \mu} {g}^{\tau \rho} {g}^{\kappa \epsilon} \ \ \ \ \ \ \ } - \frac{1}{16}\, {S}^{\lambda \nu} {g}^{\delta \sigma} {g}^{\alpha \gamma} {g}^{\beta \rho} {g}^{\tau \mu} {g}^{\kappa \epsilon} - \frac{1}{8}\, {S}^{\lambda \nu} {g}^{\delta \sigma} {g}^{\alpha \tau} {g}^{\beta \gamma} {g}^{\kappa \mu} {g}^{\rho \epsilon} + \frac{1}{8}\, {S}^{\lambda \nu} {g}^{\delta \sigma} {g}^{\alpha \tau} {g}^{\beta \gamma} {g}^{\kappa \rho} {g}^{\mu \epsilon} + \frac{1}{8}\, {S}^{\lambda \nu} {g}^{\delta \sigma} {g}^{\alpha \tau} {g}^{\beta \mu} {g}^{\kappa \gamma} {g}^{\rho \epsilon} - \frac{1}{8}\, {S}^{\lambda \nu} {g}^{\delta \sigma} {g}^{\alpha \tau} {g}^{\beta \rho} {g}^{\kappa \gamma} {g}^{\mu \epsilon} - \frac{1}{16}\, {S}^{\lambda \nu} {g}^{\delta \tau} {g}^{\alpha \beta} {g}^{\kappa \gamma} {g}^{\mu \epsilon} {g}^{\rho \sigma} + \frac{1}{16}\, {S}^{\lambda \nu} {g}^{\delta \tau} {g}^{\alpha \beta} {g}^{\kappa \gamma} {g}^{\mu \sigma} {g}^{\rho \epsilon} - \frac{1}{16}\, {S}^{\lambda \nu} {g}^{\delta \tau} {g}^{\alpha \beta} {g}^{\kappa \sigma} {g}^{\mu \gamma} {g}^{\rho \epsilon}
- \frac{1}{4}\, {S}^{\lambda \nu} {g}^{\delta \tau} {g}^{\alpha \gamma} {g}^{\beta \epsilon} {g}^{\kappa \mu} {g}^{\rho \sigma} + \frac{1}{4}\, {S}^{\lambda \nu} {g}^{\delta \tau} {g}^{\alpha \gamma} {g}^{\beta \epsilon} {g}^{\kappa \rho} {g}^{\mu \sigma} + \frac{1}{4}\, {S}^{\lambda \nu} {g}^{\delta \tau} {g}^{\alpha \gamma} {g}^{\beta \mu} {g}^{\kappa \epsilon} {g}^{\rho \sigma} - \frac{1}{4}\, {S}^{\lambda \nu} {g}^{\delta \tau} {g}^{\alpha \gamma} {g}^{\beta \rho} {g}^{\kappa \epsilon} {g}^{\mu \sigma} + \frac{1}{16}\, {S}^{\lambda \nu} {g}^{\delta \tau} {g}^{\alpha \kappa} {g}^{\beta \gamma} {g}^{\mu \epsilon} {g}^{\rho \sigma} - \frac{1}{16}\, {S}^{\lambda \nu} {g}^{\delta \tau} {g}^{\alpha \kappa} {g}^{\beta \gamma} {g}^{\mu \sigma} {g}^{\rho \epsilon} + \frac{1}{16}\, {S}^{\lambda \nu} {g}^{\delta \tau} {g}^{\alpha \kappa} {g}^{\beta \sigma} {g}^{\mu \gamma} {g}^{\rho \epsilon} + \frac{1}{8}\, {S}^{\lambda \nu} {g}^{\delta \tau} {g}^{\alpha \sigma} {g}^{\beta \gamma} {g}^{\kappa \mu} {g}^{\rho \epsilon} - \frac{1}{8}\, {S}^{\lambda \nu} {g}^{\delta \tau} {g}^{\alpha \sigma} {g}^{\beta \gamma} {g}^{\kappa \rho} {g}^{\mu \epsilon} - \frac{1}{8}\, {S}^{\lambda \nu} {g}^{\delta \tau} {g}^{\alpha \sigma} {g}^{\beta \mu} {g}^{\kappa \gamma} {g}^{\rho \epsilon} + \frac{1}{8}\, {S}^{\lambda \nu} {g}^{\delta \tau} {g}^{\alpha \sigma} {g}^{\beta \rho} {g}^{\kappa \gamma} {g}^{\mu \epsilon} - \frac{1}{8}\, {S}^{\lambda \tau} {g}^{\delta \beta} {g}^{\alpha \mu} {g}^{\kappa \gamma} {g}^{\nu \sigma} {g}^{\rho \epsilon} + \frac{1}{32}\, {S}^{\lambda \tau} {g}^{\delta \beta} {g}^{\alpha \mu} {g}^{\kappa \sigma} {g}^{\nu \gamma} {g}^{\rho \epsilon} + \frac{1}{16}\, {S}^{\lambda \tau} {g}^{\delta \beta} {g}^{\alpha \nu} {g}^{\kappa \gamma} {g}^{\mu \epsilon} {g}^{\rho \sigma} + \frac{1}{16}\, {S}^{\lambda \tau} {g}^{\delta \beta} {g}^{\alpha \nu} {g}^{\kappa \gamma} {g}^{\mu \sigma} {g}^{\rho \epsilon} + \frac{1}{8}\, {S}^{\lambda \tau} {g}^{\delta \kappa} {g}^{\alpha \mu} {g}^{\beta \gamma} {g}^{\nu \sigma} {g}^{\rho \epsilon} - \frac{1}{16}\, {S}^{\lambda \tau} {g}^{\delta \kappa} {g}^{\alpha \mu} {g}^{\beta \sigma} {g}^{\nu \gamma} {g}^{\rho \epsilon} - \frac{1}{8}\, {S}^{\lambda \tau} {g}^{\delta \kappa} {g}^{\alpha \nu} {g}^{\beta \gamma} {g}^{\mu \epsilon} {g}^{\rho \sigma} - \frac{1}{8}\, {S}^{\lambda \tau} {g}^{\delta \kappa} {g}^{\alpha \nu} {g}^{\beta \gamma} {g}^{\mu \sigma} {g}^{\rho \epsilon} + \frac{1}{8}\, {S}^{\lambda \tau} {g}^{\delta \kappa} {g}^{\alpha \nu} {g}^{\beta \sigma} {g}^{\mu \gamma} {g}^{\rho \epsilon} + \frac{1}{8}\, {S}^{\lambda \tau} {g}^{\delta \mu} {g}^{\alpha \gamma} {g}^{\beta \epsilon} {g}^{\kappa \nu} {g}^{\rho \sigma} - \frac{1}{8}\, {S}^{\lambda \tau} {g}^{\delta \mu} {g}^{\alpha \gamma} {g}^{\beta \nu} {g}^{\kappa \epsilon} {g}^{\rho \sigma} + \frac{1}{4}\, {S}^{\lambda \tau} {g}^{\delta \mu} {g}^{\alpha \gamma} {g}^{\beta \nu} {g}^{\kappa \sigma} {g}^{\rho \epsilon} - \frac{1}{8}\, {S}^{\lambda \tau} {g}^{\delta \mu} {g}^{\alpha \gamma} {g}^{\beta \sigma} {g}^{\kappa \nu} {g}^{\rho \epsilon} - \frac{1}{8}\, {S}^{\lambda \tau} {g}^{\delta \mu} {g}^{\alpha \kappa} {g}^{\beta \gamma} {g}^{\nu \sigma} {g}^{\rho \epsilon} + \frac{1}{32}\, {S}^{\lambda \tau} {g}^{\delta \mu} {g}^{\alpha \kappa} {g}^{\beta \sigma} {g}^{\nu \gamma} {g}^{\rho \epsilon} + \frac{1}{16}\, {S}^{\lambda \tau} {g}^{\delta \mu} {g}^{\alpha \nu} {g}^{\beta \gamma} {g}^{\kappa \epsilon} {g}^{\rho \sigma}  + \frac{1}{16}\, {S}^{\lambda \tau} {g}^{\delta \mu} {g}^{\alpha \nu} {g}^{\beta \gamma} {g}^{\kappa \sigma} {g}^{\rho \epsilon} - \frac{1}{32}\, {S}^{\lambda \tau} {g}^{\delta \mu} {g}^{\alpha \nu} {g}^{\beta \sigma} {g}^{\kappa \gamma} {g}^{\rho \epsilon} - \frac{1}{8}\, {S}^{\lambda \tau} {g}^{\delta \nu} {g}^{\alpha \gamma} {g}^{\beta \epsilon} {g}^{\kappa \rho} {g}^{\mu \sigma} + \frac{1}{8}\, {S}^{\lambda \tau} {g}^{\delta \nu} {g}^{\alpha \gamma} {g}^{\beta \rho} {g}^{\kappa \epsilon} {g}^{\mu \sigma} + \frac{1}{16}\, {S}^{\lambda \tau} {g}^{\delta \nu} {g}^{\alpha \kappa} {g}^{\beta \gamma} {g}^{\mu \epsilon} {g}^{\rho \sigma} + \frac{1}{16}\, {S}^{\lambda \tau} {g}^{\delta \nu} {g}^{\alpha \kappa} {g}^{\beta \gamma} {g}^{\mu \sigma} {g}^{\rho \epsilon} - \frac{1}{16}\, {S}^{\lambda \tau} {g}^{\delta \nu} {g}^{\alpha \mu} {g}^{\beta \gamma} {g}^{\kappa \epsilon} {g}^{\rho \sigma} - \frac{1}{32}\, {S}^{\lambda \tau} {g}^{\delta \nu} {g}^{\alpha \mu} {g}^{\beta \gamma} {g}^{\kappa \sigma} {g}^{\rho \epsilon} - \frac{1}{32}\, {S}^{\lambda \tau} {g}^{\delta \nu} {g}^{\alpha \rho} {g}^{\beta \gamma} {g}^{\kappa \epsilon} {g}^{\mu \sigma} - \frac{1}{8}\, {S}^{\lambda \tau} {g}^{\delta \nu} {g}^{\alpha \sigma} {g}^{\beta \gamma} {g}^{\kappa \mu} {g}^{\rho \epsilon} + \frac{1}{8}\, {S}^{\lambda \tau} {g}^{\delta \nu} {g}^{\alpha \sigma} {g}^{\beta \gamma} {g}^{\kappa \rho} {g}^{\mu \epsilon} - \frac{1}{8}\, {S}^{\lambda \tau} {g}^{\delta \nu} {g}^{\alpha \sigma} {g}^{\beta \rho} {g}^{\kappa \gamma} {g}^{\mu \epsilon} + \frac{1}{8}\, {S}^{\mu \gamma} {g}^{\delta \beta} {g}^{\lambda \tau} {g}^{\alpha \nu} {g}^{\kappa \epsilon} {g}^{\rho \sigma} - \frac{1}{8}\, {S}^{\mu \gamma} {g}^{\delta \beta} {g}^{\lambda \tau} {g}^{\alpha \nu} {g}^{\kappa \sigma} {g}^{\rho \epsilon} + \frac{1}{16}\, {S}^{\mu \gamma} {g}^{\delta \nu} {g}^{\lambda \tau} {g}^{\alpha \kappa} {g}^{\beta \epsilon} {g}^{\rho \sigma} - \frac{1}{16}\, {S}^{\mu \gamma} {g}^{\delta \nu} {g}^{\lambda \tau} {g}^{\alpha \kappa} {g}^{\beta \sigma} {g}^{\rho \epsilon} - \frac{1}{16}\, {S}^{\mu \gamma} {g}^{\delta \nu} {g}^{\lambda \tau} {g}^{\alpha \rho} {g}^{\beta \epsilon} {g}^{\kappa \sigma} + \frac{1}{16}\, {S}^{\mu \gamma} {g}^{\delta \nu} {g}^{\lambda \tau} {g}^{\alpha \rho} {g}^{\beta \sigma} {g}^{\kappa \epsilon} - \frac{1}{8}\, {S}^{\mu \gamma} {g}^{\delta \tau} {g}^{\lambda \beta} {g}^{\alpha \nu} {g}^{\kappa \epsilon} {g}^{\rho \sigma} + \frac{1}{8}\, {S}^{\mu \gamma} {g}^{\delta \tau} {g}^{\lambda \beta} {g}^{\alpha \nu} {g}^{\kappa \sigma} {g}^{\rho \epsilon}  + \frac{1}{8}\, {S}^{\mu \gamma} {g}^{\delta \tau} {g}^{\lambda \kappa} {g}^{\alpha \epsilon} {g}^{\beta \nu} {g}^{\rho \sigma} + \frac{1}{8}\, {S}^{\mu \gamma} {g}^{\delta \tau} {g}^{\lambda \kappa} {g}^{\alpha \nu} {g}^{\beta \epsilon} {g}^{\rho \sigma} - \frac{1}{8}\, {S}^{\mu \gamma} {g}^{\delta \tau} {g}^{\lambda \kappa} {g}^{\alpha \nu} {g}^{\beta \sigma} {g}^{\rho \epsilon} - \frac{1}{8}\, {S}^{\mu \gamma} {g}^{\delta \tau} {g}^{\lambda \kappa} {g}^{\alpha \sigma} {g}^{\beta \nu} {g}^{\rho \epsilon} - \frac{1}{16}\, {S}^{\mu \gamma} {g}^{\delta \tau} {g}^{\lambda \sigma} {g}^{\alpha \epsilon} {g}^{\beta \nu} {g}^{\kappa \rho} + \frac{1}{16}\, {S}^{\mu \sigma} {g}^{\delta \beta} {g}^{\lambda \nu} {g}^{\alpha \gamma} {g}^{\tau \rho} {g}^{\kappa \epsilon} + \frac{1}{16}\, {S}^{\mu \sigma} {g}^{\delta \beta} {g}^{\lambda \tau} {g}^{\alpha \nu} {g}^{\kappa \gamma} {g}^{\rho \epsilon} - \frac{1}{8}\, {S}^{\mu \sigma} {g}^{\delta \nu} {g}^{\lambda \tau} {g}^{\alpha \gamma} {g}^{\beta \epsilon} {g}^{\kappa \rho} + \frac{1}{8}\, {S}^{\mu \sigma} {g}^{\delta \nu} {g}^{\lambda \tau} {g}^{\alpha \gamma} {g}^{\beta \rho} {g}^{\kappa \epsilon} + \frac{1}{32}\, {S}^{\mu \sigma} {g}^{\delta \nu} {g}^{\lambda \tau} {g}^{\alpha \kappa} {g}^{\beta \gamma} {g}^{\rho \epsilon} - \frac{1}{32}\, {S}^{\mu \sigma} {g}^{\delta \nu} {g}^{\lambda \tau} {g}^{\alpha \rho} {g}^{\beta \gamma} {g}^{\kappa \epsilon} - \frac{1}{16}\, {S}^{\mu \sigma} {g}^{\delta \tau} {g}^{\lambda \beta} {g}^{\alpha \nu} {g}^{\kappa \gamma} {g}^{\rho \epsilon} + \frac{1}{8}\, {S}^{\mu \sigma} {g}^{\delta \tau} {g}^{\lambda \kappa} {g}^{\alpha \nu} {g}^{\beta \gamma} {g}^{\rho \epsilon} - \frac{1}{4}\, {S}^{\mu \sigma} {g}^{\delta \tau} {g}^{\lambda \nu} {g}^{\alpha \gamma} {g}^{\beta \rho} {g}^{\kappa \epsilon} - \frac{1}{32}\, {S}^{\nu \gamma} {g}^{\delta \beta} {g}^{\lambda \mu} {g}^{\alpha \rho} {g}^{\tau \sigma} {g}^{\kappa \epsilon} - \frac{1}{8}\, {S}^{\nu \gamma} {g}^{\delta \beta} {g}^{\lambda \tau} {g}^{\alpha \mu} {g}^{\kappa \epsilon} {g}^{\rho \sigma} + \frac{1}{16}\, {S}^{\nu \gamma} {g}^{\delta \beta} {g}^{\lambda \tau} {g}^{\alpha \mu} {g}^{\kappa \sigma} {g}^{\rho \epsilon} + \frac{1}{16}\, {S}^{\nu \gamma} {g}^{\delta \beta} {g}^{\lambda \tau} {g}^{\alpha \rho} {g}^{\kappa \epsilon} {g}^{\mu \sigma} - \frac{1}{16}\, {S}^{\nu \gamma} {g}^{\delta \mu} {g}^{\lambda \tau} {g}^{\alpha \kappa} {g}^{\beta \epsilon} {g}^{\rho \sigma} + \frac{1}{32}\, {S}^{\nu \gamma} {g}^{\delta \mu} {g}^{\lambda \tau} {g}^{\alpha \kappa} {g}^{\beta \sigma} {g}^{\rho \epsilon}  + \frac{1}{16}\, {S}^{\nu \gamma} {g}^{\delta \mu} {g}^{\lambda \tau} {g}^{\alpha \rho} {g}^{\beta \epsilon} {g}^{\kappa \sigma} - \frac{1}{32}\, {S}^{\nu \gamma} {g}^{\delta \mu} {g}^{\lambda \tau} {g}^{\alpha \rho} {g}^{\beta \sigma} {g}^{\kappa \epsilon} + \frac{1}{32}\, {S}^{\nu \gamma} {g}^{\delta \rho} {g}^{\lambda \tau} {g}^{\alpha \kappa} {g}^{\beta \epsilon} {g}^{\mu \sigma} - \frac{1}{32}\, {S}^{\nu \gamma} {g}^{\delta \rho} {g}^{\lambda \tau} {g}^{\alpha \mu} {g}^{\beta \epsilon} {g}^{\kappa \sigma} + \frac{1}{8}\, {S}^{\nu \gamma} {g}^{\delta \tau} {g}^{\lambda \beta} {g}^{\alpha \mu} {g}^{\kappa \epsilon} {g}^{\rho \sigma} - \frac{1}{16}\, {S}^{\nu \gamma} {g}^{\delta \tau} {g}^{\lambda \beta} {g}^{\alpha \mu} {g}^{\kappa \sigma} {g}^{\rho \epsilon} - \frac{1}{16}\, {S}^{\nu \gamma} {g}^{\delta \tau} {g}^{\lambda \beta} {g}^{\alpha \rho} {g}^{\kappa \epsilon} {g}^{\mu \sigma} - \frac{1}{8}\, {S}^{\nu \gamma} {g}^{\delta \tau} {g}^{\lambda \kappa} {g}^{\alpha \mu} {g}^{\beta \epsilon} {g}^{\rho \sigma} + \frac{1}{8}\, {S}^{\nu \gamma} {g}^{\delta \tau} {g}^{\lambda \kappa} {g}^{\alpha \mu} {g}^{\beta \sigma} {g}^{\rho \epsilon} + \frac{1}{8}\, {S}^{\nu \gamma} {g}^{\delta \tau} {g}^{\lambda \kappa} {g}^{\alpha \rho} {g}^{\beta \epsilon} {g}^{\mu \sigma} - \frac{1}{8}\, {S}^{\nu \gamma} {g}^{\delta \tau} {g}^{\lambda \kappa} {g}^{\alpha \rho} {g}^{\beta \sigma} {g}^{\mu \epsilon} - \frac{1}{16}\, {S}^{\nu \gamma} {g}^{\delta \tau} {g}^{\lambda \mu} {g}^{\alpha \rho} {g}^{\beta \epsilon} {g}^{\kappa \sigma} + \frac{1}{16}\, {S}^{\nu \gamma} {g}^{\delta \tau} {g}^{\lambda \mu} {g}^{\alpha \rho} {g}^{\beta \sigma} {g}^{\kappa \epsilon}\left.\right) + {\partial}_{\mu}{{g}_{\nu \rho}}\,  {\partial}_{\sigma}{{g}_{\gamma \epsilon}}\,  {\partial}_{\delta}{{g}_{\lambda \alpha}}\,  {\partial}_{\beta}{{g}_{\tau \kappa}}\,  \left(\right. - \frac{1}{16}\, {S}^{\mu \gamma} {g}^{\nu \delta} {g}^{\rho \lambda} {g}^{\sigma \beta} {g}^{\epsilon \tau} {g}^{\alpha \kappa} + \frac{1}{16}\, {S}^{\mu \gamma} {g}^{\nu \delta} {g}^{\rho \lambda} {g}^{\sigma \tau} {g}^{\epsilon \beta} {g}^{\alpha \kappa} + \frac{1}{16}\, {S}^{\mu \gamma} {g}^{\nu \delta} {g}^{\rho \lambda} {g}^{\sigma \tau} {g}^{\epsilon \kappa} {g}^{\alpha \beta} - \frac{1}{16}\, {S}^{\mu \gamma} {g}^{\nu \epsilon} {g}^{\rho \delta} {g}^{\sigma \beta} {g}^{\lambda \tau} {g}^{\alpha \kappa} + \frac{1}{8}\, {S}^{\mu \gamma} {g}^{\nu \epsilon} {g}^{\rho \delta} {g}^{\sigma \tau} {g}^{\lambda \beta} {g}^{\alpha \kappa} + \frac{1}{8}\, {S}^{\mu \gamma} {g}^{\nu \epsilon} {g}^{\rho \lambda} {g}^{\sigma \beta} {g}^{\delta \tau} {g}^{\alpha \kappa} - \frac{1}{8}\, {S}^{\mu \gamma} {g}^{\nu \epsilon} {g}^{\rho \lambda} {g}^{\sigma \tau} {g}^{\delta \beta} {g}^{\alpha \kappa} - \frac{1}{8}\, {S}^{\mu \gamma} {g}^{\nu \epsilon} {g}^{\rho \lambda} {g}^{\sigma \tau} {g}^{\delta \kappa} {g}^{\alpha \beta} + \frac{1}{32}\, {S}^{\mu \gamma} {g}^{\nu \lambda} {g}^{\rho \alpha} {g}^{\sigma \beta} {g}^{\epsilon \tau} {g}^{\delta \kappa} - \frac{1}{32}\, {S}^{\mu \gamma} {g}^{\nu \lambda} {g}^{\rho \alpha} {g}^{\sigma \tau} {g}^{\epsilon \beta} {g}^{\delta \kappa} - \frac{1}{32}\, {S}^{\mu \gamma} {g}^{\nu \lambda} {g}^{\rho \alpha} {g}^{\sigma \tau} {g}^{\epsilon \kappa} {g}^{\delta \beta} + \frac{1}{4}\, {S}^{\mu \gamma} {g}^{\nu \lambda} {g}^{\rho \tau} {g}^{\sigma \alpha} {g}^{\epsilon \beta} {g}^{\delta \kappa} - \frac{1}{4}\, {S}^{\mu \gamma} {g}^{\nu \lambda} {g}^{\rho \tau} {g}^{\sigma \delta} {g}^{\epsilon \kappa} {g}^{\alpha \beta} + \frac{1}{16}\, {S}^{\mu \gamma} {g}^{\nu \sigma} {g}^{\rho \delta} {g}^{\epsilon \beta} {g}^{\lambda \tau} {g}^{\alpha \kappa} - \frac{1}{8}\, {S}^{\mu \gamma} {g}^{\nu \sigma} {g}^{\rho \delta} {g}^{\epsilon \tau} {g}^{\lambda \beta} {g}^{\alpha \kappa} - \frac{1}{8}\, {S}^{\mu \gamma} {g}^{\nu \sigma} {g}^{\rho \lambda} {g}^{\epsilon \beta} {g}^{\delta \tau} {g}^{\alpha \kappa} + \frac{1}{8}\, {S}^{\mu \gamma} {g}^{\nu \sigma} {g}^{\rho \lambda} {g}^{\epsilon \tau} {g}^{\delta \beta} {g}^{\alpha \kappa} + \frac{1}{8}\, {S}^{\mu \gamma} {g}^{\nu \sigma} {g}^{\rho \lambda} {g}^{\epsilon \tau} {g}^{\delta \kappa} {g}^{\alpha \beta} + \frac{1}{32}\, {S}^{\mu \sigma} {g}^{\nu \delta} {g}^{\rho \lambda} {g}^{\gamma \beta} {g}^{\epsilon \tau} {g}^{\alpha \kappa} - \frac{1}{32}\, {S}^{\mu \sigma} {g}^{\nu \delta} {g}^{\rho \lambda} {g}^{\gamma \tau} {g}^{\epsilon \kappa} {g}^{\alpha \beta} + \frac{1}{8}\, {S}^{\mu \sigma} {g}^{\nu \delta} {g}^{\rho \tau} {g}^{\gamma \lambda} {g}^{\epsilon \kappa} {g}^{\alpha \beta} + \frac{1}{32}\, {S}^{\mu \sigma} {g}^{\nu \gamma} {g}^{\rho \delta} {g}^{\epsilon \beta} {g}^{\lambda \tau} {g}^{\alpha \kappa} - \frac{1}{8}\, {S}^{\mu \sigma} {g}^{\nu \gamma} {g}^{\rho \delta} {g}^{\epsilon \tau} {g}^{\lambda \beta} {g}^{\alpha \kappa} + \frac{1}{16}\, {S}^{\mu \sigma} {g}^{\nu \gamma} {g}^{\rho \lambda} {g}^{\epsilon \tau} {g}^{\delta \beta} {g}^{\alpha \kappa} + \frac{1}{16}\, {S}^{\mu \sigma} {g}^{\nu \gamma} {g}^{\rho \lambda} {g}^{\epsilon \tau} {g}^{\delta \kappa} {g}^{\alpha \beta} + \frac{1}{128}\, {S}^{\mu \sigma} {g}^{\nu \lambda} {g}^{\rho \alpha} {g}^{\gamma \tau} {g}^{\epsilon \kappa} {g}^{\delta \beta} + \frac{1}{32}\, {S}^{\nu \gamma} {g}^{\mu \delta} {g}^{\rho \epsilon} {g}^{\sigma \beta} {g}^{\lambda \tau} {g}^{\alpha \kappa} - \frac{1}{8}\, {S}^{\nu \gamma} {g}^{\mu \delta} {g}^{\rho \epsilon} {g}^{\sigma \tau} {g}^{\lambda \beta} {g}^{\alpha \kappa} + \frac{1}{32}\, {S}^{\nu \gamma} {g}^{\mu \delta} {g}^{\rho \lambda} {g}^{\sigma \beta} {g}^{\epsilon \tau} {g}^{\alpha \kappa} - \frac{1}{16}\, {S}^{\nu \gamma} {g}^{\mu \delta} {g}^{\rho \lambda} {g}^{\sigma \tau} {g}^{\epsilon \beta} {g}^{\alpha \kappa} - \frac{1}{16}\, {S}^{\nu \gamma} {g}^{\mu \delta} {g}^{\rho \lambda} {g}^{\sigma \tau} {g}^{\epsilon \kappa} {g}^{\alpha \beta} - \frac{1}{16}\, {S}^{\nu \gamma} {g}^{\mu \epsilon} {g}^{\rho \delta} {g}^{\sigma \beta} {g}^{\lambda \tau} {g}^{\alpha \kappa} + \frac{1}{8}\, {S}^{\nu \gamma} {g}^{\mu \epsilon} {g}^{\rho \delta} {g}^{\sigma \tau} {g}^{\lambda \beta} {g}^{\alpha \kappa} + \frac{1}{8}\, {S}^{\nu \gamma} {g}^{\mu \epsilon} {g}^{\rho \lambda} {g}^{\sigma \beta} {g}^{\delta \tau} {g}^{\alpha \kappa} - \frac{1}{8}\, {S}^{\nu \gamma} {g}^{\mu \epsilon} {g}^{\rho \lambda} {g}^{\sigma \tau} {g}^{\delta \beta} {g}^{\alpha \kappa} - \frac{1}{8}\, {S}^{\nu \gamma} {g}^{\mu \epsilon} {g}^{\rho \lambda} {g}^{\sigma \tau} {g}^{\delta \kappa} {g}^{\alpha \beta} + \frac{1}{32}\, {S}^{\nu \gamma} {g}^{\mu \lambda} {g}^{\rho \alpha} {g}^{\sigma \tau} {g}^{\epsilon \kappa} {g}^{\delta \beta} + \frac{1}{32}\, {S}^{\nu \gamma} {g}^{\mu \lambda} {g}^{\rho \delta} {g}^{\sigma \tau} {g}^{\epsilon \beta} {g}^{\alpha \kappa} + \frac{1}{16}\, {S}^{\nu \gamma} {g}^{\mu \lambda} {g}^{\rho \delta} {g}^{\sigma \tau} {g}^{\epsilon \kappa} {g}^{\alpha \beta} + \frac{1}{16}\, {S}^{\nu \gamma} {g}^{\mu \lambda} {g}^{\rho \epsilon} {g}^{\sigma \tau} {g}^{\delta \beta} {g}^{\alpha \kappa} + \frac{1}{16}\, {S}^{\nu \gamma} {g}^{\mu \lambda} {g}^{\rho \epsilon} {g}^{\sigma \tau} {g}^{\delta \kappa} {g}^{\alpha \beta} + \frac{1}{8}\, {S}^{\nu \gamma} {g}^{\mu \lambda} {g}^{\rho \tau} {g}^{\sigma \alpha} {g}^{\epsilon \kappa} {g}^{\delta \beta} - \frac{1}{8}\, {S}^{\nu \gamma} {g}^{\mu \lambda} {g}^{\rho \tau} {g}^{\sigma \kappa} {g}^{\epsilon \alpha} {g}^{\delta \beta} + \frac{1}{32}\, {S}^{\nu \gamma} {g}^{\mu \sigma} {g}^{\rho \delta} {g}^{\epsilon \beta} {g}^{\lambda \tau} {g}^{\alpha \kappa} - \frac{1}{8}\, {S}^{\nu \gamma} {g}^{\mu \sigma} {g}^{\rho \delta} {g}^{\epsilon \tau} {g}^{\lambda \beta} {g}^{\alpha \kappa} + \frac{1}{16}\, {S}^{\nu \gamma} {g}^{\mu \sigma} {g}^{\rho \lambda} {g}^{\epsilon \tau} {g}^{\delta \beta} {g}^{\alpha \kappa} + \frac{1}{16}\, {S}^{\nu \gamma} {g}^{\mu \sigma} {g}^{\rho \lambda} {g}^{\epsilon \tau} {g}^{\delta \kappa} {g}^{\alpha \beta}\left.\right)
\end{dmath*}
\end{dgroup*}
\endgroup

The second order field redefinition for the duality covariant two-form $\Delta_2 \widehat B_{\mu \nu}$ in (\ref{DeltaB}) also has a shorter expression when contracted with some antisymmetric tensor $A^{\mu \nu}$
\be
\Delta_2  B = A^{\mu \nu} \Delta_2  B_{\mu \nu}\ . \label{Delta2tildeBa}
\ee
The result is
\begingroup\makeatletter\def\f@size{7}\check@mathfonts
\begin{dgroup*}
\begin{dmath*}
\Delta_2  B= {\partial}_{\mu}{{g}_{\nu \rho}}\,  {\partial}_{\sigma}{{B}_{\gamma \epsilon}}\,  {\partial}_{\delta}{{B}_{\lambda \alpha}}\,  {\partial}_{\beta}{{B}_{\tau \kappa}}\,  \left(\right.\frac{1}{16}\, {A}^{\gamma \lambda} {g}^{\sigma \beta} {g}^{\epsilon \tau} {g}^{\delta \mu} {g}^{\alpha \nu} {g}^{\kappa \rho} - \frac{1}{16}\, {A}^{\gamma \lambda} {g}^{\sigma \beta} {g}^{\epsilon \tau} {g}^{\delta \nu} {g}^{\alpha \mu} {g}^{\kappa \rho} - \frac{1}{16}\, {A}^{\gamma \lambda} {g}^{\sigma \beta} {g}^{\epsilon \tau} {g}^{\delta \nu} {g}^{\alpha \rho} {g}^{\kappa \mu} - \frac{1}{16}\, {A}^{\gamma \lambda} {g}^{\sigma \tau} {g}^{\epsilon \beta} {g}^{\delta \mu} {g}^{\alpha \nu} {g}^{\kappa \rho} + \frac{1}{16}\, {A}^{\gamma \lambda} {g}^{\sigma \tau} {g}^{\epsilon \beta} {g}^{\delta \nu} {g}^{\alpha \mu} {g}^{\kappa \rho} + \frac{1}{16}\, {A}^{\gamma \lambda} {g}^{\sigma \tau} {g}^{\epsilon \beta} {g}^{\delta \nu} {g}^{\alpha \rho} {g}^{\kappa \mu} + \frac{1}{16}\, {A}^{\gamma \lambda} {g}^{\sigma \tau} {g}^{\epsilon \kappa} {g}^{\delta \mu} {g}^{\alpha \nu} {g}^{\beta \rho} - \frac{1}{16}\, {A}^{\gamma \lambda} {g}^{\sigma \tau} {g}^{\epsilon \kappa} {g}^{\delta \nu} {g}^{\alpha \mu} {g}^{\beta \rho} - \frac{1}{16}\, {A}^{\gamma \lambda} {g}^{\sigma \tau} {g}^{\epsilon \kappa} {g}^{\delta \nu} {g}^{\alpha \rho} {g}^{\beta \mu} - \frac{1}{16}\, {A}^{\gamma \mu} {g}^{\sigma \delta} {g}^{\epsilon \lambda} {g}^{\alpha \tau} {g}^{\beta \nu} {g}^{\kappa \rho} - \frac{1}{16}\, {A}^{\gamma \mu} {g}^{\sigma \lambda} {g}^{\epsilon \alpha} {g}^{\delta \tau} {g}^{\beta \nu} {g}^{\kappa \rho} + \frac{1}{16}\, {A}^{\gamma \mu} {g}^{\sigma \lambda} {g}^{\epsilon \delta} {g}^{\alpha \tau} {g}^{\beta \nu} {g}^{\kappa \rho} - \frac{1}{16}\, {A}^{\gamma \nu} {g}^{\sigma \delta} {g}^{\epsilon \lambda} {g}^{\alpha \beta} {g}^{\tau \mu} {g}^{\kappa \rho} + \frac{1}{16}\, {A}^{\gamma \nu} {g}^{\sigma \delta} {g}^{\epsilon \lambda} {g}^{\alpha \tau} {g}^{\beta \mu} {g}^{\kappa \rho} - \frac{1}{16}\, {A}^{\gamma \nu} {g}^{\sigma \delta} {g}^{\epsilon \lambda} {g}^{\alpha \tau} {g}^{\beta \rho} {g}^{\kappa \mu} - \frac{1}{16}\, {A}^{\gamma \nu} {g}^{\sigma \lambda} {g}^{\epsilon \alpha} {g}^{\delta \beta} {g}^{\tau \mu} {g}^{\kappa \rho} + \frac{1}{16}\, {A}^{\gamma \nu} {g}^{\sigma \lambda} {g}^{\epsilon \alpha} {g}^{\delta \tau} {g}^{\beta \mu} {g}^{\kappa \rho} - \frac{1}{16}\, {A}^{\gamma \nu} {g}^{\sigma \lambda} {g}^{\epsilon \alpha} {g}^{\delta \tau} {g}^{\beta \rho} {g}^{\kappa \mu} + \frac{1}{16}\, {A}^{\gamma \nu} {g}^{\sigma \lambda} {g}^{\epsilon \delta} {g}^{\alpha \beta} {g}^{\tau \mu} {g}^{\kappa \rho} - \frac{1}{16}\, {A}^{\gamma \nu} {g}^{\sigma \lambda} {g}^{\epsilon \delta} {g}^{\alpha \tau} {g}^{\beta \mu} {g}^{\kappa \rho} + \frac{1}{16}\, {A}^{\gamma \nu} {g}^{\sigma \lambda} {g}^{\epsilon \delta} {g}^{\alpha \tau} {g}^{\beta \rho} {g}^{\kappa \mu} + \frac{1}{16}\, {A}^{\sigma \delta} {g}^{\gamma \beta} {g}^{\epsilon \tau} {g}^{\lambda \mu} {g}^{\alpha \nu} {g}^{\kappa \rho} + \frac{1}{32}\, {A}^{\sigma \delta} {g}^{\gamma \tau} {g}^{\epsilon \kappa} {g}^{\lambda \mu} {g}^{\alpha \nu} {g}^{\beta \rho} - \frac{1}{16}\, {A}^{\sigma \lambda} {g}^{\gamma \beta} {g}^{\epsilon \tau} {g}^{\delta \mu} {g}^{\alpha \nu} {g}^{\kappa \rho} + \frac{1}{16}\, {A}^{\sigma \lambda} {g}^{\gamma \beta} {g}^{\epsilon \tau} {g}^{\delta \nu} {g}^{\alpha \mu} {g}^{\kappa \rho} + \frac{1}{16}\, {A}^{\sigma \lambda} {g}^{\gamma \beta} {g}^{\epsilon \tau} {g}^{\delta \nu} {g}^{\alpha \rho} {g}^{\kappa \mu} + \frac{1}{16}\, {A}^{\sigma \lambda} {g}^{\gamma \mu} {g}^{\epsilon \nu} {g}^{\delta \beta} {g}^{\alpha \tau} {g}^{\kappa \rho} - \frac{1}{16}\, {A}^{\sigma \lambda} {g}^{\gamma \mu} {g}^{\epsilon \nu} {g}^{\delta \tau} {g}^{\alpha \beta} {g}^{\kappa \rho} + \frac{1}{16}\, {A}^{\sigma \lambda} {g}^{\gamma \mu} {g}^{\epsilon \nu} {g}^{\delta \tau} {g}^{\alpha \kappa} {g}^{\beta \rho} - \frac{1}{32}\, {A}^{\sigma \lambda} {g}^{\gamma \tau} {g}^{\epsilon \kappa} {g}^{\delta \mu} {g}^{\alpha \nu} {g}^{\beta \rho} + \frac{1}{32}\, {A}^{\sigma \lambda} {g}^{\gamma \tau} {g}^{\epsilon \kappa} {g}^{\delta \nu} {g}^{\alpha \mu} {g}^{\beta \rho} + \frac{1}{32}\, {A}^{\sigma \lambda} {g}^{\gamma \tau} {g}^{\epsilon \kappa} {g}^{\delta \nu} {g}^{\alpha \rho} {g}^{\beta \mu} + \frac{1}{16}\, {A}^{\sigma \mu} {g}^{\gamma \delta} {g}^{\epsilon \lambda} {g}^{\alpha \tau} {g}^{\beta \nu} {g}^{\kappa \rho} + \frac{1}{32}\, {A}^{\sigma \mu} {g}^{\gamma \lambda} {g}^{\epsilon \alpha} {g}^{\delta \tau} {g}^{\beta \nu} {g}^{\kappa \rho} + \frac{1}{16}\, {A}^{\sigma \nu} {g}^{\gamma \delta} {g}^{\epsilon \lambda} {g}^{\alpha \beta} {g}^{\tau \mu} {g}^{\kappa \rho} - \frac{1}{16}\, {A}^{\sigma \nu} {g}^{\gamma \delta} {g}^{\epsilon \lambda} {g}^{\alpha \tau} {g}^{\beta \mu} {g}^{\kappa \rho} + \frac{1}{16}\, {A}^{\sigma \nu} {g}^{\gamma \delta} {g}^{\epsilon \lambda} {g}^{\alpha \tau} {g}^{\beta \rho} {g}^{\kappa \mu} + \frac{1}{32}\, {A}^{\sigma \nu} {g}^{\gamma \lambda} {g}^{\epsilon \alpha} {g}^{\delta \beta} {g}^{\tau \mu} {g}^{\kappa \rho} - \frac{1}{32}\, {A}^{\sigma \nu} {g}^{\gamma \lambda} {g}^{\epsilon \alpha} {g}^{\delta \tau} {g}^{\beta \mu} {g}^{\kappa \rho}  + \frac{1}{32}\, {A}^{\sigma \nu} {g}^{\gamma \lambda} {g}^{\epsilon \alpha} {g}^{\delta \tau} {g}^{\beta \rho} {g}^{\kappa \mu}\left.\right) + {\partial}_{\mu}{{g}_{\nu \rho}}\,  {\partial}_{\sigma}{{g}_{\gamma \epsilon}}\,  {\partial}_{\delta}{{g}_{\lambda \alpha}}\,  {\partial}_{\beta}{{B}_{\tau \kappa}}\,  \left(\right.\frac{1}{16}\, {A}^{\beta \mu} {g}^{\tau \sigma} {g}^{\kappa \gamma} {g}^{\nu \delta} {g}^{\rho \lambda} {g}^{\epsilon \alpha} - \frac{1}{32}\, {A}^{\beta \mu} {g}^{\tau \sigma} {g}^{\kappa \gamma} {g}^{\nu \lambda} {g}^{\rho \alpha} {g}^{\epsilon \delta} - \frac{1}{16}\, {A}^{\beta \nu} {g}^{\tau \sigma} {g}^{\kappa \gamma} {g}^{\mu \delta} {g}^{\rho \lambda} {g}^{\epsilon \alpha} + \frac{1}{16}\, {A}^{\beta \nu} {g}^{\tau \sigma} {g}^{\kappa \gamma} {g}^{\mu \lambda} {g}^{\rho \alpha} {g}^{\epsilon \delta} + \frac{1}{16}\, {A}^{\beta \nu} {g}^{\tau \sigma} {g}^{\kappa \gamma} {g}^{\mu \lambda} {g}^{\rho \delta} {g}^{\epsilon \alpha} + \frac{1}{32}\, {A}^{\mu \gamma} {g}^{\beta \delta} {g}^{\tau \sigma} {g}^{\kappa \epsilon} {g}^{\nu \lambda} {g}^{\rho \alpha} - \frac{1}{32}\, {A}^{\mu \gamma} {g}^{\beta \epsilon} {g}^{\tau \sigma} {g}^{\kappa \delta} {g}^{\nu \lambda} {g}^{\rho \alpha} + \frac{1}{16}\, {A}^{\mu \gamma} {g}^{\beta \epsilon} {g}^{\tau \sigma} {g}^{\kappa \lambda} {g}^{\nu \delta} {g}^{\rho \alpha} - \frac{1}{16}\, {A}^{\mu \gamma} {g}^{\beta \lambda} {g}^{\tau \sigma} {g}^{\kappa \epsilon} {g}^{\nu \delta} {g}^{\rho \alpha} - \frac{1}{16}\, {A}^{\mu \gamma} {g}^{\beta \nu} {g}^{\tau \rho} {g}^{\kappa \delta} {g}^{\sigma \lambda} {g}^{\epsilon \alpha} - \frac{1}{16}\, {A}^{\mu \gamma} {g}^{\beta \nu} {g}^{\tau \rho} {g}^{\kappa \lambda} {g}^{\sigma \alpha} {g}^{\epsilon \delta} + \frac{1}{16}\, {A}^{\mu \gamma} {g}^{\beta \nu} {g}^{\tau \rho} {g}^{\kappa \lambda} {g}^{\sigma \delta} {g}^{\epsilon \alpha} + \frac{1}{32}\, {A}^{\mu \gamma} {g}^{\beta \sigma} {g}^{\tau \epsilon} {g}^{\kappa \delta} {g}^{\nu \lambda} {g}^{\rho \alpha} - \frac{1}{16}\, {A}^{\mu \gamma} {g}^{\beta \sigma} {g}^{\tau \epsilon} {g}^{\kappa \lambda} {g}^{\nu \delta} {g}^{\rho \alpha} + \frac{1}{32}\, {A}^{\mu \sigma} {g}^{\beta \nu} {g}^{\tau \rho} {g}^{\kappa \delta} {g}^{\gamma \lambda} {g}^{\epsilon \alpha} - \frac{1}{16}\, {A}^{\mu \sigma} {g}^{\beta \nu} {g}^{\tau \rho} {g}^{\kappa \lambda} {g}^{\gamma \delta} {g}^{\epsilon \alpha} + \frac{1}{16}\, {A}^{\nu \gamma} {g}^{\beta \delta} {g}^{\tau \mu} {g}^{\kappa \rho} {g}^{\sigma \lambda} {g}^{\epsilon \alpha} + \frac{1}{16}\, {A}^{\nu \gamma} {g}^{\beta \lambda} {g}^{\tau \mu} {g}^{\kappa \rho} {g}^{\sigma \alpha} {g}^{\epsilon \delta} - \frac{1}{16}\, {A}^{\nu \gamma} {g}^{\beta \lambda} {g}^{\tau \mu} {g}^{\kappa \rho} {g}^{\sigma \delta} {g}^{\epsilon \alpha}  + \frac{1}{16}\, {A}^{\nu \gamma} {g}^{\beta \mu} {g}^{\tau \rho} {g}^{\kappa \delta} {g}^{\sigma \lambda} {g}^{\epsilon \alpha} + \frac{1}{16}\, {A}^{\nu \gamma} {g}^{\beta \mu} {g}^{\tau \rho} {g}^{\kappa \lambda} {g}^{\sigma \alpha} {g}^{\epsilon \delta} - \frac{1}{16}\, {A}^{\nu \gamma} {g}^{\beta \mu} {g}^{\tau \rho} {g}^{\kappa \lambda} {g}^{\sigma \delta} {g}^{\epsilon \alpha} - \frac{1}{16}\, {A}^{\nu \gamma} {g}^{\beta \rho} {g}^{\tau \mu} {g}^{\kappa \delta} {g}^{\sigma \lambda} {g}^{\epsilon \alpha} - \frac{1}{16}\, {A}^{\nu \gamma} {g}^{\beta \rho} {g}^{\tau \mu} {g}^{\kappa \lambda} {g}^{\sigma \alpha} {g}^{\epsilon \delta} + \frac{1}{16}\, {A}^{\nu \gamma} {g}^{\beta \rho} {g}^{\tau \mu} {g}^{\kappa \lambda} {g}^{\sigma \delta} {g}^{\epsilon \alpha} + \frac{1}{16}\, {A}^{\tau \mu} {g}^{\beta \gamma} {g}^{\kappa \epsilon} {g}^{\nu \delta} {g}^{\rho \lambda} {g}^{\sigma \alpha} - \frac{1}{32}\, {A}^{\tau \mu} {g}^{\beta \gamma} {g}^{\kappa \epsilon} {g}^{\nu \lambda} {g}^{\rho \alpha} {g}^{\sigma \delta} + \frac{1}{16}\, {A}^{\tau \mu} {g}^{\beta \gamma} {g}^{\kappa \sigma} {g}^{\nu \delta} {g}^{\rho \lambda} {g}^{\epsilon \alpha} - \frac{1}{32}\, {A}^{\tau \mu} {g}^{\beta \gamma} {g}^{\kappa \sigma} {g}^{\nu \lambda} {g}^{\rho \alpha} {g}^{\epsilon \delta} - \frac{1}{16}\, {A}^{\tau \mu} {g}^{\beta \sigma} {g}^{\kappa \gamma} {g}^{\nu \delta} {g}^{\rho \lambda} {g}^{\epsilon \alpha} + \frac{1}{32}\, {A}^{\tau \mu} {g}^{\beta \sigma} {g}^{\kappa \gamma} {g}^{\nu \lambda} {g}^{\rho \alpha} {g}^{\epsilon \delta} - \frac{1}{16}\, {A}^{\tau \nu} {g}^{\beta \gamma} {g}^{\kappa \epsilon} {g}^{\mu \delta} {g}^{\rho \lambda} {g}^{\sigma \alpha} + \frac{1}{16}\, {A}^{\tau \nu} {g}^{\beta \gamma} {g}^{\kappa \epsilon} {g}^{\mu \lambda} {g}^{\rho \alpha} {g}^{\sigma \delta} + \frac{1}{16}\, {A}^{\tau \nu} {g}^{\beta \gamma} {g}^{\kappa \epsilon} {g}^{\mu \lambda} {g}^{\rho \delta} {g}^{\sigma \alpha} - \frac{1}{16}\, {A}^{\tau \nu} {g}^{\beta \gamma} {g}^{\kappa \sigma} {g}^{\mu \delta} {g}^{\rho \lambda} {g}^{\epsilon \alpha} + \frac{1}{16}\, {A}^{\tau \nu} {g}^{\beta \gamma} {g}^{\kappa \sigma} {g}^{\mu \lambda} {g}^{\rho \alpha} {g}^{\epsilon \delta} + \frac{1}{16}\, {A}^{\tau \nu} {g}^{\beta \gamma} {g}^{\kappa \sigma} {g}^{\mu \lambda} {g}^{\rho \delta} {g}^{\epsilon \alpha} + \frac{1}{16}\, {A}^{\tau \nu} {g}^{\beta \sigma} {g}^{\kappa \gamma} {g}^{\mu \delta} {g}^{\rho \lambda} {g}^{\epsilon \alpha} - \frac{1}{16}\, {A}^{\tau \nu} {g}^{\beta \sigma} {g}^{\kappa \gamma} {g}^{\mu \lambda} {g}^{\rho \alpha} {g}^{\epsilon \delta}  - \frac{1}{16}\, {A}^{\tau \nu} {g}^{\beta \sigma} {g}^{\kappa \gamma} {g}^{\mu \lambda} {g}^{\rho \delta} {g}^{\epsilon \alpha}\left.\right)
\end{dmath*}
\end{dgroup*}
\endgroup

The last result we show here is the functional dependence of $F_2[{\cal H}]$ on the generalized metric. Here, we really mean $F_2^{\rm non-cov}$, but for simplicity we drop the supralabel. Again, it shortens the result to contract $F_2$ with some symmetric tensor. However, since $F_2$ is projected $F_2 = \{F_2\}$, in turns out to be convenient to introduce two symmetric tensors with projections $\underline{S}^{M N} = P^M{}_I P^N{}_J \underline{S}^{I J}$ and $\overline{S}^{M N} = \bar P^M{}_I \bar P^N{}_J \overline{S}^{I J}$, such that
\be
F_2 = \underline{S}^{M N} F_{2 MN} + \overline{S}^{M N} F_{2 MN} \ . \label{F2Resulta}
\ee
The result is given by
\begingroup\makeatletter\def\f@size{7}\check@mathfonts
\begin{dgroup*}
\begin{dmath*}
F_2 = {\partial}_{M N}{{\cal H}_{P Q}}\,  {\partial}_{K L}{{\cal H}_{I J}}\,  \left(\right. - \frac{1}{8}\, {\cal H}^{M K} {\cal H}^{N L} {P}^{P I} {\overline{S}}^{Q J} + \frac{1}{8}\, {\cal H}^{M K} {\cal H}^{N L} {\bar P}^{P I} {\underline{S}}^{Q J} - \frac{1}{4}\, {\cal H}^{M K} {P}^{N I} {P}^{P L} {\overline{S}}^{Q J} + \frac{1}{2}\, {\cal H}^{M K} {P}^{N I} {\bar P}^{P J} {\underline{S}}^{Q L} - \frac{1}{8}\, {\cal H}^{M K} {P}^{P I} {\bar P}^{Q J} {\underline{S}}^{N L} + \frac{1}{2}\, {\cal H}^{M K} {P}^{P I} {\bar P}^{N J} {\overline{S}}^{Q L} - \frac{1}{8}\, {\cal H}^{M K} {P}^{P I} {\bar P}^{Q J} {\overline{S}}^{N L} - \frac{1}{4}\, {\cal H}^{M K} {\bar P}^{N I} {\bar P}^{P L} {\underline{S}}^{Q J} + \frac{1}{4}\, {P}^{M I} {P}^{P K} {\bar P}^{Q J} {\underline{S}}^{N L} + {P}^{M I} {P}^{P K} {\bar P}^{N J} {\overline{S}}^{Q L} - \frac{1}{4}\, {P}^{M I} {P}^{P K} {\bar P}^{Q J} {\overline{S}}^{N L} - {P}^{M I} {\bar P}^{N J} {\bar P}^{P K} {\underline{S}}^{Q L} + \frac{1}{4}\, {P}^{P I} {\bar P}^{M J} {\bar P}^{Q K} {\underline{S}}^{N L} - \frac{1}{4}\, {P}^{P I} {\bar P}^{M J} {\bar P}^{Q K} {\overline{S}}^{N L}\left.\right) + {\partial}_{M}{{\cal H}_{N P}}\,  {\partial}_{Q}{{\cal H}_{K L}}\,  {\partial}_{I J}{{\cal H}_{R S}}\,  \left(\right.\frac{1}{4}\, {\cal H}^{M I} {P}^{N J} {P}^{K R} {\bar P}^{P L} {\overline{S}}^{Q S} - \frac{1}{4}\, {\cal H}^{M I} {P}^{N J} {P}^{K R} {\bar P}^{P Q} {\overline{S}}^{L S} - \frac{1}{4}\, {\cal H}^{M I} {P}^{N K} {P}^{Q R} {\bar P}^{L J} {\overline{S}}^{P S} + \frac{1}{4}\, {\cal H}^{M I} {P}^{N K} {P}^{Q R} {\bar P}^{L S} {\overline{S}}^{P J} + \frac{1}{4}\, {\cal H}^{M I} {P}^{N K} {P}^{Q R} {\bar P}^{P J} {\overline{S}}^{L S} - \frac{1}{4}\, {\cal H}^{M I} {P}^{N K} {P}^{Q R} {\bar P}^{P S} {\overline{S}}^{L J} + \frac{1}{2}\, {\cal H}^{M I} {P}^{N Q} {P}^{K R} {\bar P}^{L J} {\overline{S}}^{P S} - \frac{1}{4}\, {\cal H}^{M I} {P}^{N Q} {P}^{K R} {\bar P}^{L S} {\overline{S}}^{P J} - \frac{1}{4}\, {\cal H}^{M I} {P}^{N Q} {P}^{K R} {\bar P}^{P J} {\overline{S}}^{L S} - \frac{1}{4}\, {\cal H}^{M I} {P}^{N R} {P}^{K J} {\bar P}^{P L} {\overline{S}}^{Q S} + \frac{1}{2}\, {\cal H}^{M I} {P}^{N R} {P}^{K J} {\bar P}^{P Q} {\overline{S}}^{L S} + \frac{1}{4}\, {\cal H}^{M Q} {P}^{N I} {P}^{K R} {\bar P}^{P J} {\overline{S}}^{L S} - \frac{1}{2}\, {\cal H}^{M Q} {P}^{N R} {P}^{K I} {\bar P}^{P J} {\overline{S}}^{L S} + \frac{1}{4}\, {\cal H}^{M Q} {P}^{N R} {P}^{K I} {\bar P}^{P S} {\overline{S}}^{L J} + \frac{1}{4}\, {\cal H}^{M I} {P}^{K J} {\bar P}^{N L} {\bar P}^{Q R} {\underline{S}}^{P S} - \frac{1}{2}\, {\cal H}^{M I} {P}^{K J} {\bar P}^{N Q} {\bar P}^{L R} {\underline{S}}^{P S} - \frac{1}{4}\, {\cal H}^{M I} {P}^{K R} {\bar P}^{N L} {\bar P}^{Q S} {\underline{S}}^{P J} + \frac{1}{4}\, {\cal H}^{M I} {P}^{K R} {\bar P}^{N Q} {\bar P}^{L S} {\underline{S}}^{P J} - \frac{1}{4}\, {\cal H}^{M I} {P}^{N J} {\bar P}^{P K} {\bar P}^{Q R} {\underline{S}}^{L S}  + \frac{1}{4}\, {\cal H}^{M I} {P}^{N J} {\bar P}^{P Q} {\bar P}^{K R} {\underline{S}}^{L S} - \frac{1}{4}\, {\cal H}^{M I} {P}^{N K} {\bar P}^{P J} {\bar P}^{L R} {\underline{S}}^{Q S} + \frac{1}{4}\, {\cal H}^{M I} {P}^{N K} {\bar P}^{P R} {\bar P}^{L J} {\underline{S}}^{Q S} + \frac{1}{4}\, {\cal H}^{M I} {P}^{N Q} {\bar P}^{P J} {\bar P}^{K R} {\underline{S}}^{L S} - \frac{1}{2}\, {\cal H}^{M I} {P}^{N Q} {\bar P}^{P R} {\bar P}^{K J} {\underline{S}}^{L S} + \frac{1}{4}\, {\cal H}^{M I} {P}^{N R} {\bar P}^{P K} {\bar P}^{Q S} {\underline{S}}^{L J} - \frac{1}{4}\, {\cal H}^{M Q} {P}^{N I} {\bar P}^{P J} {\bar P}^{K R} {\underline{S}}^{L S} + \frac{1}{2}\, {\cal H}^{M Q} {P}^{N I} {\bar P}^{P R} {\bar P}^{K J} {\underline{S}}^{L S} - \frac{1}{4}\, {\cal H}^{M Q} {P}^{N R} {\bar P}^{P S} {\bar P}^{K I} {\underline{S}}^{L J} + \frac{1}{2}\, {P}^{M K} {P}^{N I} {P}^{Q R} {\bar P}^{P J} {\overline{S}}^{L S} - \frac{1}{2}\, {P}^{M K} {P}^{N I} {P}^{Q R} {\bar P}^{P S} {\overline{S}}^{L J} + \frac{1}{2}\, {P}^{M R} {P}^{N I} {P}^{K J} {\bar P}^{P Q} {\overline{S}}^{L S} - \frac{1}{2}\, {P}^{M R} {P}^{N Q} {P}^{K I} {\bar P}^{P J} {\overline{S}}^{L S} - \frac{1}{2}\, {P}^{M K} {P}^{N I} {\bar P}^{P J} {\bar P}^{L R} {\underline{S}}^{Q S} + \frac{1}{2}\, {P}^{M K} {P}^{N I} {\bar P}^{P R} {\bar P}^{L J} {\underline{S}}^{Q S} + \frac{1}{4}\, {P}^{M K} {P}^{N R} {\bar P}^{P I} {\bar P}^{L S} {\underline{S}}^{Q J} - \frac{3}{8}\, {P}^{M K} {P}^{N R} {\bar P}^{P S} {\bar P}^{L I} {\underline{S}}^{Q J} - \frac{1}{2}\, {P}^{M R} {P}^{K I} {\bar P}^{N Q} {\bar P}^{L S} {\underline{S}}^{P J} - \frac{1}{2}\, {P}^{M R} {P}^{K I} {\bar P}^{N S} {\bar P}^{L J} {\underline{S}}^{P Q} + \frac{1}{2}\, {P}^{M R} {P}^{N I} {\bar P}^{P Q} {\bar P}^{K S} {\underline{S}}^{L J} - \frac{1}{2}\, {P}^{M R} {P}^{N K} {\bar P}^{P I} {\bar P}^{L S} {\underline{S}}^{Q J} + \frac{1}{2}\, {P}^{M R} {P}^{N K} {\bar P}^{P S} {\bar P}^{L I} {\underline{S}}^{Q J} + \frac{1}{2}\, {P}^{M R} {P}^{N Q} {\bar P}^{P I} {\bar P}^{K S} {\underline{S}}^{L J} - {P}^{M R} {P}^{N Q} {\bar P}^{P S} {\bar P}^{K I} {\underline{S}}^{L J} - \frac{1}{2}\, {P}^{N I} {P}^{K J} {\bar P}^{M L} {\bar P}^{P R} {\underline{S}}^{Q S} + \frac{1}{4}\, {P}^{N I} {P}^{K R} {\bar P}^{M L} {\bar P}^{P S} {\underline{S}}^{Q J} + \frac{1}{4}\, {P}^{N I} {P}^{K R} {\bar P}^{P J} {\bar P}^{L S} {\underline{S}}^{M Q} + \frac{1}{8}\, {P}^{N R} {P}^{K I} {\bar P}^{M L} {\bar P}^{P S} {\underline{S}}^{Q J} - \frac{1}{2}\, {P}^{M K} {P}^{N R} {\bar P}^{P I} {\bar P}^{L J} {\overline{S}}^{Q S} + \frac{1}{4}\, {P}^{M K} {P}^{N R} {\bar P}^{P I} {\bar P}^{L S} {\overline{S}}^{Q J} - \frac{1}{2}\, {P}^{M K} {P}^{N R} {\bar P}^{P I} {\bar P}^{Q S} {\overline{S}}^{L J} + \frac{1}{8}\, {P}^{M K} {P}^{N R} {\bar P}^{P S} {\bar P}^{L I} {\overline{S}}^{Q J} + \frac{1}{2}\, {P}^{M K} {P}^{N R} {\bar P}^{Q S} {\bar P}^{L I} {\overline{S}}^{P J} - \frac{1}{2}\, {P}^{N I} {P}^{K R} {\bar P}^{M L} {\bar P}^{P J} {\overline{S}}^{Q S} + \frac{1}{4}\, {P}^{N I} {P}^{K R} {\bar P}^{M L} {\bar P}^{P S} {\overline{S}}^{Q J} - \frac{1}{2}\, {P}^{N I} {P}^{K R} {\bar P}^{M S} {\bar P}^{P L} {\overline{S}}^{Q J}
 \end{dmath*}
\begin{dmath*}
{ + \frac{1}{2}\, {P}^{N I} {P}^{K R} {\bar P}^{M S} {\bar P}^{P Q} {\overline{S}}^{L J} + \frac{1}{4}\, {P}^{N I} {P}^{K R} {\bar P}^{P J} {\bar P}^{L S} {\overline{S}}^{M Q} + \frac{1}{2}\, {P}^{N R} {P}^{K I} {\bar P}^{M L} {\bar P}^{P J} {\overline{S}}^{Q S} - \frac{3}{8}\, {P}^{N R} {P}^{K I} {\bar P}^{M L} {\bar P}^{P S} {\overline{S}}^{Q J}  - \frac{1}{2}\, {P}^{N R} {P}^{K I} {\bar P}^{M S} {\bar P}^{L J} {\overline{S}}^{P Q} \ \ \ \ \ \ }+ \frac{1}{2}\, {P}^{N R} {P}^{K I} {\bar P}^{M S} {\bar P}^{P L} {\overline{S}}^{Q J} - {P}^{N R} {P}^{K I} {\bar P}^{M S} {\bar P}^{P Q} {\overline{S}}^{L J} + \frac{1}{2}\, {P}^{M K} {\bar P}^{N I} {\bar P}^{Q R} {\bar P}^{L J} {\underline{S}}^{P S} + \frac{1}{2}\, {P}^{N I} {\bar P}^{M K} {\bar P}^{P J} {\bar P}^{Q R} {\underline{S}}^{L S} - \frac{1}{2}\, {P}^{N I} {\bar P}^{M R} {\bar P}^{P Q} {\bar P}^{K J} {\underline{S}}^{L S} - \frac{1}{2}\, {P}^{N R} {\bar P}^{M K} {\bar P}^{P I} {\bar P}^{Q S} {\underline{S}}^{L J}\left.\right) + {\partial}_{M}{{\cal H}_{N P}}\,  {\partial}_{Q}{{\cal H}_{K L}}\,  {\partial}_{I}{{\cal H}_{J R}}\,  {\partial}_{S}{{\cal H}_{T U}}\,  \left(\right.\frac{1}{32}\, {\cal H}^{M Q} {\cal H}^{I S} {P}^{N J} {P}^{K T} {\bar P}^{P R} {\overline{S}}^{L U} - \frac{1}{16}\, {\cal H}^{M Q} {\cal H}^{I S} {P}^{N J} {P}^{K T} {\bar P}^{P U} {\overline{S}}^{L R} + \frac{1}{32}\, {\cal H}^{M Q} {\cal H}^{I S} {P}^{N K} {P}^{J T} {\bar P}^{P R} {\overline{S}}^{L U} - \frac{1}{32}\, {\cal H}^{M Q} {\cal H}^{I S} {P}^{N J} {\bar P}^{P K} {\bar P}^{R T} {\underline{S}}^{L U} - \frac{1}{32}\, {\cal H}^{M Q} {\cal H}^{I S} {P}^{N J} {\bar P}^{P R} {\bar P}^{K T} {\underline{S}}^{L U} + \frac{1}{16}\, {\cal H}^{M Q} {\cal H}^{I S} {P}^{N J} {\bar P}^{P T} {\bar P}^{K R} {\underline{S}}^{L U} + \frac{1}{4}\, {\cal H}^{M Q} {P}^{N I} {P}^{K S} {P}^{J T} {\bar P}^{P U} {\overline{S}}^{L R} - \frac{1}{16}\, {\cal H}^{M Q} {P}^{N I} {P}^{K S} {P}^{J T} {\bar P}^{R U} {\overline{S}}^{P L} - \frac{1}{4}\, {\cal H}^{M Q} {P}^{N I} {P}^{K T} {P}^{J S} {\bar P}^{P U} {\overline{S}}^{L R} + \frac{1}{8}\, {\cal H}^{M Q} {P}^{N K} {P}^{I T} {P}^{J S} {\bar P}^{P R} {\overline{S}}^{L U} - \frac{1}{4}\, {\cal H}^{M Q} {P}^{N T} {P}^{K I} {P}^{J S} {\bar P}^{P R} {\overline{S}}^{L U} + \frac{1}{8}\, {\cal H}^{M Q} {P}^{N T} {P}^{K I} {P}^{J S} {\bar P}^{P U} {\overline{S}}^{L R} + \frac{1}{16}\, {\cal H}^{M Q} {P}^{I T} {P}^{J S} {\bar P}^{N R} {\bar P}^{K U} {\underline{S}}^{P L} - \frac{1}{8}\, {\cal H}^{M Q} {P}^{N I} {P}^{J S} {\bar P}^{P K} {\bar P}^{R T} {\underline{S}}^{L U} - \frac{1}{4}\, {\cal H}^{M Q} {P}^{N I} {P}^{J S} {\bar P}^{P R} {\bar P}^{K T} {\underline{S}}^{L U} + \frac{1}{8}\, {\cal H}^{M Q} {P}^{N I} {P}^{J S} {\bar P}^{P T} {\bar P}^{K R} {\underline{S}}^{L U} + \frac{1}{16}\, {\cal H}^{M Q} {P}^{N I} {P}^{J T} {\bar P}^{P K} {\bar P}^{R U} {\underline{S}}^{L S} + \frac{1}{4}\, {\cal H}^{M Q} {P}^{N I} {P}^{J T} {\bar P}^{P R} {\bar P}^{K U} {\underline{S}}^{L S} - \frac{1}{8}\, {\cal H}^{M Q} {P}^{N I} {P}^{J T} {\bar P}^{P U} {\bar P}^{K R} {\underline{S}}^{L S} + \frac{1}{16}\, {\cal H}^{M Q} {P}^{N I} {P}^{K S} {\bar P}^{P J} {\bar P}^{L T} {\underline{S}}^{R U} + \frac{1}{8}\, {\cal H}^{M Q} {P}^{N I} {P}^{K S} {\bar P}^{P L} {\bar P}^{J T} {\underline{S}}^{R U} - \frac{1}{16}\, {\cal H}^{M Q} {P}^{N I} {P}^{K S} {\bar P}^{P T} {\bar P}^{L J} {\underline{S}}^{R U} - \frac{1}{16}\, {\cal H}^{M Q} {P}^{N I} {P}^{K T} {\bar P}^{P J} {\bar P}^{L U} {\underline{S}}^{R S} - \frac{1}{4}\, {\cal H}^{M Q} {P}^{N I} {P}^{K T} {\bar P}^{P L} {\bar P}^{J U} {\underline{S}}^{R S} + \frac{1}{8}\, {\cal H}^{M Q} {P}^{N I} {P}^{K T} {\bar P}^{P U} {\bar P}^{L J} {\underline{S}}^{R S} - \frac{1}{8}\, {\cal H}^{M Q} {P}^{N J} {P}^{I T} {\bar P}^{P K} {\bar P}^{R U} {\underline{S}}^{L S} - \frac{1}{8}\, {\cal H}^{M Q} {P}^{N J} {P}^{I T} {\bar P}^{P R} {\bar P}^{K U} {\underline{S}}^{L S} + \frac{1}{8}\, {\cal H}^{M Q} {P}^{N J} {P}^{I T} {\bar P}^{P U} {\bar P}^{K R} {\underline{S}}^{L S} + \frac{3}{32}\, {\cal H}^{M Q} {P}^{N J} {P}^{K T} {\bar P}^{P L} {\bar P}^{R U} {\underline{S}}^{I S} + \frac{1}{64}\, {\cal H}^{M Q} {P}^{N J} {P}^{K T} {\bar P}^{P R} {\bar P}^{L U} {\underline{S}}^{I S} - \frac{1}{16}\, {\cal H}^{M Q} {P}^{N J} {P}^{K T} {\bar P}^{P U} {\bar P}^{L R} {\underline{S}}^{I S} + \frac{1}{8}\, {\cal H}^{M Q} {P}^{N K} {P}^{I T} {\bar P}^{P J} {\bar P}^{L U} {\underline{S}}^{R S} - \frac{1}{32}\, {\cal H}^{M Q} {P}^{N K} {P}^{J T} {\bar P}^{P R} {\bar P}^{L U} {\underline{S}}^{I S} + \frac{1}{16}\, {\cal H}^{M Q} {P}^{N J} {P}^{K T} {\bar P}^{I U} {\bar P}^{R S} {\overline{S}}^{P L} + \frac{1}{16}\, {\cal H}^{M Q} {P}^{N J} {P}^{K T} {\bar P}^{P I} {\bar P}^{L S} {\overline{S}}^{R U} - \frac{1}{16}\, {\cal H}^{M Q} {P}^{N J} {P}^{K T} {\bar P}^{P I} {\bar P}^{L U} {\overline{S}}^{R S} - \frac{1}{4}\, {\cal H}^{M Q} {P}^{N J} {P}^{K T} {\bar P}^{P I} {\bar P}^{R S} {\overline{S}}^{L U} + \frac{1}{4}\, {\cal H}^{M Q} {P}^{N J} {P}^{K T} {\bar P}^{P I} {\bar P}^{R U} {\overline{S}}^{L S} + \frac{1}{8}\, {\cal H}^{M Q} {P}^{N J} {P}^{K T} {\bar P}^{P L} {\bar P}^{I U} {\overline{S}}^{R S} - \frac{1}{32}\, {\cal H}^{M Q} {P}^{N J} {P}^{K T} {\bar P}^{P L} {\bar P}^{R U} {\overline{S}}^{I S} - \frac{1}{8}\, {\cal H}^{M Q} {P}^{N J} {P}^{K T} {\bar P}^{P R} {\bar P}^{I U} {\overline{S}}^{L S} + \frac{1}{64}\, {\cal H}^{M Q} {P}^{N J} {P}^{K T} {\bar P}^{P R} {\bar P}^{L U} {\overline{S}}^{I S} - \frac{1}{16}\, {\cal H}^{M Q} {P}^{N J} {P}^{K T} {\bar P}^{P U} {\bar P}^{L R} {\overline{S}}^{I S} + \frac{1}{8}\, {\cal H}^{M Q} {P}^{N K} {P}^{J T} {\bar P}^{P I} {\bar P}^{L S} {\overline{S}}^{R U} - \frac{1}{4}\, {\cal H}^{M Q} {P}^{N K} {P}^{J T} {\bar P}^{P I} {\bar P}^{L U} {\overline{S}}^{R S} - \frac{1}{8}\, {\cal H}^{M Q} {P}^{N K} {P}^{J T} {\bar P}^{P I} {\bar P}^{R S} {\overline{S}}^{L U} + \frac{1}{16}\, {\cal H}^{M Q} {P}^{N K} {P}^{J T} {\bar P}^{P I} {\bar P}^{R U} {\overline{S}}^{L S} - \frac{1}{8}\, {\cal H}^{M Q} {P}^{N K} {P}^{J T} {\bar P}^{P R} {\bar P}^{I U} {\overline{S}}^{L S} + \frac{3}{32}\, {\cal H}^{M Q} {P}^{N K} {P}^{J T} {\bar P}^{P R} {\bar P}^{L U} {\overline{S}}^{I S} - \frac{1}{16}\, {\cal H}^{M Q} {P}^{N T} {P}^{K J} {\bar P}^{P I} {\bar P}^{L S} {\overline{S}}^{R U} + \frac{1}{8}\, {\cal H}^{M Q} {P}^{N T} {P}^{K J} {\bar P}^{P I} {\bar P}^{L U} {\overline{S}}^{R S} + \frac{1}{8}\, {\cal H}^{M Q} {P}^{N T} {P}^{K J} {\bar P}^{P I} {\bar P}^{R S} {\overline{S}}^{L U} - \frac{1}{8}\, {\cal H}^{M Q} {P}^{N T} {P}^{K J} {\bar P}^{P I} {\bar P}^{R U} {\overline{S}}^{L S} + \frac{1}{8}\, {\cal H}^{M Q} {P}^{N T} {P}^{K J} {\bar P}^{P R} {\bar P}^{I U} {\overline{S}}^{L S} - \frac{1}{16}\, {\cal H}^{M Q} {P}^{J T} {\bar P}^{N I} {\bar P}^{K S} {\bar P}^{R U} {\underline{S}}^{P L} + \frac{1}{8}\, {\cal H}^{M Q} {P}^{N J} {\bar P}^{P K} {\bar P}^{I T} {\bar P}^{R S} {\underline{S}}^{L U} - \frac{1}{4}\, {\cal H}^{M Q} {P}^{N J} {\bar P}^{P T} {\bar P}^{K I} {\bar P}^{R S} {\underline{S}}^{L U} + \frac{1}{4}\, {\cal H}^{M Q} {P}^{N T} {\bar P}^{P I} {\bar P}^{K S} {\bar P}^{J U} {\underline{S}}^{L R} - \frac{1}{4}\, {\cal H}^{M Q} {P}^{N T} {\bar P}^{P I} {\bar P}^{K U} {\bar P}^{J S} {\underline{S}}^{L R}  + \frac{1}{8}\, {\cal H}^{M Q} {P}^{N T} {\bar P}^{P U} {\bar P}^{K I} {\bar P}^{J S} {\underline{S}}^{L R} + \frac{1}{8}\, {P}^{M K} {P}^{N Q} {P}^{I T} {P}^{J S} {\bar P}^{P R} {\overline{S}}^{L U} - \frac{1}{4}\, {P}^{M T} {P}^{N Q} {P}^{K I} {P}^{J S} {\bar P}^{P R} {\overline{S}}^{L U} - \frac{1}{8}\, {P}^{M J} {P}^{N K} {P}^{Q T} {\bar P}^{P L} {\bar P}^{R U} {\underline{S}}^{I S} - \frac{1}{8}\, {P}^{M J} {P}^{N K} {P}^{Q T} {\bar P}^{P R} {\bar P}^{L U} {\underline{S}}^{I S} + \frac{3}{16}\, {P}^{M J} {P}^{N K} {P}^{Q T} {\bar P}^{P U} {\bar P}^{L R} {\underline{S}}^{I S} + \frac{3}{8}\, {P}^{M J} {P}^{N Q} {P}^{K T} {\bar P}^{P L} {\bar P}^{R U} {\underline{S}}^{I S} + \frac{1}{8}\, {P}^{M J} {P}^{N Q} {P}^{K T} {\bar P}^{P R} {\bar P}^{L U} {\underline{S}}^{I S} - \frac{1}{4}\, {P}^{M J} {P}^{N Q} {P}^{K T} {\bar P}^{P U} {\bar P}^{L R} {\underline{S}}^{I S} + \frac{1}{4}\, {P}^{M K} {P}^{N Q} {P}^{I T} {\bar P}^{P J} {\bar P}^{L U} {\underline{S}}^{R S} - \frac{1}{16}\, {P}^{M K} {P}^{N Q} {P}^{J T} {\bar P}^{P R} {\bar P}^{L U} {\underline{S}}^{I S} + \frac{1}{4}\, {P}^{M T} {P}^{N Q} {P}^{K I} {\bar P}^{P J} {\bar P}^{L U} {\underline{S}}^{R S} - \frac{1}{2}\, {P}^{M T} {P}^{N Q} {P}^{K I} {\bar P}^{P L} {\bar P}^{J U} {\underline{S}}^{R S} - \frac{1}{4}\, {P}^{M T} {P}^{N Q} {P}^{K I} {\bar P}^{P U} {\bar P}^{L J} {\underline{S}}^{R S} - \frac{1}{8}\, {P}^{M J} {P}^{N K} {P}^{Q T} {\bar P}^{I U} {\bar P}^{R S} {\overline{S}}^{P L} + \frac{1}{4}\, {P}^{M J} {P}^{N K} {P}^{Q T} {\bar P}^{P I} {\bar P}^{R S} {\overline{S}}^{L U}
 \end{dmath*}
\begin{dmath*}
{ - \frac{1}{4}\, {P}^{M J} {P}^{N K} {P}^{Q T} {\bar P}^{P I} {\bar P}^{R U} {\overline{S}}^{L S} + \frac{1}{4}\, {P}^{M J} {P}^{N K} {P}^{Q T} {\bar P}^{P R} {\bar P}^{I U} {\overline{S}}^{L S} + \frac{1}{16}\, {P}^{M J} {P}^{N K} {P}^{Q T} {\bar P}^{P U} {\bar P}^{L R} {\overline{S}}^{I S} + \frac{1}{4}\, {P}^{M J} {P}^{N Q} {P}^{K T} {\bar P}^{I U} {\bar P}^{R S} {\overline{S}}^{P L} \ \ \ \ \ \ } - \frac{1}{4}\, {P}^{M J} {P}^{N Q} {P}^{K T} {\bar P}^{P I} {\bar P}^{R S} {\overline{S}}^{L U}  + \frac{1}{4}\, {P}^{M J} {P}^{N Q} {P}^{K T} {\bar P}^{P I} {\bar P}^{R U} {\overline{S}}^{L S} + \frac{1}{4}\, {P}^{M J} {P}^{N Q} {P}^{K T} {\bar P}^{P L} {\bar P}^{I U} {\overline{S}}^{R S} - \frac{1}{8}\, {P}^{M J} {P}^{N Q} {P}^{K T} {\bar P}^{P L} {\bar P}^{R U} {\overline{S}}^{I S} - \frac{1}{4}\, {P}^{M J} {P}^{N Q} {P}^{K T} {\bar P}^{P U} {\bar P}^{L I} {\overline{S}}^{R S} - \frac{1}{2}\, {P}^{M K} {P}^{N J} {P}^{Q T} {\bar P}^{P I} {\bar P}^{R S} {\overline{S}}^{L U} + \frac{1}{4}\, {P}^{M K} {P}^{N J} {P}^{Q T} {\bar P}^{P I} {\bar P}^{R U} {\overline{S}}^{L S} - \frac{1}{4}\, {P}^{M K} {P}^{N J} {P}^{Q T} {\bar P}^{P R} {\bar P}^{I U} {\overline{S}}^{L S} + \frac{1}{8}\, {P}^{M K} {P}^{N Q} {P}^{J T} {\bar P}^{P I} {\bar P}^{L S} {\overline{S}}^{R U} - \frac{1}{4}\, {P}^{M K} {P}^{N Q} {P}^{J T} {\bar P}^{P I} {\bar P}^{L U} {\overline{S}}^{R S} - \frac{1}{4}\, {P}^{M K} {P}^{N Q} {P}^{J T} {\bar P}^{P I} {\bar P}^{R S} {\overline{S}}^{L U} + \frac{1}{8}\, {P}^{M K} {P}^{N Q} {P}^{J T} {\bar P}^{P I} {\bar P}^{R U} {\overline{S}}^{L S} - \frac{1}{4}\, {P}^{M K} {P}^{N Q} {P}^{J T} {\bar P}^{P R} {\bar P}^{I U} {\overline{S}}^{L S} + \frac{1}{16}\, {P}^{M K} {P}^{N Q} {P}^{J T} {\bar P}^{P R} {\bar P}^{L U} {\overline{S}}^{I S} + \frac{1}{4}\, {P}^{M K} {P}^{N T} {P}^{Q J} {\bar P}^{P I} {\bar P}^{R S} {\overline{S}}^{L U} - \frac{1}{4}\, {P}^{M K} {P}^{N T} {P}^{Q J} {\bar P}^{P I} {\bar P}^{R U} {\overline{S}}^{L S} + \frac{1}{2}\, {P}^{M K} {P}^{N T} {P}^{Q J} {\bar P}^{P R} {\bar P}^{I U} {\overline{S}}^{L S} - \frac{1}{4}\, {P}^{M T} {P}^{N K} {P}^{Q J} {\bar P}^{P I} {\bar P}^{L S} {\overline{S}}^{R U} - \frac{1}{4}\, {P}^{M T} {P}^{N K} {P}^{Q J} {\bar P}^{P I} {\bar P}^{R S} {\overline{S}}^{L U} + \frac{1}{4}\, {P}^{M T} {P}^{N K} {P}^{Q J} {\bar P}^{P I} {\bar P}^{R U} {\overline{S}}^{L S} - \frac{1}{2}\, {P}^{M T} {P}^{N K} {P}^{Q J} {\bar P}^{P R} {\bar P}^{I U} {\overline{S}}^{L S} + \frac{1}{4}\, {P}^{M T} {P}^{N Q} {P}^{K J} {\bar P}^{P I} {\bar P}^{L S} {\overline{S}}^{R U} + \frac{1}{4}\, {P}^{M T} {P}^{N Q} {P}^{K J} {\bar P}^{P I} {\bar P}^{R S} {\overline{S}}^{L U} + \frac{1}{4}\, {P}^{M T} {P}^{N Q} {P}^{K J} {\bar P}^{P L} {\bar P}^{I U} {\overline{S}}^{R S} + \frac{1}{4}\, {P}^{M J} {P}^{K T} {\bar P}^{N Q} {\bar P}^{L I} {\bar P}^{R U} {\underline{S}}^{P S} - \frac{1}{4}\, {P}^{M J} {P}^{N Q} {\bar P}^{P I} {\bar P}^{K S} {\bar P}^{R T} {\underline{S}}^{L U} + \frac{1}{4}\, {P}^{M J} {P}^{N Q} {\bar P}^{P I} {\bar P}^{K T} {\bar P}^{R S} {\underline{S}}^{L U} + \frac{1}{4}\, {P}^{M J} {P}^{N Q} {\bar P}^{P K} {\bar P}^{I T} {\bar P}^{R S} {\underline{S}}^{L U} - \frac{1}{4}\, {P}^{M J} {P}^{N Q} {\bar P}^{P T} {\bar P}^{K I} {\bar P}^{R S} {\underline{S}}^{L U} + \frac{1}{4}\, {P}^{M J} {P}^{N T} {\bar P}^{P Q} {\bar P}^{K I} {\bar P}^{R U} {\underline{S}}^{L S} - \frac{1}{4}\, {P}^{M J} {P}^{N T} {\bar P}^{P Q} {\bar P}^{K R} {\bar P}^{I U} {\underline{S}}^{L S} - \frac{1}{8}\, {P}^{M J} {P}^{Q T} {\bar P}^{N K} {\bar P}^{I U} {\bar P}^{R S} {\underline{S}}^{P L} + \frac{1}{4}\, {P}^{M K} {P}^{J T} {\bar P}^{N U} {\bar P}^{Q R} {\bar P}^{L I} {\underline{S}}^{P S} + \frac{1}{8}\, {P}^{M K} {P}^{N Q} {\bar P}^{P I} {\bar P}^{L S} {\bar P}^{J T} {\underline{S}}^{R U} - \frac{1}{4}\, {P}^{M K} {P}^{N Q} {\bar P}^{P I} {\bar P}^{L T} {\bar P}^{J S} {\underline{S}}^{R U} - \frac{1}{4}\, {P}^{M K} {P}^{N T} {\bar P}^{P L} {\bar P}^{Q J} {\bar P}^{I U} {\underline{S}}^{R S} - \frac{1}{4}\, {P}^{M K} {P}^{N T} {\bar P}^{P Q} {\bar P}^{L I} {\bar P}^{J U} {\underline{S}}^{R S} + \frac{1}{4}\, {P}^{M K} {P}^{N T} {\bar P}^{P Q} {\bar P}^{L J} {\bar P}^{I U} {\underline{S}}^{R S} - \frac{1}{8}\, {P}^{M K} {P}^{N T} {\bar P}^{P U} {\bar P}^{Q J} {\bar P}^{L I} {\underline{S}}^{R S} - \frac{1}{4}\, {P}^{M T} {P}^{K J} {\bar P}^{N L} {\bar P}^{Q R} {\bar P}^{I U} {\underline{S}}^{P S}%
 - \frac{1}{4}\, {P}^{M T} {P}^{K J} {\bar P}^{N Q} {\bar P}^{L I} {\bar P}^{R U} {\underline{S}}^{P S} - \frac{1}{2}\, {P}^{M T} {P}^{N J} {\bar P}^{P Q} {\bar P}^{K I} {\bar P}^{R U} {\underline{S}}^{L S} + \frac{1}{2}\, {P}^{M T} {P}^{N J} {\bar P}^{P Q} {\bar P}^{K R} {\bar P}^{I U} {\underline{S}}^{L S} + \frac{1}{4}\, {P}^{M T} {P}^{N K} {\bar P}^{P L} {\bar P}^{Q J} {\bar P}^{I U} {\underline{S}}^{R S} - \frac{1}{4}\, {P}^{M T} {P}^{N K} {\bar P}^{P Q} {\bar P}^{L J} {\bar P}^{I U} {\underline{S}}^{R S} + \frac{1}{4}\, {P}^{M T} {P}^{N K} {\bar P}^{P U} {\bar P}^{Q J} {\bar P}^{L I} {\underline{S}}^{R S} + \frac{1}{4}\, {P}^{M T} {P}^{N Q} {\bar P}^{P I} {\bar P}^{K S} {\bar P}^{J U} {\underline{S}}^{L R} - \frac{1}{4}\, {P}^{M T} {P}^{N Q} {\bar P}^{P I} {\bar P}^{K U} {\bar P}^{J S} {\underline{S}}^{L R} + \frac{1}{2}\, {P}^{M T} {P}^{N Q} {\bar P}^{P U} {\bar P}^{K I} {\bar P}^{J S} {\underline{S}}^{L R} + \frac{1}{4}\, {P}^{M T} {P}^{Q J} {\bar P}^{N I} {\bar P}^{K S} {\bar P}^{R U} {\underline{S}}^{P L} - \frac{1}{4}\, {P}^{M T} {P}^{Q J} {\bar P}^{N I} {\bar P}^{K U} {\bar P}^{R S} {\underline{S}}^{P L} - \frac{1}{16}\, {P}^{N J} {P}^{K T} {\bar P}^{M L} {\bar P}^{P Q} {\bar P}^{R U} {\underline{S}}^{I S} - \frac{1}{16}\, {P}^{N J} {P}^{K T} {\bar P}^{M U} {\bar P}^{P L} {\bar P}^{Q R} {\underline{S}}^{I S} + \frac{1}{8}\, {P}^{N K} {P}^{J T} {\bar P}^{M R} {\bar P}^{P Q} {\bar P}^{L U} {\underline{S}}^{I S} - \frac{1}{4}\, {P}^{N J} {P}^{K T} {\bar P}^{M L} {\bar P}^{P Q} {\bar P}^{I U} {\overline{S}}^{R S} + \frac{1}{16}\, {P}^{N J} {P}^{K T} {\bar P}^{M L} {\bar P}^{P Q} {\bar P}^{R U} {\overline{S}}^{I S} + \frac{1}{8}\, {P}^{N J} {P}^{K T} {\bar P}^{M R} {\bar P}^{P L} {\bar P}^{Q U} {\overline{S}}^{I S} - \frac{1}{8}\, {P}^{N J} {P}^{K T} {\bar P}^{M R} {\bar P}^{P Q} {\bar P}^{L U} {\overline{S}}^{I S} - \frac{3}{16}\, {P}^{N J} {P}^{K T} {\bar P}^{M U} {\bar P}^{P L} {\bar P}^{Q R} {\overline{S}}^{I S} - \frac{1}{4}\, {P}^{N J} {P}^{K T} {\bar P}^{M U} {\bar P}^{P Q} {\bar P}^{L I} {\overline{S}}^{R S}
 + \frac{1}{4}\, {P}^{N J} {P}^{K T} {\bar P}^{M U} {\bar P}^{P Q} {\bar P}^{L R} {\overline{S}}^{I S} + \frac{1}{8}\, {P}^{N K} {P}^{J T} {\bar P}^{M R} {\bar P}^{P L} {\bar P}^{Q U} {\overline{S}}^{I S} + \frac{1}{4}\, {P}^{N K} {P}^{J T} {\bar P}^{M R} {\bar P}^{P Q} {\bar P}^{I U} {\overline{S}}^{L S} - \frac{3}{8}\, {P}^{N K} {P}^{J T} {\bar P}^{M R} {\bar P}^{P Q} {\bar P}^{L U} {\overline{S}}^{I S} + \frac{1}{2}\, {P}^{N K} {P}^{J T} {\bar P}^{M U} {\bar P}^{P Q} {\bar P}^{L I} {\overline{S}}^{R S} - \frac{1}{8}\, {P}^{N J} {\bar P}^{M K} {\bar P}^{P Q} {\bar P}^{I T} {\bar P}^{R S} {\underline{S}}^{L U} + \frac{1}{4}\, {P}^{N J} {\bar P}^{M T} {\bar P}^{P Q} {\bar P}^{K I} {\bar P}^{R S} {\underline{S}}^{L U}\left.\right)
\end{dmath*}
\end{dgroup*}
\endgroup

\end{appendix}


\newpage
\begin{thebibliography}{98}\label{SEC::Refs}

\bibitem{Siegel:1993xq}
 W.~Siegel, ``Superspace duality in low-energy superstrings,'' Phys.\ Rev.\ D {\bf 48} (1993) 2826 [hep-th/9305073].

 W.~Siegel, ``Two vierbein formalism for string inspired axionic gravity,'' Phys.\ Rev.\ D {\bf 47} (1993) 5453 [hep-th/9302036].

\bibitem{Hull:2009mi}
  C.~Hull and B.~Zwiebach, ``Double Field Theory,'' JHEP {\bf 0909} (2009) 099 [arXiv:0904.4664 [hep-th]].

  C.~Hull and B.~Zwiebach, ``The Gauge algebra of double field theory and Courant brackets,'' JHEP {\bf 0909} (2009) 090 [arXiv:0908.1792 [hep-th]].

  O.~Hohm, C.~Hull and B.~Zwiebach, ``Background independent action for double field theory,'' JHEP {\bf 1007} (2010) 016 [arXiv:1003.5027 [hep-th]].

  O.~Hohm, C.~Hull and B.~Zwiebach, ``Generalized metric formulation of double field theory,''  JHEP {\bf 1008} (2010) 008  [arXiv:1006.4823 [hep-th]].

\bibitem{reviews}
  G.~Aldazabal, D.~Marques and C.~Nunez,
  ``Double Field Theory: A Pedagogical Review,''
  Class.\ Quant.\ Grav.\  {\bf 30}, 163001 (2013)
  [arXiv:1305.1907 [hep-th]].

   O.~Hohm, D.~Lüst and B.~Zwiebach,
  ``The Spacetime of Double Field Theory: Review, Remarks, and Outlook,''
  Fortsch.\ Phys.\  {\bf 61}, 926 (2013)
  [arXiv:1309.2977 [hep-th]].

   D.~S.~Berman and D.~C.~Thompson,
  ``Duality Symmetric String and M-Theory,''
  Phys.\ Rept.\  {\bf 566}, 1 (2014)
  [arXiv:1306.2643 [hep-th]].

  \bibitem{Bedoya:2014pma}
  O.~A.~Bedoya, D.~Marques and C.~Nunez, ``Heterotic $\alpha$'-corrections in Double Field Theory,'' JHEP {\bf 1412} (2014) 074 [arXiv:1407.0365 [hep-th]].

  O.~Hohm and B.~Zwiebach, ``Double field theory at order $\alpha'$,'' JHEP {\bf 1411} (2014) 075 [arXiv:1407.3803 [hep-th]].

  A.~Coimbra, R.~Minasian, H.~Triendl and D.~Waldram, ``Generalised geometry for string corrections,'' JHEP {\bf 1411} (2014) 160 [arXiv:1407.7542 [hep-th]].

  K.~Lee, ``Quadratic $\alpha$'-corrections to heterotic double field theory,''
  Nucl.\ Phys.\ B {\bf 899} (2015) 594
  doi:10.1016/j.nuclphysb.2015.08.013
  [arXiv:1504.00149 [hep-th]].

   O.~Hohm, A.~Sen and B.~Zwiebach,
  ``Heterotic Effective Action and Duality Symmetries Revisited,''
  JHEP {\bf 1502} (2015) 079
  doi:10.1007/JHEP02(2015)079
  [arXiv:1411.5696 [hep-th]].

\bibitem{Marques:2015vua}
  D.~Marques and C.~A.~Nunez,
  ``T-duality and $\alpha$'-corrections,''
  JHEP {\bf 1510} (2015) 084
  doi:10.1007/JHEP10(2015)084
  [arXiv:1507.00652 [hep-th]].

\bibitem{Metsaev:1987zx}
  R.~R.~Metsaev and A.~A.~Tseytlin, ``Order alpha-prime (Two Loop) Equivalence of the String Equations of Motion and the Sigma Model Weyl Invariance Conditions: Dependence on the Dilaton and the Antisymmetric Tensor,''  Nucl.\ Phys.\ B {\bf 293} (1987) 385.


\bibitem{Bergshoeff:1988nn}
  E.~Bergshoeff and M.~de Roo, ``Supersymmetric Chern-simons Terms in Ten-dimensions,'' Phys.\ Lett.\ B {\bf 218} (1989) 210.

  E.~A.~Bergshoeff and M.~de Roo, ``The Quartic Effective Action of the Heterotic String and Supersymmetry,'' Nucl.\ Phys.\ B {\bf 328} (1989) 439.


\bibitem{Hohm:2013jaa}
  O.~Hohm, W.~Siegel and B.~Zwiebach, ``Doubled $\alpha'$-geometry,'' JHEP {\bf 1402} (2014) 065 [arXiv:1306.2970 [hep-th]].

\bibitem{Hohm:2016lim}
  O.~Hohm, U.~Naseer and B.~Zwiebach,
  ``On the curious spectrum of duality invariant higher-derivative gravity,''
  JHEP {\bf 1608} (2016) 173
  doi:10.1007/JHEP08(2016)173
  [arXiv:1607.01784 [hep-th]].

\bibitem{Huang:2016bdd}
  Y.~t.~Huang, W.~Siegel and E.~Y.~Yuan,
  ``Factorization of Chiral String Amplitudes,''
  JHEP {\bf 1609} (2016) 101
  doi:10.1007/JHEP09(2016)101
  [arXiv:1603.02588 [hep-th]].

  M.~M.~Leite and W.~Siegel,
  ``Chiral Closed strings: Four massless states scattering amplitude,''
  arXiv:1610.02052 [hep-th].

  \bibitem{Hohm:2015mka}
  O.~Hohm and B.~Zwiebach,
  ``Double metric, generalized metric, and $\alpha'$-deformed double field theory,''
  Phys.\ Rev.\ D {\bf 93} (2016) no.6,  064035
  doi:10.1103/PhysRevD.93.064035
  [arXiv:1509.02930 [hep-th]].

  \bibitem{Green:1984sg}
  M.~B.~Green and J.~H.~Schwarz, ``Anomaly Cancellation in Supersymmetric D=10 Gauge Theory and Superstring Theory,'' Phys.\ Lett.\ B {\bf 149} (1984) 117.

\bibitem{Hohm:2014eba}
  O.~Hohm and B.~Zwiebach, ``Green-Schwarz mechanism and $\alpha'$-deformed Courant brackets,'' JHEP {\bf 1501} (2015) 012 [arXiv:1407.0708 [hep-th]].

\bibitem{Naseer:2016izx}
  U.~Naseer and B.~Zwiebach,
  ``Three-point Functions in Duality-Invariant Higher-Derivative Gravity,''
  JHEP {\bf 1603} (2016) 147
  doi:10.1007/JHEP03(2016)147
  [arXiv:1602.01101 [hep-th]].

\bibitem{Metsaev:1986yb}
  R.~R.~Metsaev and A.~A.~Tseytlin, ``Curvature Cubed Terms in String Theory Effective Actions,''
  Phys.\ Lett.\ B {\bf 185} (1987) 52.
  doi:10.1016/0370-2693(87)91527-9

\bibitem{Buscher:1987qj}
  T.~H.~Buscher, ``Path Integral Derivation of Quantum Duality in Nonlinear Sigma Models,''  Phys.\ Lett.\ B {\bf 201} (1988) 466.

  T.~H.~Buscher, ``A Symmetry of the String Background Field Equations,'' Phys.\ Lett.\ B {\bf 194} (1987) 59.

\bibitem{Hohm:2015doa}
  O.~Hohm and B.~Zwiebach,
  ``T-duality Constraints on Higher Derivatives Revisited,''
  JHEP {\bf 1604} (2016) 101
  doi:10.1007/JHEP04(2016)101
  [arXiv:1510.00005 [hep-th]].

\bibitem{Meissner:1996sa}
  K.~A.~Meissner, ``Symmetries of higher order string gravity actions,'' Phys.\ Lett.\ B {\bf 392} (1997) 298 [hep-th/9610131].

\bibitem{godazgar}   H.~Godazgar and M.~Godazgar,
  ``Duality completion of higher derivative corrections,''
  JHEP {\bf 1309}, 140 (2013)
  [arXiv:1306.4918 [hep-th]].

  \bibitem{Camanho:2014apa}
  X.~O.~Camanho, J.~D.~Edelstein, J.~Maldacena and A.~Zhiboedov,
  ``Causality Constraints on Corrections to the Graviton Three-Point Coupling,''
  JHEP {\bf 1602} (2016) 020
  doi:10.1007/JHEP02(2016)020
  [arXiv:1407.5597 [hep-th]].

  \bibitem{D'Appollonio:2015gpa}
  G.~D'Appollonio, P.~Di Vecchia, R.~Russo and G.~Veneziano,
  ``Regge behavior saves String Theory from causality violations,''
  JHEP {\bf 1505} (2015) 144
  doi:10.1007/JHEP05(2015)144
  [arXiv:1502.01254 [hep-th]].


\bibitem{Jones:1988hk}
  D.~R.~T.~Jones and A.~M.~Lawrence,
  ``Field Redefinition Dependence of the Low-energy String Effective Action,''
  Z.\ Phys.\ C {\bf 42}, 153 (1989).
  doi:10.1007/BF01565137

\bibitem{Hohm:2015ugy}
  O.~Hohm,
  ``On factorizations in perturbative quantum gravity,''
  JHEP {\bf 1104} (2011) 103
  doi:10.1007/JHEP04(2011)103
  [arXiv:1103.0032 [hep-th]].

  O.~Hohm and D.~Marques,
  ``Perturbative Double Field Theory on General Backgrounds,''
  Phys.\ Rev.\ D {\bf 93} (2016) no.2,  025032
  doi:10.1103/PhysRevD.93.025032
  [arXiv:1512.02658 [hep-th]].

\bibitem{Peeters:2007wn} K.~Peeters, ``Introducing Cadabra: A Symbolic computer algebra system for field theory problems,''
  hep-th/0701238 [HEP-TH].

\bibitem{Ricci}
  I.~Jeon, K.~Lee and J.~H.~Park,
  ``Differential geometry with a projection: Application to double field theory,''
  JHEP {\bf 1104} (2011) 014
  doi:10.1007/JHEP04(2011)014
  [arXiv:1011.1324 [hep-th]].

  O.~Hohm and B.~Zwiebach,
  ``On the Riemann Tensor in Double Field Theory,''
  JHEP {\bf 1205} (2012) 126
  doi:10.1007/JHEP05(2012)126
  [arXiv:1112.5296 [hep-th]].


\end{thebibliography}
\end{document}